%% file: main.tex
\documentclass[10pt,twocolumn,letterpaper]{article}

\input{sec/notation}

\usepackage{cvpr}              % To produce the CAMERA-READY version
\usepackage{url}
\input{preamble}

\definecolor{cvprblue}{rgb}{0.21,0.49,0.74}
\usepackage[pagebackref,breaklinks,colorlinks,allcolors=cvprblue]{hyperref}
\usepackage{bm}
\usepackage{array}
\usepackage{siunitx}
\usepackage{comment}
\usepackage{threeparttable}
\usepackage{cases}

\usepackage{tikz}
\usetikzlibrary{decorations.fractals, spy}

\newlength{\imagewidth}

\def\paperID{11958} % *** Enter the Paper ID here
\def\confName{CVPR}
\def\confYear{2025}

\title{RainyGS: Efficient Rain Synthesis with Physically-Based Gaussian Splatting}

\author{Qiyu Dai$^{1}$\textsuperscript{*} \qquad Xingyu Ni$^{2}$\textsuperscript{*} \qquad Qianfan Shen$^{3}$\\  Wenzheng Chen$^{4,5}$\textsuperscript{\dag} \qquad Baoquan Chen$^{1,6}$\textsuperscript{\dag} \qquad Mengyu Chu$^{1,6}$\textsuperscript{\dag} \vspace{0.3em} \\
{\normalsize $^1$School of Intelligence Science and Technology, Peking University} \quad
{\normalsize $^2$School of Computer Science, Peking University} \\
{\normalsize $^3$School of EECS, Peking University} \quad {\normalsize $^4$Wangxuan Institute of Computer Technology, Peking University}\\
{\normalsize $^5$State Key Laboratory of Multimedia Information Processing, Peking University} \\ {\normalsize $^6$State Key Laboratory of General Artificial Intelligence, Peking University} \vspace{-1.5em}
}

\begin{document}

\twocolumn[{%
    \renewcommand\twocolumn[1][]{#1}%
    \maketitle  
    \begin{center}
        \centering
        \includegraphics[width=\textwidth]{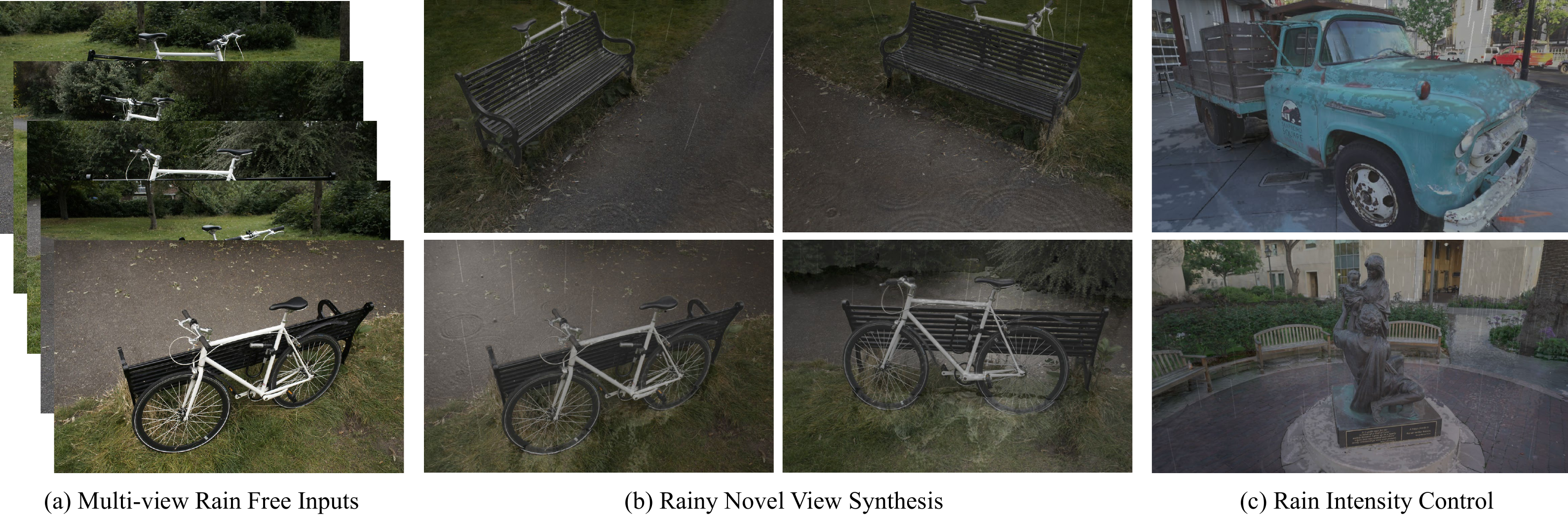} 
            \vspace*{-0.4cm}
        \captionof{figure}{ Using multi-view images as input (a),  {\name}  constructs 3D scenes with physically-based {\gs} techniques to enable efficient and photorealistic rain synthesis (b). Our approach provides users with flexible control over rain intensity, from light drizzle to heavy downpour (c), achieving high-quality, realistic rain effects in a computationally efficient manner. Zoom in for better visual effects.}
    \end{center}
}]

\renewcommand{\thefootnote}{*}\footnotetext{Joint first authors.}
\renewcommand{\thefootnote}{\dag}\footnotetext{Corresponding authors.}

\input{sec/0_abstract}    
\input{sec/1_intro_v3}
\input{sec/2_related_work_v2}

\input{sec/3_method}
\input{sec/4_experiments}

\input{sec/5_conclusions}

\section*{Acknowledgement}
This work was supported in part by the National Key R\&D Program of China under Grants 2022ZD0160802 (Xingyu Ni and Mengyu Chu) and 2022ZD0160801 (Qiyu Dai).
Wenzheng Chen also thanks the support from State Key Laboratory of General Artificial Intelligence.

{
    \small
    \bibliographystyle{ieeenat_fullname}
    \bibliography{main}
}


\end{document}

\title{RainyGS: Efficient Rain Synthesis with Physically-Based Gaussian Splatting (Supplementary Material)}

\author{Qiyu Dai}
\affiliation{
\institution{School of Intelligence Science and Technology, Peking University}
\country{China}
}
\authornote{joint first authors}

\author{Xingyu Ni}
\affiliation{
\institution{School of Computer Science, Peking University}
\country{China}
}
\authornotemark[1]

\author{Qianfan Shen}
\affiliation{
\institution{School of EECS, Peking University}
\country{China}
}

\author{Wenzheng Chen}
\affiliation{
\institution{Wangxuan Institute of Computer Technology, Peking University}
\country{China}
}
\authornotemark[2]

\author{Baoquan Chen}
\affiliation{
\institution{School of Intelligence Science and Technology, Peking University}
\country{China}
}
\authornotemark[2]

\author{Mengyu Chu}
\affiliation{
\institution{School of Intelligence Science and Technology, Peking University}
\country{China}
}
\authornote{corresponding authors}

\makeatletter
\let\@authorsaddresses\@empty
\makeatother

%  \centering
%  \includegraphics[width=\textwidth]{Figures/teaserv4.png}
%  \caption{Our new GARM-LS framework \rvv{allows} for highly detailed fluid simulation with surface tension on a relatively coarse grid. (Far Left) Rain drops falling into the pond. (Middle Left) Flow from a breaking dam. (Middle Right) A water ball falling through a Galton board. (Far Right) Splashes from a collision of two droplets.} 
% \label{fig:teaser}

\maketitle

\appendix

\input{src_sup/0_abstract}
\input{src_sup/1_aux_map}
\input{src_sup/2_rain_sim}
\input{src_sup/3_ssr}

\input{src_sup/4_results}

% Bibliography
\bibliographystyle{ACM-Reference-Format}
\bibliography{refs_sup}

\clearpage
\setcounter{page}{1}
\maketitlesupplementary

In this supplementary material, we present a

\input{sec_supp/X_suppl}

{
    \small
    \bibliographystyle{ieeenat_fullname}
    \bibliography{main}
}

%% file: sec/notation.tex
\newcommand{\name}{RainyGS}
\newcommand{\gs}{Gaussian Splatting}
\newcommand{\gsshort}{3DGS}
\newcommand{\qy}[1]{{\color{black}{#1}}}
\newcommand{\xy}[1]{{\color{black}{#1}}}

\newcommand{\methodgaussiansplattingshort}{{3DGS}}

\newcommand{\gshader}{{GaussianShader}}

\newcommand{\gofshort}{{GOF}}

\newcommand{\pgsr}{{PGSR}}

\newcommand{\curView}{V}
\newcommand{\rgbSrc}{I_s}
\newcommand{\depthMap}{D}
\newcommand{\normalMap}{N}
\newcommand{\heightMap}{H}

\newcommand{\threeDGaussian}{G}
\newcommand{\point}{\bm{x}}
\newcommand{\centerPos}{\bm{\mu}}
\newcommand{\threeDCov}{\bm{\Sigma}}
\newcommand{\threeDScaling}{\bm{S}}
\newcommand{\threeDRotation}{\bm{R}}

\newcommand{\opacity}{o}
\newcommand{\twoDCov}{\bm{\Sigma}'}
\newcommand{\Jacobian}{\bm{J}}
\newcommand{\extrinsic}{\bm{V}}
\newcommand{\twoDGaussianFromThreeDG}{G'}
\newcommand{\pixelColor}{C}
\newcommand{\weightBlending}{\alpha}

\newcommand{\del}{\bm{\nabla}}

%% file: preamble.tex
\newcommand{\cmy}[1]{{\color{blue}#1}}

\makeatletter
\renewcommand{\paragraph}{%
  \@startsection{paragraph}{4}%
  {\z@}{0.5\baselineskip \@plus 1ex \@minus .2ex}{-1em}%
  {\normalfont\normalsize\bfseries}%
}
\makeatother

%% file: sec/0_abstract.tex
\begin{abstract}

We consider the problem of adding dynamic rain effects to in-the-wild scenes in a physically-correct manner.
Recent advances in scene modeling have made significant progress, with NeRF and {\gs} techniques emerging as powerful tools for reconstructing complex scenes.
However, while effective for novel view synthesis, these methods typically struggle with challenging scene editing tasks, such as physics-based rain simulation.
%
% In contrast, traditional physics-based simulations can generate realistic rain effects, like raindrops and splashes, but they generally rely on skilled artists to carefully set up high-fidelity scenes. 
%
% This process lacks flexibility and diversity, limiting its scope for broader applications.
%
In contrast, traditional physics-based simulations can generate realistic rain effects, such as raindrops and splashes, but they often rely on skilled artists to carefully set up high-fidelity scenes. 
This process lacks flexibility and scalability, limiting its applicability to broader, open-world environments.
In this work, we introduce RainyGS, a novel approach that leverages the strengths of both physics-based modeling and {\gs} to generate photorealistic, dynamic rain effects in open-world scenes with physical accuracy.
At the core of our method is the integration of physically-based raindrop and shallow water simulation techniques within the fast {\gs} rendering framework, enabling realistic and efficient simulations of raindrop behavior, splashes, and reflections.
% 
% Our method supports efficient rainy effects synthesis at more than 10 fps, providing users with a flexible way to control rain intensity—from light drizzles to heavy downpours.
Our method supports synthesizing rain effects at over \qy{30} fps, offering users flexible control over rain intensity—from light drizzles to heavy downpours.
We demonstrate that RainyGS performs effectively for both real-world outdoor scenes and large-scale driving scenarios, delivering more photorealistic and physically-accurate rain effects compared to state-of-the-art methods. \qy{Project page can be found at \url{https://pku-vcl-geometry.github.io/RainyGS/}.}

\end{abstract}

%% file: sec/1_intro_v3.tex
\section{Introduction}
\label{sec:intro}

In this paper, we investigate how to add photorealistic rain effects to in-the-wild scenes, \ie, given multi-view images of an open-world scene captured under sunny or cloudy conditions, we aim to synthesize its corresponding rainy conditions in a photorealistic and physically-accurate manner.
Rain synthesis for open-world scenes plays a critical role in a wide range of applications, including AR/VR, gaming, robotics, and autonomous driving~\cite{visionpro,sun2020scalability}. 
%
% For example, it not only enables AR glasses to display realistic rain effects in virtual outdoor environments, enriching applications in entertainment and travel, but also allows embodied robots and autonomous vehicles to be trained in challenging rainy conditions, enhancing their safety and reliability~\cite{visionpro,sun2020scalability}.
%
However, achieving realistic rain effects is highly challenging, as it involves simulating complex, high-order physical and rendering phenomena. This includes the formation of rain streaks in the sky, water accumulation on surfaces, and realistic reflection and refraction effects, \etc.
Furthermore, these elements must appear concurrently, evolve dynamically over time, and be simulated efficiently to meet the demands of broader applications---further increasing complexity and making this problem largely underexplored.

The problem of in-the-wild rain synthesis is closely related to two important research domains: scene modeling and physics-based simulations.
In recent years, 3D scene modeling has made significant progress: techniques such as Neural Radiance Fields (NeRF) and 3D {\gs} ({\gsshort})~\cite{mildenhall2020nerf,kerbl20233d} have emerged as powerful tools, not only advancing novel view synthesis~\cite{barron2021mip,barron2022mip,barron2023zip} but also enabling more complex tasks like scene editing~\cite{wang2023fegr,wang2022neural}. 
%
%However, most existing works~\cite{boss2021nerd,munkberg2022extracting,jiang2023gaussianshader,R3DG2023} focus primarily on inverse rendering-based illumination editing, falling short of more advanced, physics-based editing tasks such as rain simulation.
\qy{However, most existing works fall short of more advanced, physics-based editing tasks such as rain simulation.}
Alternatively, physics-based simulations use physical principles to generate realistic rain effects \cite{li2023garm,da2016surface}, including raindrops and splashes. These methods, however, often rely on skilled artists to manually set up high-fidelity scenes, which limits their flexibility and applicability for in-the-wild contexts.
 Recently, ClimateNeRF~\cite{Li2023ClimateNeRF} and Gaussian Splashing~\cite{feng2024gaussian} have pioneered the integration of scene modeling and physics-based simulations to synthesize extreme weather or fluid effects, achieving impressive editing results. Nevertheless, the former focuses mainly on static weather modeling, % with  volumetric ray tracing, 
 while the latter employs inefficient particle simulations, making both approaches unsuitable for efficient and dynamic rain synthesis.

\begin{comment}
	
	To fill this gap, we propose {\name}, a physically-based 3D Gaussian Splatting  simulation framework that enables  effecient and photo-realisitc rain synthesis for in-the-wild scenes. 
	%
	{\name} leverages the strength of accurate physics-based rain simulation methods as well as efficient Gaussian Splatting Rendering framework, achving physically-accurate and fast rain simulation results.
	%
	Specifically, given multi-view images as input, we first apply 3DGS to model the scene and disentangle the corresponding geometry material, and illumination conditions. 
	%
	We then apply physics modelsing appraoches to add realsict rain effects, including the flying rain streaks, accumulated rain woater on the rgroud, and high-order effects such as relfection, reflections and specular rendering effects.
	%
	We carefully choose proper and efficient simulation techniques, utlizing shock wave to model the surface water and xxxx to model reflections.  
	% 
	Consequenclty, it supports precisely model all the necessary rainy effects like drops, water on the ground, splashes, reflection  jointly, providing free-view photorealisic modeling results for open-world scenes.
\end{comment}

To fill this gap, we propose {\name}, a physics-based 3D Gaussian Splatting simulation framework that enables efficient and photorealistic rain synthesis for in-the-wild scenes.
{\name} combines the accuracy of physics-based rain simulation methods with the efficiency of the {\gs} rendering framework, achieving both physically accurate and fast rain simulation results.
Specifically, given multi-view images as input, we first \qy{use 3DGS to reconstruct the scene's appearance and geometry.}
%use 3DGS to model the scene and disentangle its corresponding geometry, material, and illumination conditions.
% 
Next, we apply physics-based models to add rain effects, including flying rain streaks, accumulated water on the ground, and high-order effects such as reflections, refractions, and specular rendering, ensuring photorealistic and physically accurate renderings.  
%
\begin{comment}

\cmy{
	This is challenging due to Gaussian's limitation on lacking accurate geometry and its unsuitability for rendering reflections, owing to its splatting technology.
	%
	Our approach addresses these challenges by thoughtfully integrating high-fidelity and efficient simulation and rendering techniques.
	For the dynamics, we adapt shallow-water simulation \cite{chentanez2010real}, whose physical assumptions are particularly well-suited for rain scenarios and provide more accurate results than position-based dynamics. Utilizing height fields to model surface water dynamics not only circumvents the geometric limitations of GS but also ensures real-time efficiency.
	For rendering, we integrate screen-space ray tracing \cite{mcguire2014ray} with 3D GS to accurately render reflections of visible parts, which are most critical to visual experience and well-suited for rasterization pipelines.
}
% I feel it is important to explain why integrating shallow water and screen-space rendering is not a simple A+B, but a perfect fit for RainyGS
\end{comment}

Notably, these effects are extremely challenging in the vanilla 3DGS frameworks, 
while we re-implement them by carefully integrating high-fidelity and efficient simulation and rendering techniques.
For the dynamics, we adapt shallow-water simulation \cite{chentanez2010real}, whose physical assumptions are particularly well-suited for rain scenarios and provide more accurate results than position-based dynamics. Utilizing height fields to model surface water dynamics not only circumvents GS’s lack of internal geometric detail but also ensures efficiency.
For rendering, we integrate screen-space ray tracing \cite{mcguire2014ray} with 3D GS to accurately render reflections of visible parts, which are most critical to visual experience and well-suited for rasterization pipelines.
Consequently, {\name} enables precise modeling of all necessary rainy effects—such as raindrops, water on the ground, splashes, and reflections—providing photorealistic, free-view rain synthesis for in-the-wild scenes.

To the best of our knowledge, {\name} for the first time provides a unified, open-world rain synthesis framework, enabling efficient and realistic rain effects with physical accuracy.
Notably, it renders rain effects at a fast speed of more than \qy{30} fps and allows users to freely control rain intensity, from light drizzle to heavy downpour, achieving high-quality, realistic results in a computationally efficient manner.
We validate our method across diverse in-the-wild scenes, including MipNeRF360, Tanks and Temples dataset, and Waymo driving scenarios~\cite{Knapitsch2017,barron2022mip,sun2020scalability}, demonstrating superior simulation performance, high-fidelity preservation, and better user control compared to 2D-based editing methods and video generation baselines.

%% file: sec/2_related_work_v2.tex
\section{Related Work}
\label{sec:related_work}

\begin{figure*}
	\centering
	\includegraphics[width=1\linewidth]{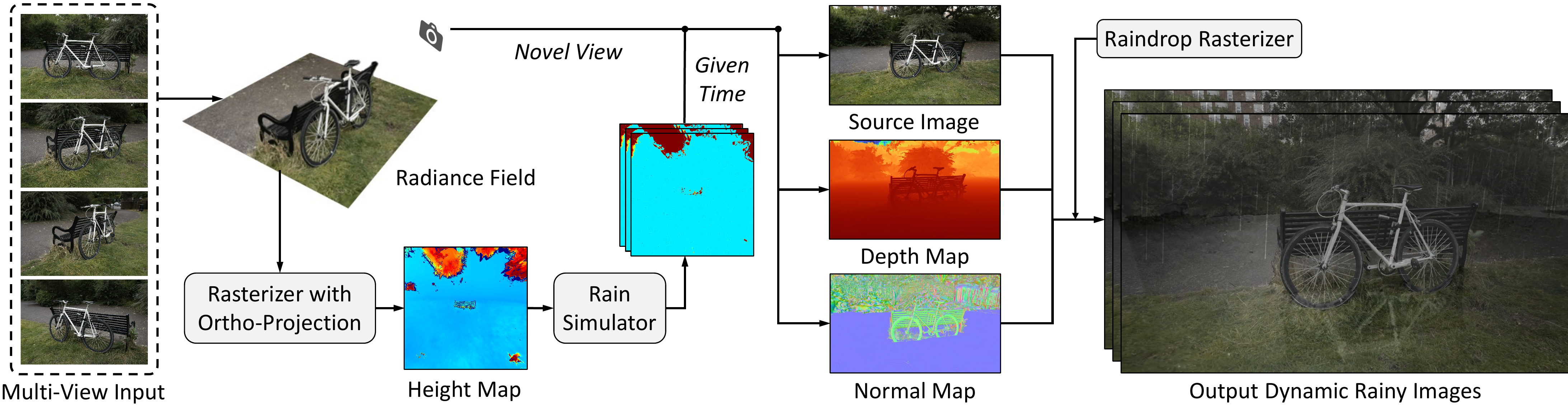}
	\caption{
    \label{fig:pipeline} 
    \textbf{The pipeline of {\name}}.  
    % Taken mulit-view images as input, we apply {\gshader}~\cite{jiang2023gaussianshader} and {\gof}~\cite{Yu2024GOF} to disentangle the appearance and lighting and extract geometry of the scene, respectively. Next, we acquire the height map and utilize the shallow-water simulation techniques to synthesize realsictic accumulated dynamic water on the ground with waves and splashes. To render novel views, we further prepare auxiliary maps, including the appreace, depth, and mormap maps, to synthesize rain streaks, reflection, and refraction effects using efficient screen-based raytracing techniques. And finally, we compose them all to present realsic, dynamic rainy effects shown in the right.
    %
    % Taken multi-view images as input, we apply {\gshader}~\cite{jiang2023gaussianshader} and {\gof}~\cite{Yu2024GOF} to disentangle the appearance and lighting, and to extract the geometry of the scene, respectively. Next, we generate the height map and use shallow-water simulation techniques to synthesize realistic dynamic water accumulation on the ground, incorporating waves and splashes. For rendering novel views, we prepare auxiliary maps, including appearance, depth, and motion maps, to synthesize rain streaks, reflections, and refraction effects using efficient screen-based ray tracing techniques. Finally, we combine all these elements to present realistic dynamic rainy effects, as shown on the right.
    Taken multi-view images as input, we apply \qy{{\pgsr}~\cite{chen2024pgsr} to recover both the appearance and geometry of the scene.} Next, we generate the height map and use shallow-water simulation techniques to synthesize realistic dynamic water accumulation on the ground, incorporating waves and splashes. For rendering novel views, we prepare auxiliary maps, including appearance, depth, and motion maps, to synthesize rain streaks, reflections, and refraction effects using efficient screen-based ray tracing techniques. Finally, we combine all these elements to present realistic dynamic rainy effects, as shown on the right.
    }
    \vspace*{-0.3cm}
\end{figure*}

\begin{comment}
% In this section, we first review the novel view synthesis (NVS) methods based on physically-based rendering. Next, we describe physically-based weather simulation. Lastly, we discuss 3D scene editing that utilizes Radiance Fields.

In this section, we review three related topics. 
We first talk about recent 3D scene modeling and editing works.
Next, we describe physically-based weather simulations. 
Lastly, we discuss their combinations: physically-based simulations meets  neural scene modeling and editing. 
\end{comment}

In this section, we review three related topics. 
First, we discuss recent advances in 3D scene modeling. 
Next, we describe physics-based weather simulations. 
Lastly, we explore the combination of these approaches: how physics-based simulations are integrated with neural scene modeling and editing.

\paragraph{3D Scene Modeling}
Recent advances in scene modeling have seen significant progress, driven by powerful 3D representations such as Neural Radiance Fields (NeRF) and 3D Gaussian Splatting (3DGS) techniques \cite{mildenhall2020nerf,kerbl20233d}.
While most of these methods are designed for novel view synthesis \cite{mildenhall2020nerf,kerbl20233d,barron2021mip,barron2022mip,barron2023zip,muller2022instant}, 
\qy{a variety of works explore surface reconstruction from implicit representations \cite{wangneus,li2023neuralangelo,guedon2024sugar,huang20242d,Yu2024GOF}, both of which are essential for physics-based applications, such as rain simulation.}
% a variety of works incorporate physics-based rendering (PBR) models to decouple the scene into geometry, material, and lighting attributes \cite{zhang2021physg,boss2021nerd,chen2021dib,munkberg2022extracting,hasselgren2022shape,jin2023tensoir,jiang2023gaussianshader,R3DG2023}, further enabling applications such as illumination editing and object insertion \cite{liang2024photorealistic,wang2023fegr,wang2022neural}.
% %
% However, these works primarily focus on inverse-rendering-based editing, leaving more advanced, physics-based applications---such as rain simulation---underexplored.
%
Alternatively, modeling dynamic rainy scenes belongs to the category of dynamics modeling \cite{attal2023hyperreel,fridovich2023k,cao2023hexplane,luiten2023dynamic,wu20234d,yang2023real,duan20244d}. However, this task is more challenging than regular 4D scenes, as rainy scenes involve various high-order physical and rendering effects, such as flying streaks and waving reflections.

\paragraph{Weather Simulation}
Climate simulations play a crucial role in a wide range of applications, from entertainment to agriculture and city modeling.
Traditional physics-based weather simulations typically rely on computer graphics techniques, such as those used in capillary wave simulations, smoke simulation, and snow simulation in wind using metaballs and fluid dynamics \cite{bridson2015fluid}. 
For example, Fournier and Reeves achieve excellent capillary wave simulations using simple Fourier transform methods \cite{fournier1986simple}, while \citet{fedkiw2001visual} simulates smoke, and others \cite{stomakhin2013material,gissler2020implicit} simulate snow in wind.
Our method applies the shallow water method \cite{chentanez2010real} due to its efficiency and demonstrates how to benefit from such simulations while retaining the strong scene modeling properties of 3D representations.
Alternatively, many works explore data-driven approaches to synthesize weather effects from image or video collections. 
For instance, \cite{Visualizing} collects climate image datasets and performs image editing with CycleGAN \cite{zhu2017unpaired}.
\cite{schmidt2022climategan} leverage depth information to estimate water masks and perform GAN-based image editing and inpainting.
Methods like \cite{hahner2019semantic} simulate fog and snow. While these approaches offer realistic effects for single images, they do not provide immersive, view-consistent climate simulations.

\paragraph{Physics-Based Scene Editing}
As noted in \cite{Li2023ClimateNeRF}, physical simulations provide accurate dynamic predictions, while neural modeling approaches excel at modeling in-the-wild scenes. 
Therefore, combining these methods offers promising performance for open-world scene editing.
Li et al. \cite{Li2023ClimateNeRF} pioneered in syntheszing harsh weather conditions using neural fields, employing physics-based weather modeling and rendering pipelines to produce accurate and photorealistic effects for floods, smog, and snow. However, their method is still limited to static weather conditions.
Recent advances in 3D Gaussian Splatting have inspired accompanying physics-based modeling works, treating each Gaussian point as a physics particle.
PhysGaussian \cite{xie2024physgaussian} integrates physically grounded Newtonian dynamics with 3D Gaussians, achieving high-quality motion synthesis.
Gaussian Splashing \cite{feng2024gaussian} combines physics-based animations of solids and fluids with 3D Gaussian Splatting to create novel fluid visual effects.
Inspired by these works, we apply physics-based modeling within Gaussian Splatting to efficiently synthesize realistic and dynamic rainy effects.

%% file: sec/3_method.tex
\section{Methods}
\label{sec:method}

\begin{comment}

In this section, we begin by reviewing 3D Gaussian Splatting (3DGS) in \S\ref{sec:recap_3dgs}.
%
Next, in \S\ref{sec:aux_map}, we introduce how to reconstruct 3DGS scenes with high accuracy geometry, which allows us to extract various auxiliary maps, including the correspinding depth, normal and height maps, prepared  for the rain simulaiton.
% 
Subsequently, in \S\ref{sec:rain_sim}, we present our proposed height map-based shallow-water simulation, which is designed for representing dynamic accumulated water on the surface for rainy conditions.
% 
Lastly, in \S\ref{sec:water_rasterize}, we talk about how to synthesize the high-order effects such as reflection, refraction, flying rain streaks with the proposed Reflection-Aware Water Rasterization method.
%
The pipeline of our methid is visualized in Fig.~\ref{fig:pipeline}.
\end{comment}

In this section, we first review 3D Gaussian Splatting (3DGS) in \S\ref{sec:recap_3dgs}.
Next, in \S\ref{sec:aux_map}, we introduce how to reconstruct 3DGS scenes with high-accuracy geometry, which enables the extraction of various auxiliary maps, including depth, normal, and height maps, prepared for rain simulation.
In \S\ref{sec:rain_sim}, we present a height map-based shallow-water simulation, which is designed to represent the dynamic accumulation of water on the surface under rainy conditions.
Finally, in \S\ref{sec:water_rasterize}, we discuss how to synthesize high-order effects such as reflections, refractions, and flying rain streaks using the proposed Reflection-Aware Water Rasterization method.
The pipeline of our method is visualized in Fig.~\ref{fig:pipeline}.

\subsection{Preliminary: 3D Gaussian Splatting}
\label{sec:recap_3dgs}

3D Gaussian Splatting ({\methodgaussiansplattingshort})~\cite{kerbl20233d} has demonstrated real-time, state-of-the-art rendering quality across a wide range of open-world scenes. 
%
%Its use of Gaussian points makes it particularly well-suited for physics-based simulations\cite{gross2011point,xie2024physgaussian,feng2024gaussian}.
% mengyu: ours is not using its particle-based representation.
%
This method represents a scene using a dense cluster of $N$ anisotropic 3D Gaussian ellipsoids. 
Each Gaussian is defined by a 3D covariance matrix $\threeDCov$ and its center position $\centerPos$: 
\begin{equation} 
	\label{eq:3DGS} 
	\threeDGaussian(\point) = e^{-\frac{1}{2} (\point-\centerPos)^\top \threeDCov^{-1} (\point-\centerPos)}. 
\end{equation} 
To ensure that the covariance matrix remains positive semi-definite during optimization, $\threeDCov$ is decomposed into a scaling matrix $\threeDScaling$ and a rotation matrix $\threeDRotation$, which describe the geometry of the 3D Gaussian: 
\begin{equation} 
	\threeDCov = \threeDRotation \threeDScaling \threeDScaling^\top \threeDRotation^\top, 
\end{equation} 
where $\threeDScaling = \text{diag}(s_x, s_y, s_z) \in \mathbb{R}^3$ and $\threeDRotation \in \mathrm{SO}(3)$ are represented by a 3D vector and quaternion, respectively. In addition to the position $\centerPos$, $\threeDScaling$, and $\threeDRotation$, each Gaussian also includes learnable parameters such as opacity $\opacity \in (0, 1)$ and spherical harmonic (SH) coefficients in $\mathbb{R}^k$, which encode view-dependent color information. % ($k$ corresponds to the SH order). %During optimization, 3DGS adaptively adjusts the Gaussian distribution by splitting and cloning Gaussians in regions with large view-space positional gradients, and by culling nearly transparent Gaussians.

Efficient rendering and parameter optimization in {\methodgaussiansplattingshort} are facilitated by a differentiable tile-based rasterizer. 
First, 3D Gaussians are projected into 2D space by computing the camera-space covariance matrix $\twoDCov= \Jacobian \extrinsic \threeDCov \extrinsic^T \Jacobian^T$, where $\Jacobian$ is the Jacobian matrix for the affine approximation of the projection transformation, and $\extrinsic$ is the extrinsic camera matrix. 
The color of each pixel on the image plane is then determined by blending Gaussians based on their depths: 
\begin{equation} 
	\pixelColor = \sum_{i=1}^N c_i \weightBlending_i \prod_{j=1}^{i-1} (1-\weightBlending_j), 
\label{eq:pixel_color}
\end{equation}
 where $c_i$ is the color of the $i$-th 3D Gaussian $\threeDGaussian_i$, and $\weightBlending_i = \opacity_i \twoDGaussianFromThreeDG_i$, with $\opacity_i$ and $\twoDGaussianFromThreeDG_i$ representing the opacity and 2D projection of $\threeDGaussian_i$, respectively. For more details, please refer to {\methodgaussiansplattingshort}~\cite{kerbl20233d}.

% While efficitive for novel view thesis, the vanilla 3DGS is not designed for inverse rendering. Moreover, its geometry quality is often undermined by the free moving Gaussian points~\cite{}. To extend its capibility, a branch of work extend it using geometric constraints and PBR formuas to extract the geometry, material and ligting from the scnenes.  Here we choose to use two state-of-the-art methods, GaussianShader~\cite{jiang2024gaussianshader} for extracting materials, SVBRFD and Gaussian Opacity Fields (GOF)~\cite{Yu2024GOF} for acqiring precise geometry, which are essential for the rain simulation.  We provide the detials in the next section. 

\qy{While effective for novel view synthesis, vanilla 3D Gaussian Splatting (3DGS) is not optimized for geometry reconstruction, as its geometric quality is often compromised by the free movement of Gaussian points. Several recent works have extended its capabilities using geometric constraints to jointly recover both geometry and appearance from scenes. 
Here we adopt the state-of-the-art method, {\pgsr}~\cite{chen2024pgsr}, to obtain high-precision geometry and high-fidelity rendering, both of which are critical for rain simulation. The details are provided in the next section.}

\subsection{Auxiliary Map Extraction}
\label{sec:aux_map}

\begin{comment}
	Realistic Rain Simulation relies on accurate scene modeling and decomposition, \ie, acqirring the precise geometry, material and lighting of the scene from the multi-view input images.  
	%
	To achive this, we adopt GaussianShader~\cite{jiang2024gaussianshader} as the appearance module to recover PBR attributes of the scene. 
	%
	However, GaussianShader is not designed for accurate geometry reconstruction, which is crucial for simulating interactions between fluids and solid objects. 
	%
	Therefore, we further employ Gaussian Opacity Fields (GOF)~\cite{Yu2024GOF}  as the geometry module to extract detailed scene geometry. 
	%
	To ensure alignment between the two modules, we utilize multi-view depth maps from the GOF model to supervise the training of GaussianShader. 
	%
	As shown in Fig.~\ref{fig:abla-gof}, GaussianShader achieves significantly improved rendering quality under supervision by GOF depth maps. 
	
\end{comment}

\paragraph{Scene Modeling}

\qy{Realistic rain simulation relies on accurate scene modeling,
\ie, recovering precise scene geometry and appearance from multiview images.
To achieve this, we adopt PGSR~\cite{chen2024pgsr} as a unified geometry and appearance module, providing both high quality and computational efficiency.
In addition, we incorporate normal priors from a pretrained monocular normal estimation model~\cite{bae2024rethinking} to supervise rendered normal maps, significantly improving geometric quality (see Supp). Based on the recovered scene, we then generate auxiliary maps for downstream simulation.

Alternatively, scene modeling can be achieved by combining decomposable radiance fields (e.g., GaussianShader\cite{jiang2024gaussianshader}) for material and lighting extraction with surface reconstruction methods (e.g., GOF \cite{Yu2024GOF}) for geometry recovery.
In comparison, PGSR offers superior quality and efficiency, although it sacrifices albedo (Environment map can still be replaced), resulting in slightly dropped relighting effects.
A comparison of both approaches is provided in Supp.}

\paragraph{Auxiliary Map Preparation}
The proposed rain simulation method requires a height map and several auxiliary maps, which are acquired from the scene modeling process.
\qy{Specifically, we rasterize PGSR from a top-down orthographic view to generate a height map, which is then used in the proposed shallow-water simulation method presented in \S\ref{sec:rain_sim}.
In addition, we render appearance, depth, and normal maps, which are leveraged to achieve physically realistic rain rendering effects, as detailed in \S\ref{sec:water_rasterize}.

To prepare the height map $\heightMap$, we align the 3DGS point cloud with the $XY$ plane, focusing on the ground plane of the scene. 
We apply Principal Component Analysis (PCA) to decide the location of the $XY$ ground plane. 
Then, using an orthographic camera facing in the negative $Z$-direction, we render a depth map from the scene to the camera.
The height map is obtained by subtracting the depth values from the camera’s distance above the $XY$ ground plane.
}

% \qy{(todo)Specifically, we extract the corresponding meshes from {\gofshort} to create the height map, which is then used in the proposed shallow-water simulation method presented in Sec.\ref{sec:rain_sim}.
% Additionally, we render the appearance, depth, and normal maps from {\gshader}, which are utilized to create physically realistic rain rendering effects (Sec.\ref{sec:water_rasterize}).

% To prepare the height map $\heightMap$, we extract the mesh and align it with the $XY$ plane, focusing on the ground plane of the scene.
% % Using Principal Component Analysis (PCA), we centralize the translation components of all camera poses, aligning their centroids with the origin.
% % The eigenvector matrix of the translation components is applied as a rotation matrix, aligning the principal components of the cameras with the $XYZ$ coordinate axes.
% We apply Principal Component Analysis (PCA) to decide the location of the $XY$ ground plane, and uniformly sample 3D points over it.
% %
% These points represent the pixels of the height map, where each pixel's height value is determined by calculating the longest distance in the $Z$ direction from the sampled point to the mesh.}
%
This process generates the height map required for simulating rain.
After rain simulation in \S\ref{sec:rain_sim}, we obtain the height map of the scene with rain. 
We then model the rain as isotropic 3D Gaussian spheres, densely and uniformly distributed over the height map. The position of each sphere corresponds to the 3D coordinates of pixels in the height map, while its radius and opacity are configurable hyperparameters. 
More details are provided in Supp.

The proposed rain rendering method (described in \S\ref{sec:water_rasterize}) requires per-pixel depth and normal maps for the current view to compute screen-space ray transmission. 
Therefore, we employ alpha blending techniques (Eq.~\eqref{eq:pixel_color}) to blend the depths and normals of the 3D Gaussian points, producing a depth map $\depthMap$ and a normal map $\normalMap$ for the current view $\curView$. These serve as the base layers for rasterizing rainy effects.

\begin{figure*}\footnotesize
  \centering
  \setlength{\imagewidth}{0.33\textwidth}
  \subcaptionbox*{\centering (a) $I_\mathrm{src}$}{\includegraphics[width=\imagewidth]{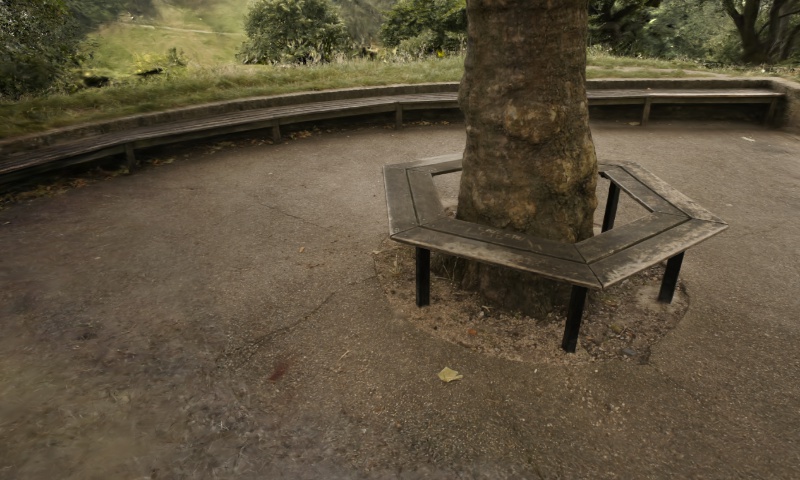}}
  \hfill
  \subcaptionbox*{\centering (c) $I_\mathrm{src}+I_\mathrm{spec}+I_\mathrm{refra}$}{\includegraphics[width=\imagewidth]{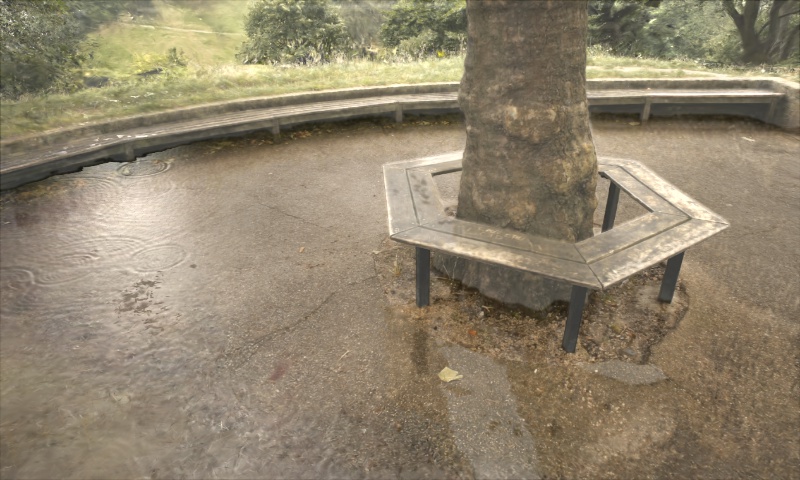}}
  \hfill
  \subcaptionbox*{\centering (e) $I_0$ in Eq.~\ref{eqn:I_0}}{\includegraphics[width=\imagewidth]{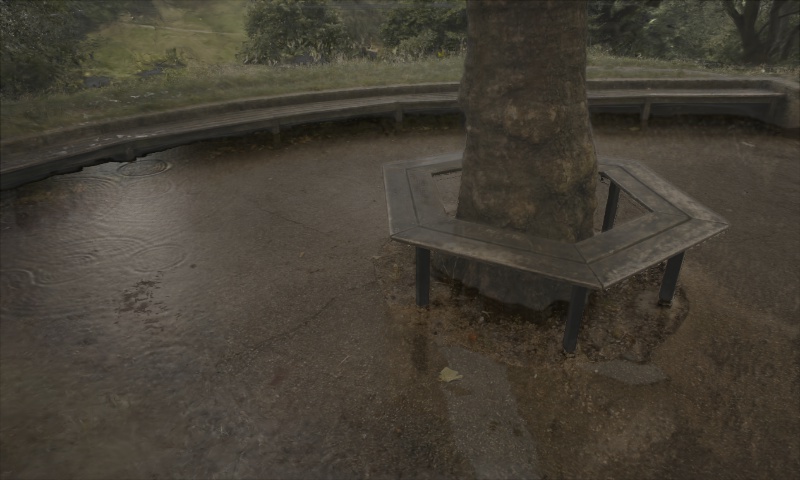}}
  \\
  \subcaptionbox*{\centering (b) $I_\mathrm{src}+I_\mathrm{spec}$}{\includegraphics[width=\imagewidth]{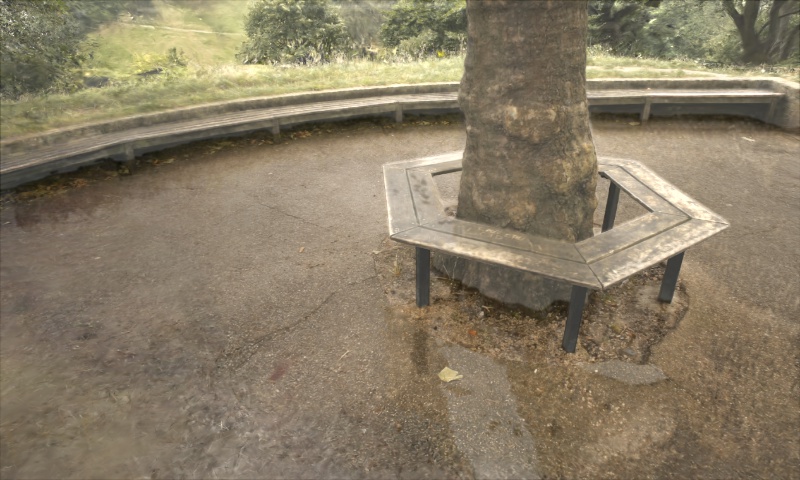}}
  \hfill
  \subcaptionbox*{\centering (d) $I_\mathrm{src}+I_\mathrm{spec}+I_\mathrm{highl}+I_\mathrm{refra}$}{\includegraphics[width=\imagewidth]{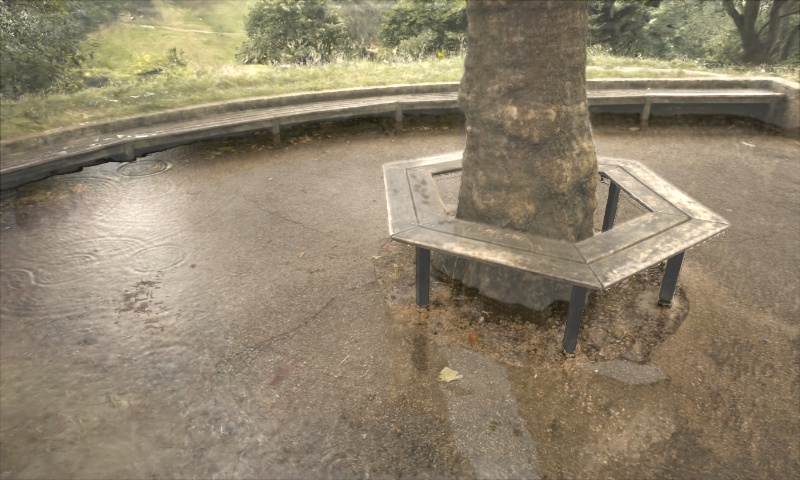}}
  \hfill
  \subcaptionbox*{\centering (f) The full method, blending $I_0$ and $I_1$, as in Eq. 8}{\includegraphics[width=\imagewidth]{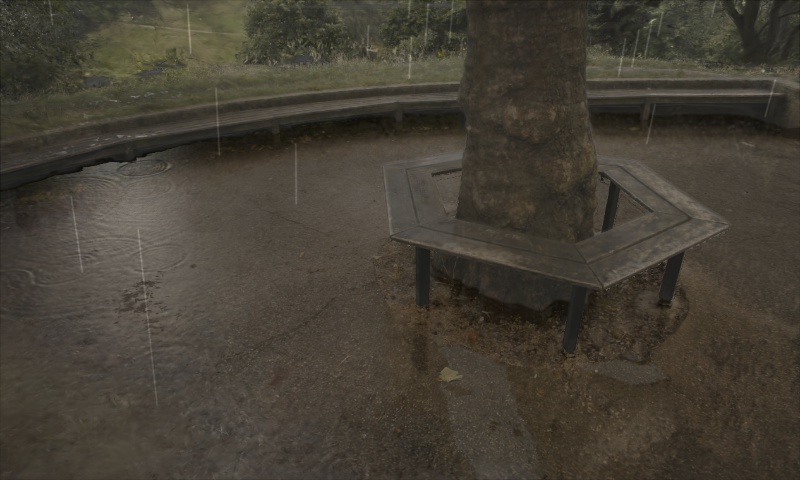}}
  \caption{\label{fig:ablation}
  %
  % We show the effects of each proposed modules. Starting with a user-specific view(a), we first utilizes hiegh map to synthesize rain water and add specular effects(b), we then add the highlight and reflection effects using screen-based ray tracing (c \& d), we also employ frenel term to adjust the rainy tones, making it dark and realistic (e). Finally, we compose with flying streaks to present  final rainy effects.
  We demonstrate the effects of each proposed module. Starting with a user-specific view (a), we first use the height map to synthesize rainwater and add specular reflection effects (b) using screen-based ray tracing. Next, we incorporate highlight (c) and refraction effects (d). We also apply the Fresnel term to adjust the rainy tones, making them darker and more realistic (e). Finally, we compose the scene with flying streaks to present the final rainy effects (f).
  }
  \vspace*{-0.3cm}
\end{figure*}

\subsection{Rain Simulation on Height Maps}
\label{sec:rain_sim}
In the context of rainy scenes, we are primarily concerned with phenomena such as water accumulation on the ground, ripples formed by falling raindrops, and large-scale flooding. In these phenomena, the water body can be well approximated using a height field. This  allows us to perform physics-based simulations of rainfall on a 2D height map,
achieving high efficiency by bypassing inner geometric details, which are often problematic due to their invisibility in 3DGS.

Concretely speaking, the simulation is based on shallow water equations (SWEs)
\begin{gather}
  \frac{\partial h}{\partial t}+(\bm{u}\cdot\del)h=-h(\del\cdot\bm{u})\text{,}\\
  \frac{\partial\bm{u}}{\partial t}+(\bm{u}\cdot\del)u=-g\del h\text{,}
\end{gather}
which is widely utilized for height-based fluid simulation in graphics \cite{bridson2015fluid}.
Here, $\bm{u}$ denotes the horizontal velocity of water surface, and $h$ is the depth of water such that $\eta=H+h$ is the height field of the scene with water covered.
To achieve the balance between efficiency and stability, we adopt the explicit numerical scheme proposed by \citet{chentanez2010real},
where the velocity is advected with the semi-lagrangian method \cite{staniforth1991semi} but the height field is treated with a first-order upwind scheme. Thus the law of mass conservation is naturally satisfied. See the supplemental material for details.

During each time step, new raindrops are randomly generated on the sky within the concerned domain, whose density, velocity, and direction are user-defined. The radius of the $i$-th raindrop $r_i$ follows an exponential distribution.
We simulate a simple fall down of raindrops and perform collision detection as follows:
\xy{\begin{enumerate}
\item Iterate over all raindrops and explicitly update the position of the $i$-th raindrop using
\begin{equation}
\bm{x}_i \gets \bm{x}_i + \bm{u}_i \Delta t\text{,}
\end{equation}
where $\bm{x}_i = (x_i, y_i, z_i)$ is the position vector, $\bm{u}_i = (u_i, v_i, w_i)$ ($w_i < 0$) is the velocity vector, and $\Delta t$ is the time step size.
\item If the raindrop's height is less than or equal to the ground height at its horizontal position ($z_i \leq \eta(x_i, y_i)$), the raindrop disappears and contributes to surface water accumulation. Its volume is added to the height field as follows:
\begin{equation}
h_{i,j} \gets h_{i,j} + \frac{4\pi r_i^3}{3(\Delta x)^2}\text{,}
\end{equation}
where $h_{i,j}$ is the height field value at the pixel corresponding to $(x_i, y_i)$, $r_i$ is the raindrop radius, and $\Delta x$ is the real-world scale of a single pixel in the height map.
\item If the raindrop's height is greater than the ground height at its horizontal position ($z_i > \eta(x_i, y_i)$) but less than or equal to the occlusion height at the same position ($z_i \leq H'(x_i, y_i)$), the raindrop disappears without contributing to the height field.
\item Repeat from Step 1 until all raindrops have been processed.
\end{enumerate}}
% \begin{itemize}
%   \item If the height of a raindrop is below the height field value 
%   (i.e., $\eta$) at its horizontal location, the raindrop vanishes, and its volume is added to the height field:
%   \begin{equation}h\gets h+4\pi r_i^3/3\Delta S\text{,}\end{equation}
%   in which $\Delta S$ is the area represented by each height value.
%   \item Otherwise, if the height of the raindrop is lower than the occlusion height field value $H'$ within a small threshold, the raindrop vanishes and makes a splash.
% \end{itemize}
Note that in the initial stage of each given scene, we obtain two height maps $H$ and $H'$ at two horizontal layers that represent the ground and occlusion, respectively.

\subsection{Reflection-Aware Water Rasterization}
\label{sec:water_rasterize}

In line with the splatting approach of 3DGS, we adopt a rasterization-based strategy for rendering the water and the rain.
The rasterization contains two passes.
The first pass renders reflection, $I_\mathrm{spec}$, and refraction, $I_\mathrm{refra}$, of the water surface together using the following equation:
% , of which the first is given as
% Equation
%% 原图segmentation
% I_{s,bg} = I_s * M_{bg}, I_{s,fluid} = I_s * M_{fluid}
% 
%% Refraction (Refra: 屏幕空间折射, Spec: bling-phong高光)
% I_{refra} = Refra(I_{s,fluid}) + Spec(N,D,V,LightDir) * M_{fluid}
% 
%% Reflection (Refle: 屏幕空间反射)
% I_{refl} = I_{0.2} + Refl(M_{fluid},I_s)
% 
%% fresnel项计算 (D深度，N法线，V视角)
% I_{fresnel} = fresnel(D,N,V) * M_{fluid}
% 
%% 反射+折射，与fresnel项加权
% I_{refl_refra} = I_{s,bg} + I_{refl} * (1 - I_{fresnel}) + I_{refra} * I_{fresnel}
% 
%% 与rain blend (depth_mask: 原始场景深度D与雨D对比大小的mask，小的为1)
% I = I_{refl_refra} * (1 - I_{depth_mask}) + I_{raindrop_rainhit} * I_{depth_mask}
% 
\begin{equation}
I_0(u,v) = (1-F)I_\mathrm{refra}+F(I_\mathrm{spec}+I_\mathrm{highl})\text{,}
\label{eqn:I_0}
\end{equation}
where $u,v$ are the horizontal coordinates of height fields, $F$ follows the Fresnel equation, and $I_\mathrm{highl}$ represents the glossy reflection of the sun, extracted from the environment map~\cite{feng2024gaussian}. 
% calculating the image with reflection and refraction.
% We denote its corresponding depth map $d_0(u,v)$.
The second pass rasterizes raindrops as $I_1(u,v)$, which includes both rain streaks and splashes.
% whose depth map is $d_1(u,v)$.
Finally, $I_0$ and $I_1$ are blended together based on their depths in the view, denoted as $d_0(u,v)$ and $d_1(u,v)$, respectively, using screen-space depth information:
\begin{subnumcases}{I(u,v)=}
  I_0(u,v)\text{,}&$d_0(u,v)<d_1(u,v)$,\\
  I_1(u,v)\text{,}&$d_0(u,v)>d_1(u,v)$.
\end{subnumcases}
We will describe the calculation of each term in the rendering passes individually.
The ablation study of these terms is demonstrated in Fig.~\ref{fig:ablation}.

\paragraph{Specular Reflection}
Wet surfaces behave like mirrors.
Considering that the surface is bumpy due to ripples, it is difficult to exactly compute the reflected colors without ray tracing, which is highly expensive when there are a lots of Gaussian ellipsoids.
However, the screen-space reflection technique can be used to address the problem to a large extent \cite{mcguire2014ray}, where the ray marching steps are only taken upon the rasterized images.
We use $I_\mathrm{spec}$ to denote the specular reflection term, which can be regarded as a function of the RGB image, depth map, normal map, and camera parameters.
See the supplemental materials for detailed algorithms.

\paragraph{Highlights}
% For each input scene, we extract the direction of sun from the environment map. 
We extract a directional light source, representing the sun, from the scene's environment map. The sun emits directional light and generates glossy highlights on the wet surface, which is calculated using the Blinn–Phong shading model \cite{blinn1977models}:
\begin{equation}
  I_\mathrm{highl}(u,v)=(\bm{n}\cdot\bm{h})^p\text{,}
\end{equation}
where $\bm{n}$ is the surface normal, $\bm{h}(u,v)$ is the half vector of the light direction $\bm{l}(u,v)$ and the view direction $\bm{v}(u,v)$. $p$ denotes the shininess.

\paragraph{Refraction}
Unlike refraction in the ocean, which is very pronounced, standing water on the rainy ground does not drastically refract light due to its shallow depth. This inspires us to use an image-based approximation \cite{pharr2005gpu}:
\begin{equation}
  I_\mathrm{refra}(u,v)=I_\mathrm{src}(u+n_uk(u,v),v+n_vk(u,v))\text{,}
\end{equation}
which distorts the source image based on the normal components projected onto the screen (i.e., $n_u$ and $n_v$).
Here $k(u,v)$ is proportional to the water depth at pixel $(u,v)$.

\paragraph{Fresnel Term}
To determine the intensity ratio of both reflection and refraction, the Fresnel--Slick equation is utilized to calculate the reflection rate \cite{schlick1994brdf}, which reads
\begin{equation}
  F(u,v)=F_0+(1-F_0)(1-\bm{h}\cdot\bm{v})^5\text{.}
\end{equation}
Here $F_0$ is the base reflectivity.

\paragraph{Raindrops}
We represent rain streaks formed by the motion of raindrops using multiple connected Gaussian ellipsoids, and represent the splash created when a raindrop hits the ground or an obstacle with a single Gaussian,
\xy{whose center is set at the raindrop’s impact point in the scene, with its scale determined by the raindrop's volume ($|\Sigma|=V^2$). The longest axis aligns with the projection of the raindrop's velocity, $v$, onto the tangent plane at the impact point ($e\propto v-(v\cdot n)n$).
This serves as a lowest-order approximation of the splash effect.} \qy{we further add rain streak color adjustment according to the base image’s normalized brightness, which enhances the realism of rain streaks under varying lighting.}

%% file: sec/4_experiments.tex
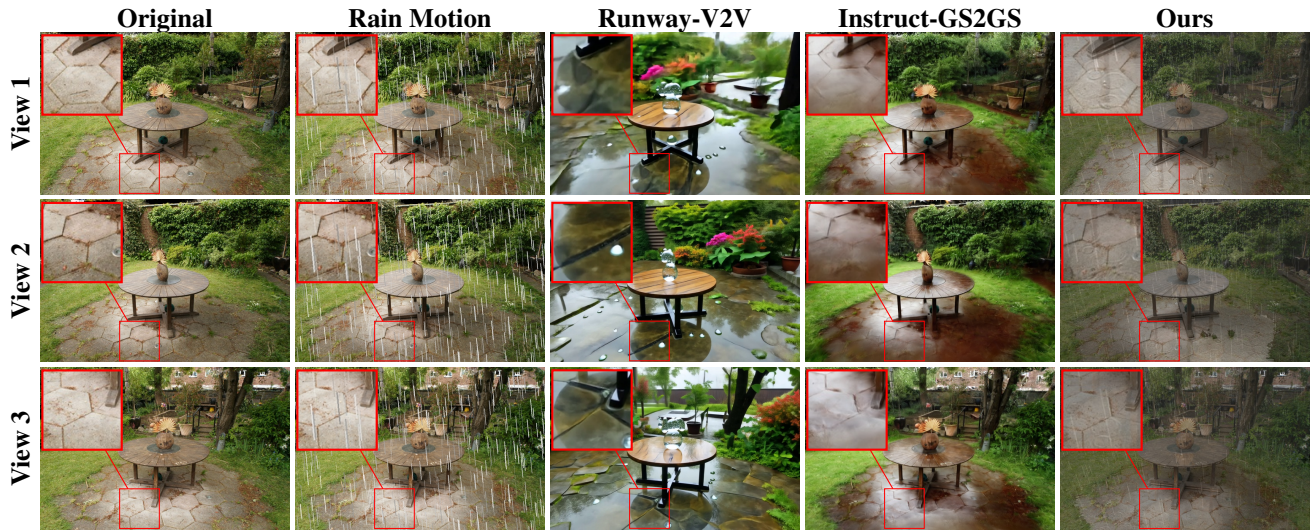
\begin{figure*}
    \centering
    \setlength{\tabcolsep}{1pt}
    \setlength{\imagewidth}{0.19\textwidth}
    \renewcommand{\arraystretch}{0.6}
    \newcommand{\formattedgraphics}[1]{%
      \begin{tikzpicture}[spy using outlines={rectangle, magnification=2, connect spies}]
        \node[anchor=south west, inner sep=0] at (0,0){\includegraphics[width=\imagewidth]{#1}};
        \spy [red,size=30pt] on (.4\imagewidth,.085\imagewidth) in node at (.17\imagewidth,.48\imagewidth);
        \end{tikzpicture}%
    }
    \begin{tabular}{m{0.38cm}<{\centering}m{\imagewidth}<{\centering}m{\imagewidth}<{\centering}m{\imagewidth}<{\centering}m{\imagewidth}<{\centering}m{\imagewidth}<{\centering}}
        & \textbf{Original} & \textbf{Rain Motion} & \textbf{Runway-V2V} & \textbf{Instruct-GS2GS} & \textbf{Ours} \\
        \rotatebox{90}{\textbf{View 1}} &
        \formattedgraphics{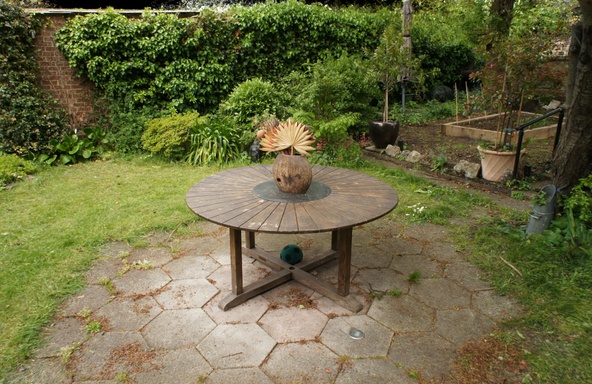} & 
        \formattedgraphics{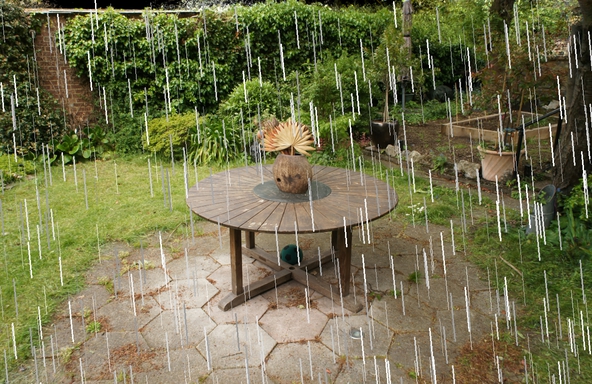} & 
        \formattedgraphics{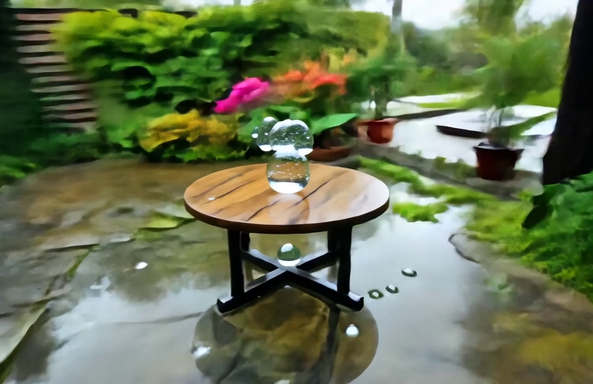} &
        \formattedgraphics{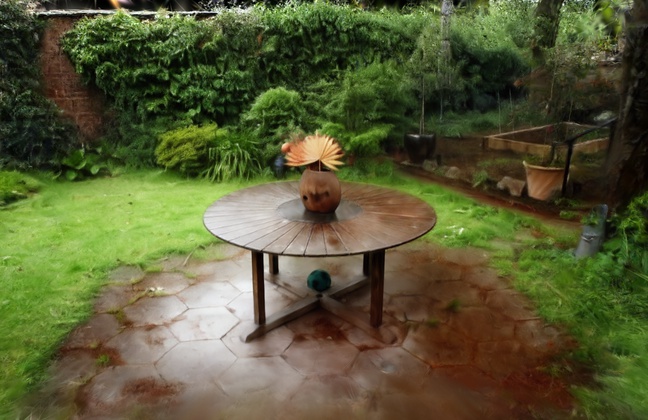} & 
        \formattedgraphics{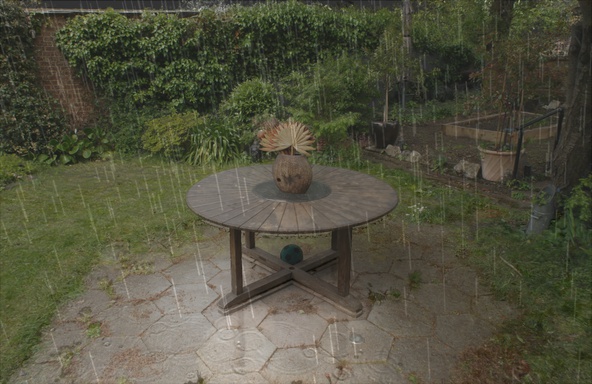} \\
        \rotatebox{90}{\textbf{View 2}} &
        \formattedgraphics{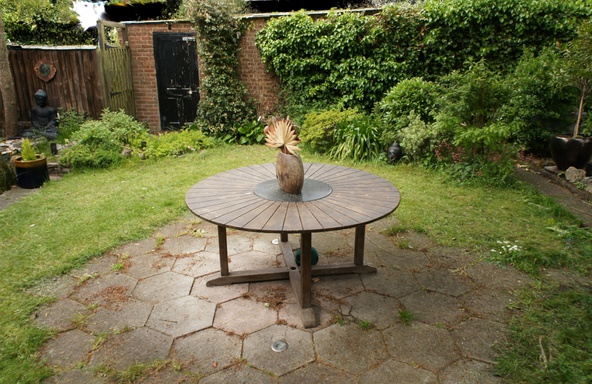} & 
        \formattedgraphics{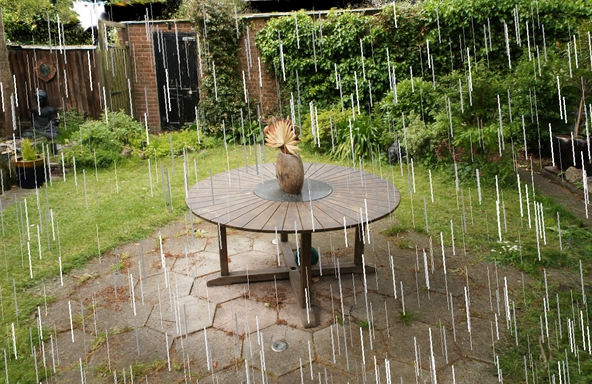} & 
        \formattedgraphics{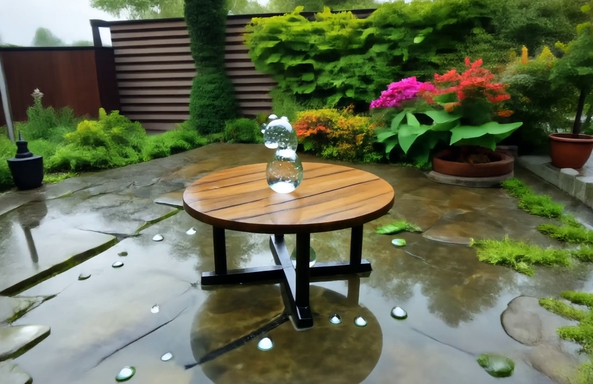} &
        \formattedgraphics{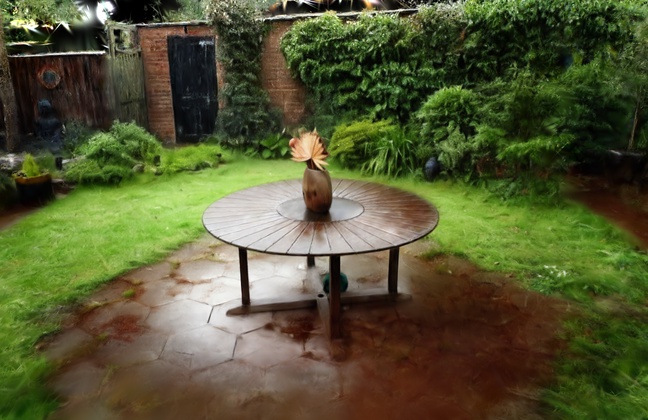} & 
        \formattedgraphics{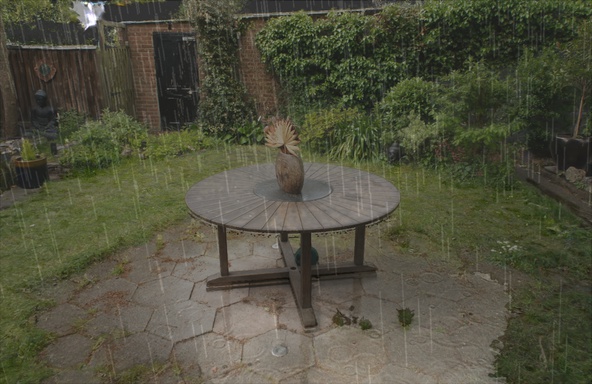} \\
        \rotatebox{90}{\textbf{View 3}} & 
        \formattedgraphics{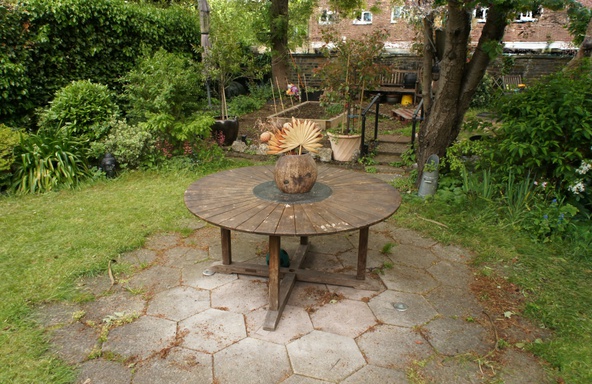} & 
        \formattedgraphics{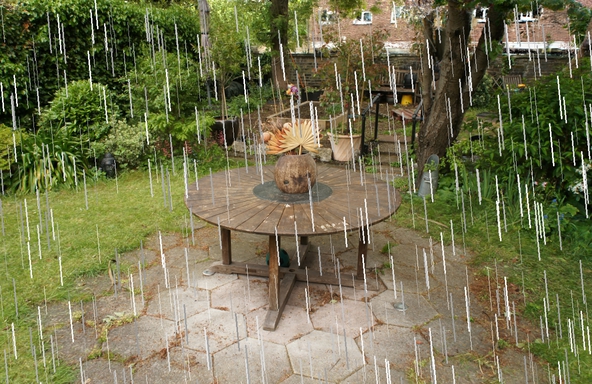} & 
        \formattedgraphics{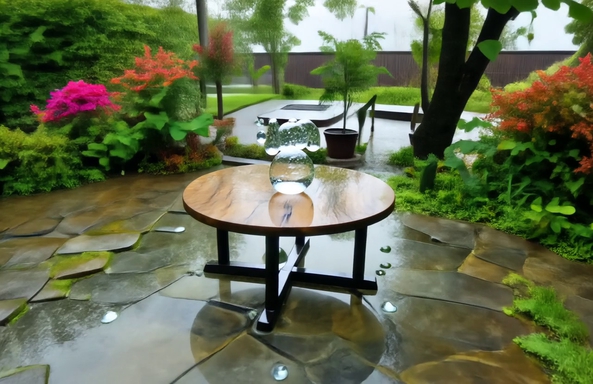} &
        \formattedgraphics{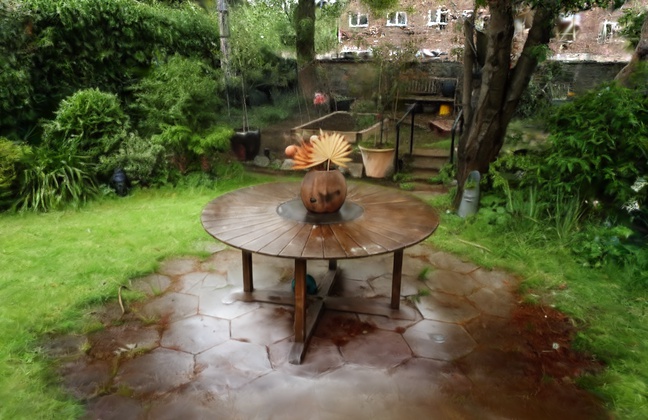} & 
        \formattedgraphics{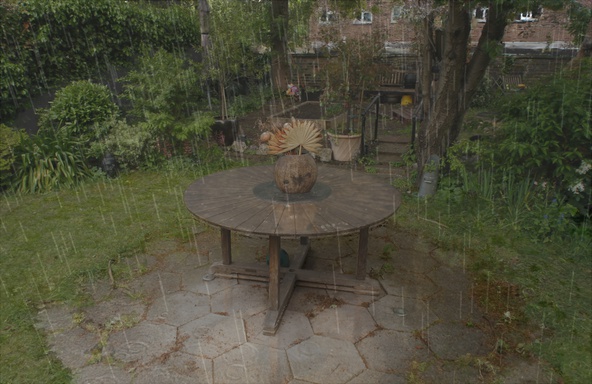} \\
    \end{tabular}
    \caption{\label{fig:fixed_time} Rain synthesis results at the same timestep from multiple views. Runway-V2V fails to maintain 3D consistency, while Rain Motion \qy{and Instruct-GS2GS} are unable to generate realistic rain streaks and puddles. RainyGS maintains 3D consistency and produces realistic rain effects.}
\end{figure*}

\begin{figure*}
    \centering
    \setlength{\tabcolsep}{1pt}
    \setlength{\imagewidth}{0.19\textwidth}
    \renewcommand{\arraystretch}{0.6}
    \newcommand{\formattedgraphics}[1]{%
      \begin{tikzpicture}[spy using outlines={rectangle, magnification=2, connect spies}]
        \node[anchor=south west, inner sep=0] at (0,0){\includegraphics[width=\imagewidth]{#1}};
        \spy [red,size=30pt] on (.4\imagewidth,.1\imagewidth) in node at (.82\imagewidth,.19\imagewidth);
        \end{tikzpicture}%
    }
    \begin{tabular}{m{0.38cm}<{\centering}m{\imagewidth}<{\centering}m{\imagewidth}<{\centering}m{\imagewidth}<{\centering}m{\imagewidth}<{\centering}m{\imagewidth}<{\centering}}
        & $t=\SI{0}{\second}$ & $t=\SI{0.24}{\second}$ & $t=\SI{0.48}{\second}$ & $t=\SI{0.72}{\second}$ & $t=\SI{0.96}{\second}$\\
        \rotatebox{90}{\textbf{Rain Motion}} &
        \formattedgraphics{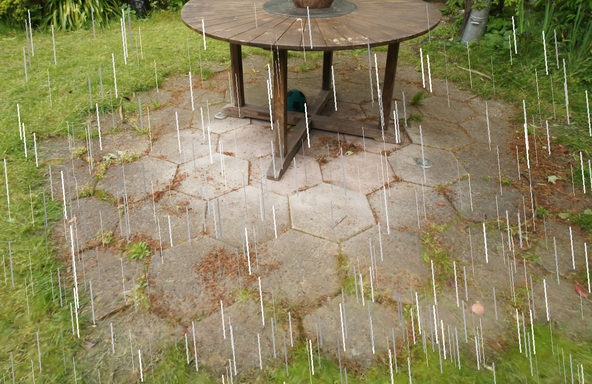} & 
        \formattedgraphics{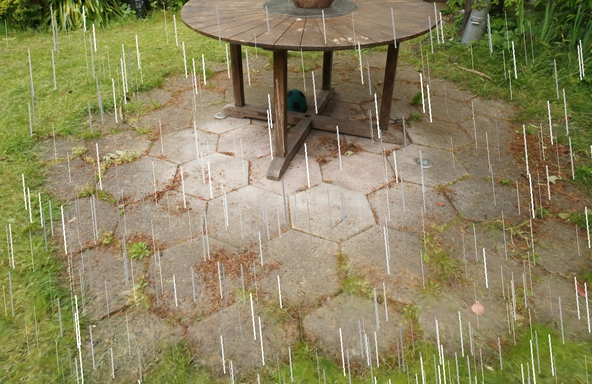} & 
        \formattedgraphics{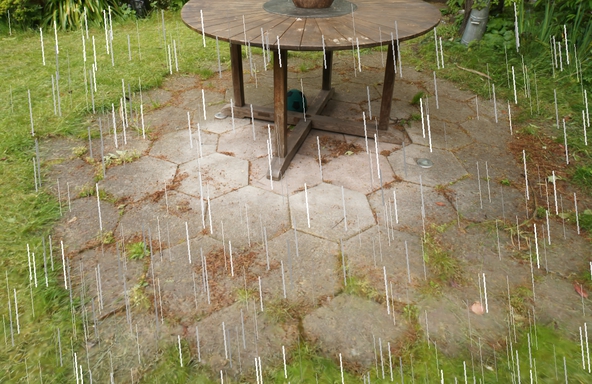} & 
        \formattedgraphics{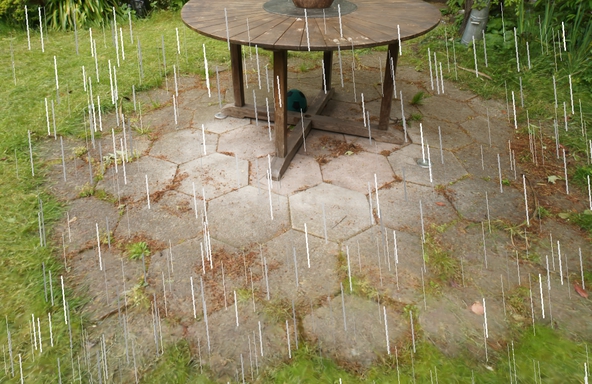} & 
        \formattedgraphics{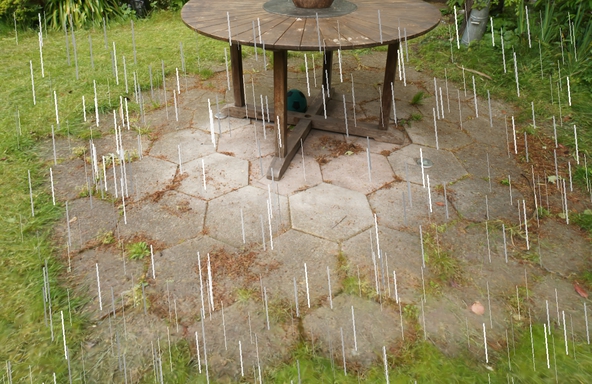}\\
        \rotatebox{90}{\textbf{Runway-V2V}} & 
        \formattedgraphics{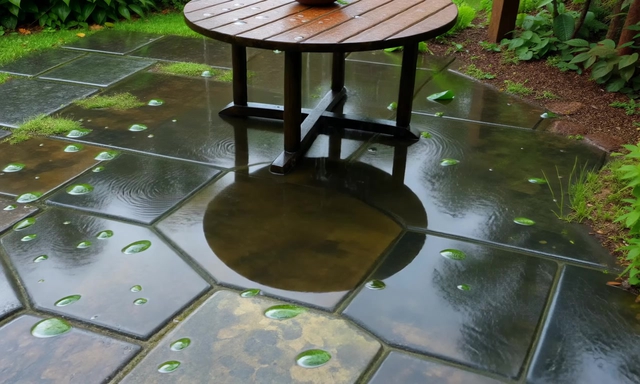} & 
        \formattedgraphics{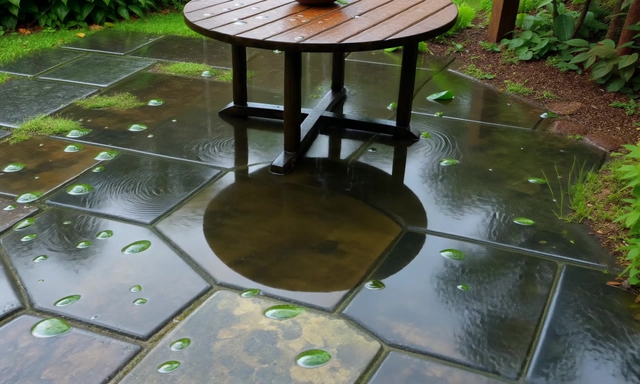} & 
        \formattedgraphics{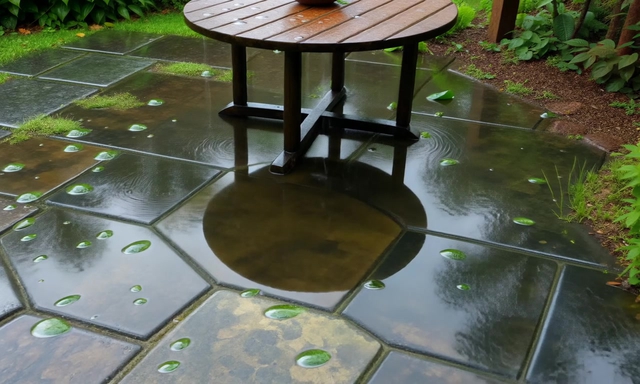} & 
        \formattedgraphics{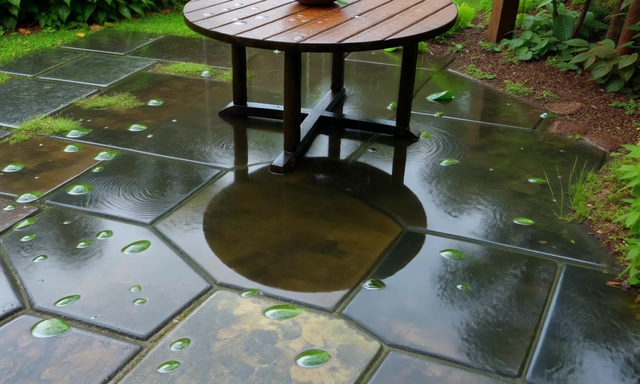} & 
        \formattedgraphics{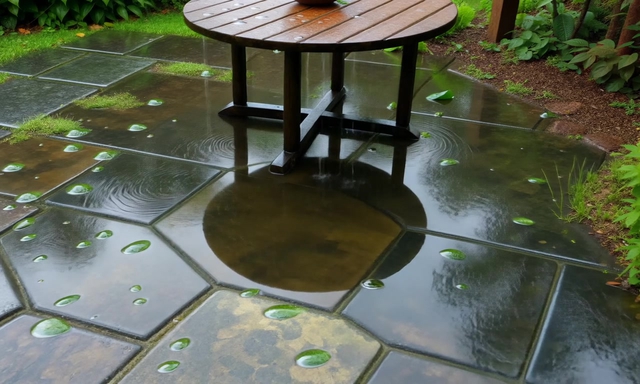}\\
        \rotatebox{90}{\textbf{Ours}} &
        \formattedgraphics{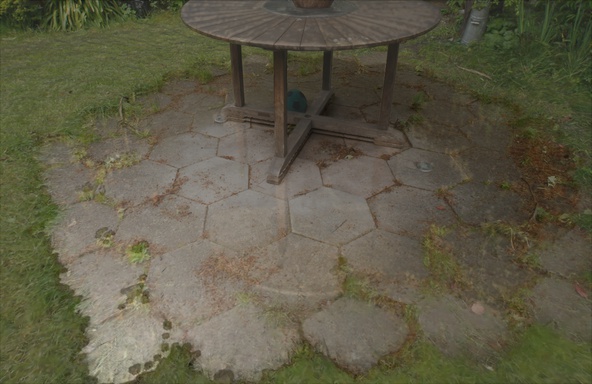} & 
        \formattedgraphics{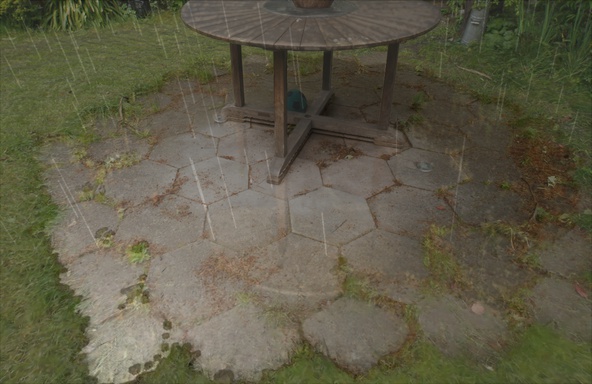} & 
        \formattedgraphics{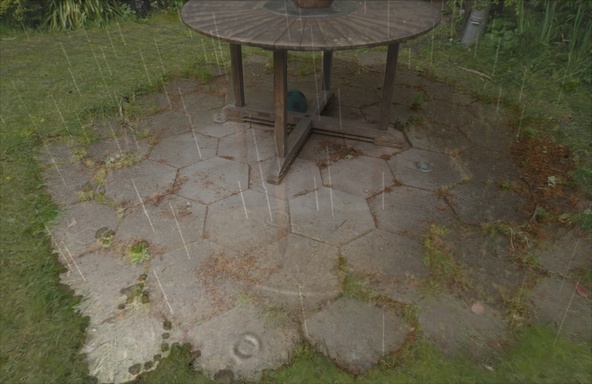} & 
        \formattedgraphics{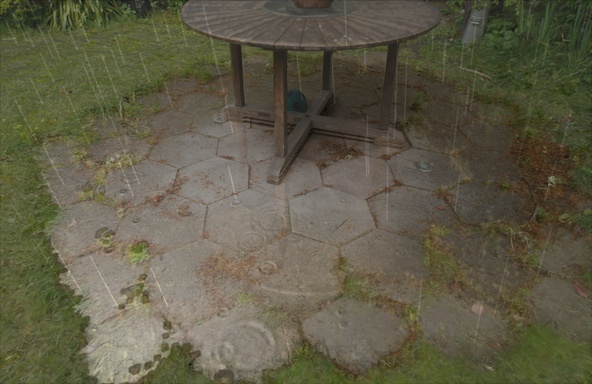} & 
        \formattedgraphics{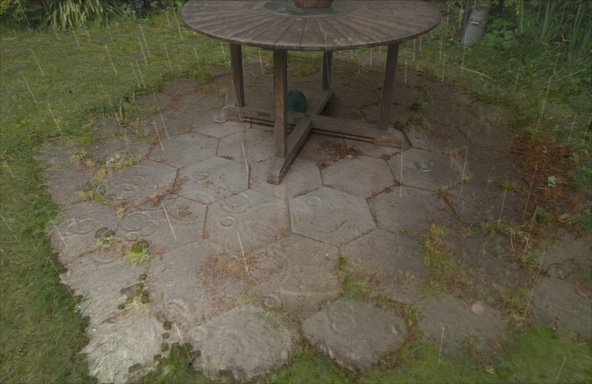}\\
    \end{tabular}
    \caption{\label{fig:fixed_view}
    %Rain synthesis results at the view point from different timesteps. Similarly, Rain Motion generates unrealistic and random rain droplets and no hydrops. Runway-V2V  generates  almost static rain content. In contrast, RainyGS produces realistic time-evloving hydrops and puddles.
    %
    Rain synthesis results at the same viewpoint from different timesteps are shown. Similarly, Rain Motion generates unrealistic and random rain droplets without hydrops. Runway-V2V produces almost static rain content. In contrast, RainyGS generates realistic time-evolving hydrops and puddles.
    }
    \vspace*{-0.3cm}
\end{figure*}
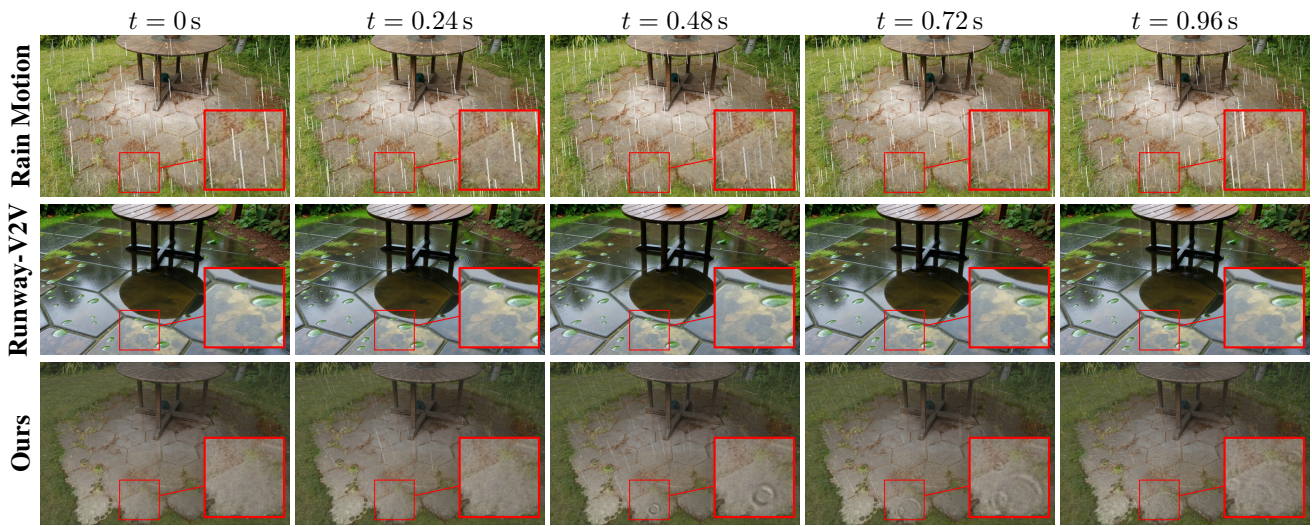

\section{Experiments}
\label{experiments}

We evaluate RainyGS across a variety of scenes to demonstrate its high-fidelity results, precise user control, and interactive performance. Our experiments encompass comparisons with video-based rain synthesis \qy{and text-driven 3D editing} methods, results on standard datasets, ablation studies on rendering effects, and performance metrics to assess computational efficiency. These evaluations highlight RainyGS's strengths in both visual fidelity and physical accuracy, as well as its ability to enhance downstream tasks, such as augmenting training scenes for autonomous driving systems. Please refer to our supplemental video for a clearer and more detailed view of the dynamic results.

\subsection{Evaluation Details}

We begin by detailing the datasets and baselines used in our evaluations. We present results on the Garden, Treehill, and Bicycle scenes from the MipNeRF360 dataset, as well as the Family and Truck scenes from the Tanks and Temples dataset.
Our baselines include Rain Motion~\cite{wang2022rethinking}, a video rain synthesis method designed to generate temporally consistent rain streaks. It adds two types of rain streaks in each frame: one for the rain streaks moving in a dominant direction, and the other for the newly appearing rain streaks in the scene. 
We also compare with Runway Gen-3 Alpha~\cite{runway2024gen3alpha, runway2024gen3alphavideo}, the leading commercial video generation model that excels in video-to-video generation. 
\qy{Additionally, We compare Instruct-GS2GS~\cite{igs2gs}, a recent text-driven 3DGS editing method.}
These methods are capable of rain synthesis and provide a solid comparison for our approach.

\subsection{Comparison with \qy{Rain Synthesis Baselines}}
%We compare RainyGS with existing video-based rain synthesis methods, such as Rain Streak, Motion Runway, and Gen-3 Alpha. 
While video generation methods create visually compelling rainy styles, they inevitably lack 3D consistency and often alter the original scene geometry, as seen in Fig.\ref{fig:fixed_time}, where Runway-V2V transforms the vase into a glass one and continuously changes the background. Furthermore, its dynamic rain is unrealistic in both simulation and rendering. As shown in Fig.\ref{fig:fixed_view}, Runway-V2V generates random ripples and raindrops but fails to produce the expected ripple effects when a specific droplet falls into the water. Additionally, the reflections and refractions of the ripples are not physically accurate, as the reflected desk leg remains straight, when it should be distorted by the ripple effects on the water surface.
Without proper stylization and 3D representation, Rain Motion falls short in both visual realism and dynamic accuracy. \qy{Instruct-GS2GS lacks advanced rain effects and only supports static editing.}
In contrast, RainyGS preserves the scene geometry and provides physically accurate rain dynamics with realistic rendering, ensuring a more consistent, efficient, and high-quality result.

\begin{figure}[htbp]
    \centering
    \setlength{\tabcolsep}{1pt}
    \setlength{\imagewidth}{0.49\linewidth}
    \begin{tabular}{cc}
        \textbf{Original} & \textbf{Rainy}\\
        \includegraphics[width=\imagewidth]{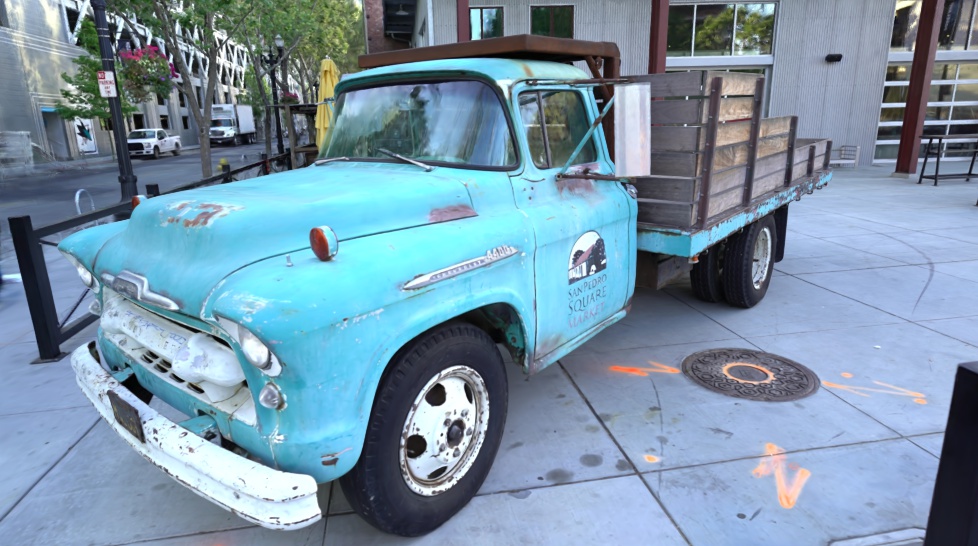} &
        \includegraphics[width=\imagewidth]{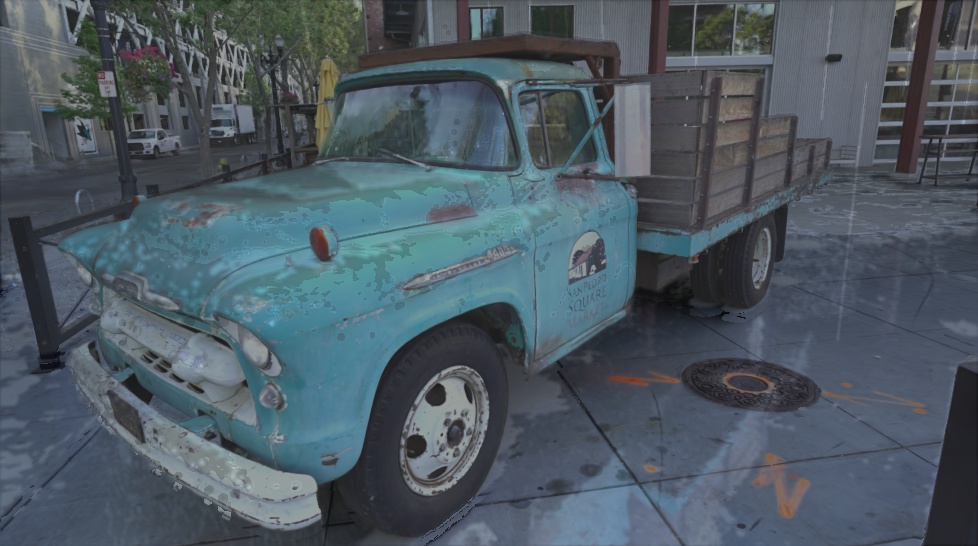}\\
        \includegraphics[width=\imagewidth]{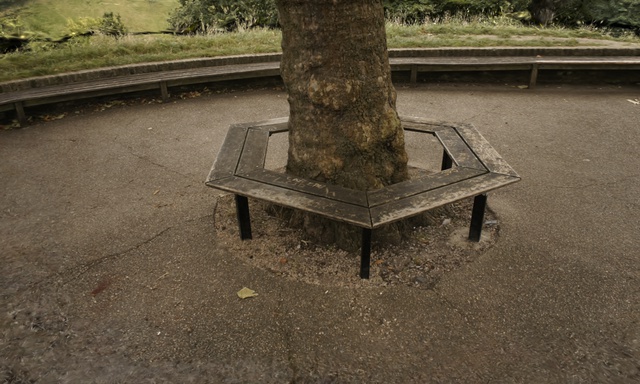} &
        \includegraphics[width=\imagewidth]{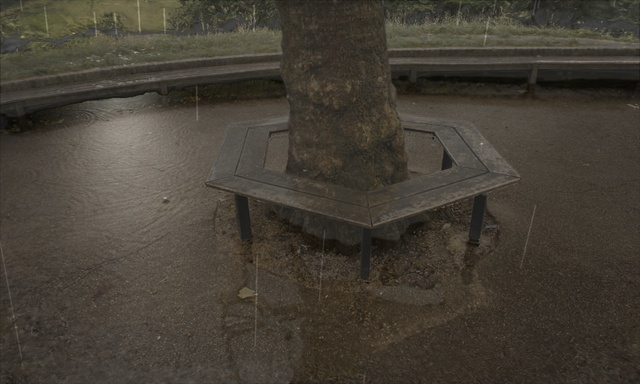}\\
        \includegraphics[width=\imagewidth]{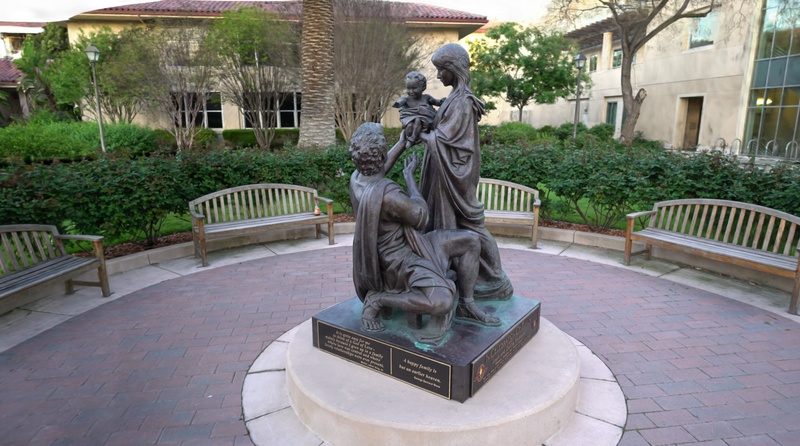} &
        \includegraphics[width=\imagewidth]{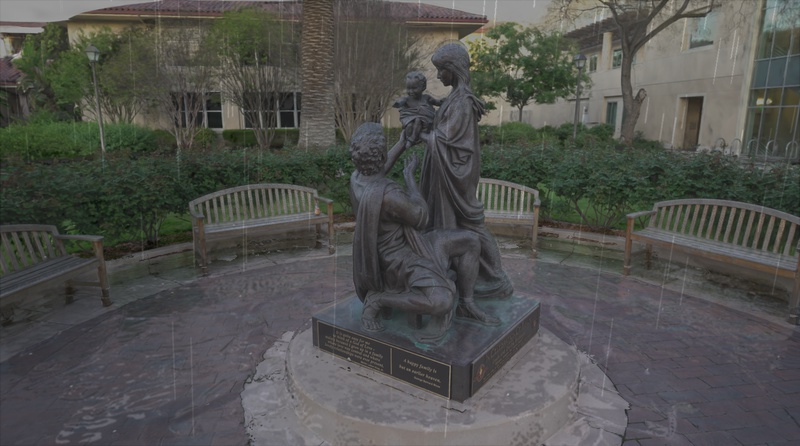}
    \end{tabular}
    \caption{\label{fig:more_scenes}
    {\name} provides users with a flexible way to control the rain intensity. From top to bottom, the levels are light drizzle, moderate rain, and heavy downpour, respectively.}
\end{figure}

% \begin{figure}
% 	\centering
% 	\includegraphics[width=.95\linewidth,height=5cm]{example-image}
% 	\caption{\label{fig:user_studies}User Studies}
% \end{figure}

\subsection{Qualitative Results}
\paragraph{High-Fidelity In-the-Wild Rainy Scene Synthesis} Fig.~\ref{fig:more_scenes} demonstrates the effectiveness of RainyGS using the MipNeRF360 and Tanks and Temples datasets, which represent complex, in-the-wild environments. 
The results confirm that RainyGS provides high visual fidelity, physical accuracy, and real-time performance in diverse scenes.
%The results confirm that RainyGS provides both high visual fidelity and physical accuracy, with real-time performance in diverse scenes.

\vspace*{-0.2cm}
 
\paragraph{Precise User Control} Our approach allows for interactive control over rain intensity, enabling users to easily generate ``light drizzle'', ``moderate rain'', and ``heavy downpour'' scenarios (as in Fig.~\ref{fig:more_scenes}). This flexibility is difficult to achieve with video-based methods but is effortlessly managed by RainyGS.

\vspace*{-0.1cm}

\paragraph{Performance} We present performance statistics to demonstrate the efficiency of RainyGS. As listed in Table~\ref{table:D-NeRF}, our method introduces only a small additional time cost compared to the underlying \qy{PGSR} method. The computation time is significantly reduced compared to \qy{PGSR} with ray tracing techniques and video-based methods, achieving responsive performance of over \qy{30} fps, while maintaining high visual fidelity. More analysis are provided in Supp.
% \qy{time}

\begin{table}[t]\footnotesize
\centering
\caption{The time and memory costs for the Garden scene. We compare our method with the original \qy{PGSR} (without rainy effects), the enhanced \qy{PGSR} (with 3D ray tracing for water rendering), and the video generator, Runway-V2V. Our method introduces a small additional time and memory cost compared to \qy{PGSR}, while being significantly faster than the other rain synthesis methods in the table.} 
\setlength{\tabcolsep}{3pt}
\begin{threeparttable}
    \mbox{}\clap{\begin{tabular}{ccc}
\specialrule{.15em}{.1em}{.1em}
\textbf{Method}\tnote{\dag} & \textbf{Time Per Frame} & \textbf{Peak Video Memory}  \\
\hline
\qy{PGSR} & \SI{0.007}{\second} & 7.989\,GB\\
\hline
\qy{PGSR} + RT & \SI{1.942}{\second} & 14.161\,GB\\
\hline
Runway-V2V & $\sim$\SI{0.4}{\second} & NA \\
\hline
Ours & \SI{0.032}{\second} & 8.561\,GB\\
\specialrule{.1em}{.05em}{.05em}
\end{tabular}}
\begin{tablenotes}
    \scriptsize
    \item[\dag] `RT' denotes ray tracing.
\end{tablenotes}
\end{threeparttable}
\vspace*{-0.4cm}
\label{table:D-NeRF}
\end{table}

\begin{comment}
    
\subsection{Ablation Studies}

In our ablation studies, we evaluate the impact of image-space rendering on reflections and refractions. Rendering the refraction and reflection of a water surface with its complex, smooth normals is challenging, as even small inaccuracies can become noticeable artifacts and cause visual discomfort. Our results demonstrate that image-space rendering effectively captures the most critical visible portions of the water surface, ensuring both realism and visual comfort. This approach allows for high-quality, real-time rain simulation while avoiding the computational cost of traditional 3D ray tracing, offering a more efficient solution for interactive applications.

\end{comment}

\subsection{Applications}

\begin{figure}[htb]
  \centering
  \setlength{\imagewidth}{.495\linewidth}
  \newcommand{\formattedgraphics}[1]{%
      \begin{tikzpicture}
        \node[anchor=south west, inner sep=0] at (0,0){\includegraphics[width=\imagewidth]{#1}};
        \draw[red] (0.45\imagewidth, 0.2\imagewidth) rectangle (0.6\imagewidth, 0.35\imagewidth);
        \end{tikzpicture}%
    }
    \subcaptionbox{\centering RainyGS Simulated Inputs}{\formattedgraphics{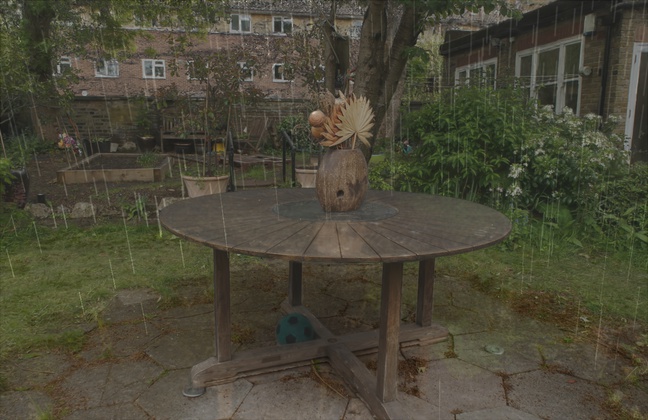}}
  \hfill
   \subcaptionbox{\centering 3DGS Reconstruction Results}{\formattedgraphics{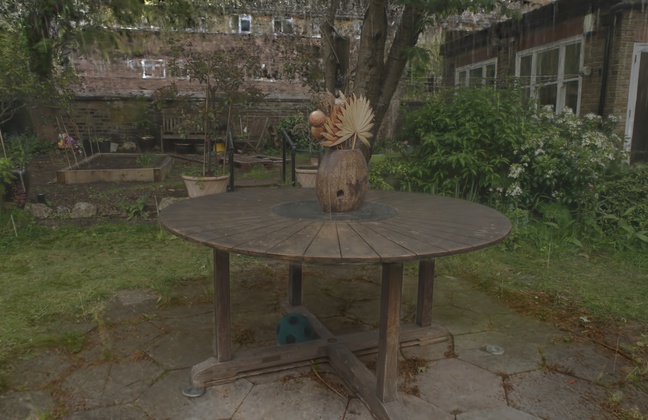}}
  \\
  \caption{\label{fig:failing}
  % We verify the scene modeling capability of 3DGS on rayiny scenes.  Taken our simulated rainy images as input(a), 3DGS reconstruct scene with degraded quality, as shown in the poor background and error rain lines.  This raises an promising future direction, further showcase the value of the proposed rain synthesis fraemwork.
  We verify the scene modeling capability of 3DGS on rainy scenes. Using our simulated rainy images as input (a), 3DGS reconstructs the scene with degraded quality, as shown by the poor background and erroneous rain lines. This highlights a promising future direction and further demonstrates the value of the proposed rain synthesis framework.
  }
\end{figure}
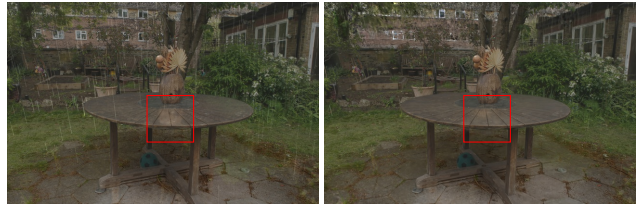

\begin{figure}[htb]
  \centering
  \setlength{\imagewidth}{.495\linewidth}
  \subcaptionbox{\centering Waymo Scene Inputs}{\includegraphics[width=\imagewidth]{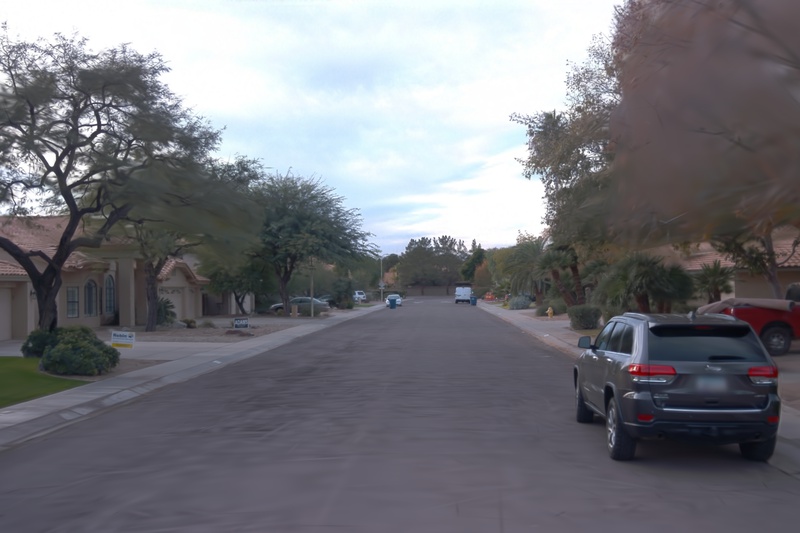}}
  \hfill
  \subcaptionbox{\centering Simulated Rainy Effects}{\includegraphics[width=\imagewidth]{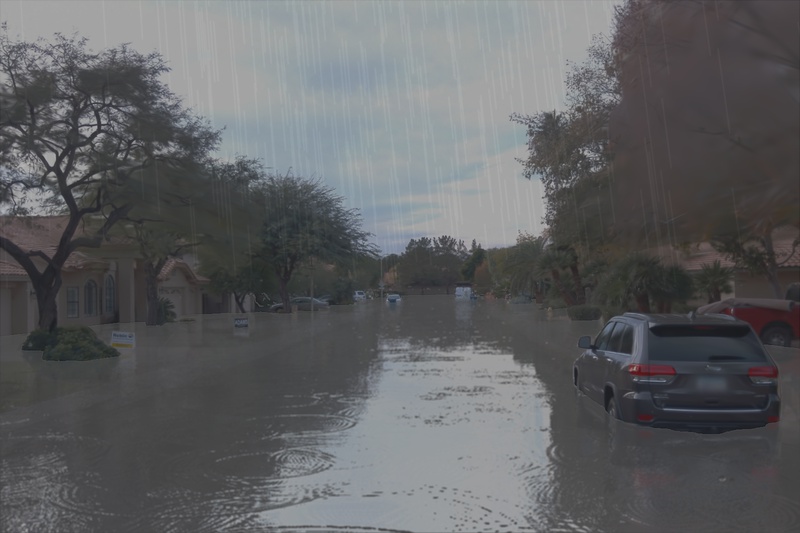}}
  \\
  \caption{\label{fig:waymo}
  % {\name} can be applied to large-scale driving scenes to synthesize corresponding raininy weather conditions, which helps creating challeing driving environments for hash rainy weathers to increase the safety fo audriving.
  {\name} can be applied to large-scale driving scenes to synthesize corresponding rainy weather conditions, helping to create challenging driving environments for harsh rainy conditions and increasing the safety of autonomous driving.
}
\vspace*{-0.3cm}
\end{figure}

\begin{comment}
    
{\name} provides a unified framework to synthesize realistic rainy effects to in-the-wild scenes, which opens a door for a variety of applications.  
%
It can not only be applied in improving challenging weather scene modeling as shown in Fi.8, but also ustilzied in  driving environments to create interactive simulation environments with more realism of real world (Fig.~\cite{}). 
%
These results highlight the scalability and versatility of our approach for real-time applications and environmental rendering, and it can be applied tp much more senarios.

\end{comment}

{\name} provides a unified framework for synthesizing realistic rainy effects in in-the-wild scenes, opening the door to a variety of applications.
It can not only help improve challenging weather scene modeling, as shown in Fig.~\ref{fig:failing}, but also be utilized in driving environments to create interactive simulations with greater realism (Fig.\ref{fig:waymo}).
These results highlight the scalability and versatility of our approach for real-time applications and environmental rendering, with potential for many more scenarios.

%% file: sec/5_conclusions.tex
\section{Conclusions}
\vspace*{-0.1cm}
\paragraph{Limitations}
A limitation of our approach is that the shallow-water dynamics, while effective for simulating surface-level rain interactions, are less accurate for fluid interactions with deep waves and are not suitable for modeling complex, multi-depth water effects, such as interactions with submerged surfaces. Additionally, due to image-space rendering, occluded objects cannot be refracted or reflected. 
\qy{Another limitation is the reliance on accurate scene modeling; limited input views may lead to imperfect geometry like irregular ground and cause water patches. Nevertheless, ground irregularities can be covered by more water amount, enabling realistic rainy effects (Fig.~\ref{fig:waymo}).}
Despite these limitations, we believe that our method captures the most critical aspects of dynamic rain simulation and rendering with interactive efficiency. We aim to address these challenges in future work to achieve even more realistic dynamics and rendering.
% Although our method is efficient, there is still significant potential for further optimization in the code and for exploring additional user interactions.
\vspace*{-0.1cm}
\paragraph{Conclusions}
In conclusion, we propose a novel rain simulator that integrates shallow-water dynamics and screen-space rendering into 3D Gaussian Splatting (3DGS) scenes. This simulator enables users to interactively generate realistic rain phenomena in complex, in-the-wild environments without the need to worry about scene geometry, lighting conditions, or other physical parameters. 
% While 3DGS offers excellent 3D consistency in novel-view synthesis and has garnered substantial interest for a variety of applications, adapting it for rain simulation presents challenges due to imperfect geometry and the entangling of lighting and material in its color representation. 

\begin{comment}
    
Each component is thoughtfully integrated in our approach. The shallow-water dynamics leverages a height field that circumvents these geometric challenges. The use of screen-space rendering further eliminates the need for exact lighting and 3D ray tracing, delivering high-quality results interactively.
When all components work in unison, the system methodically avoids unnecessary computations, operating with high efficiency that significantly surpasses ray-tracing methods.
% , such as those used in NeRF~\cite{climatenerf} and Gaussian~\cite{gaussianreflection}. 
We also demonstrate visual improvements over video-based methods. Our method offers results with enhanced realism, physical correctness, and complete 3D and temporal annotations. This provides a versatile framework with strong potential for real-time simulations and broader applications in environmental rendering.
% with substantial potential for even further optimization and broader applicability.

\end{comment}

%% file: src_sup/0_abstract.tex
The supplementary materials offer a detailed explanation of our RainyGS, covering Scene Modeling in Sec.~\ref{sec:supp_scene_modeling}, Auxiliary Map Extraction in Sec.~\ref{sec:supp_auxmap}, Rain Simulation on Height Maps in Sec.~\ref{sec:supp_rainsim}, and Screen-Space Reflection in Sec.~\ref{sec:supp_ssr}. Additionally, we present results and performance analyses from various experimental scenes in Sec.~\ref{sec:supp_exp}. To further illustrate our method, we include a video demonstrating the dynamic synthesis of rain.

%% file: src_sup/1_aux_map.tex
\section{Details of Scene Modeling}
\label{sec:supp_scene_modeling}

\subsection{Normal Prior Supervision}

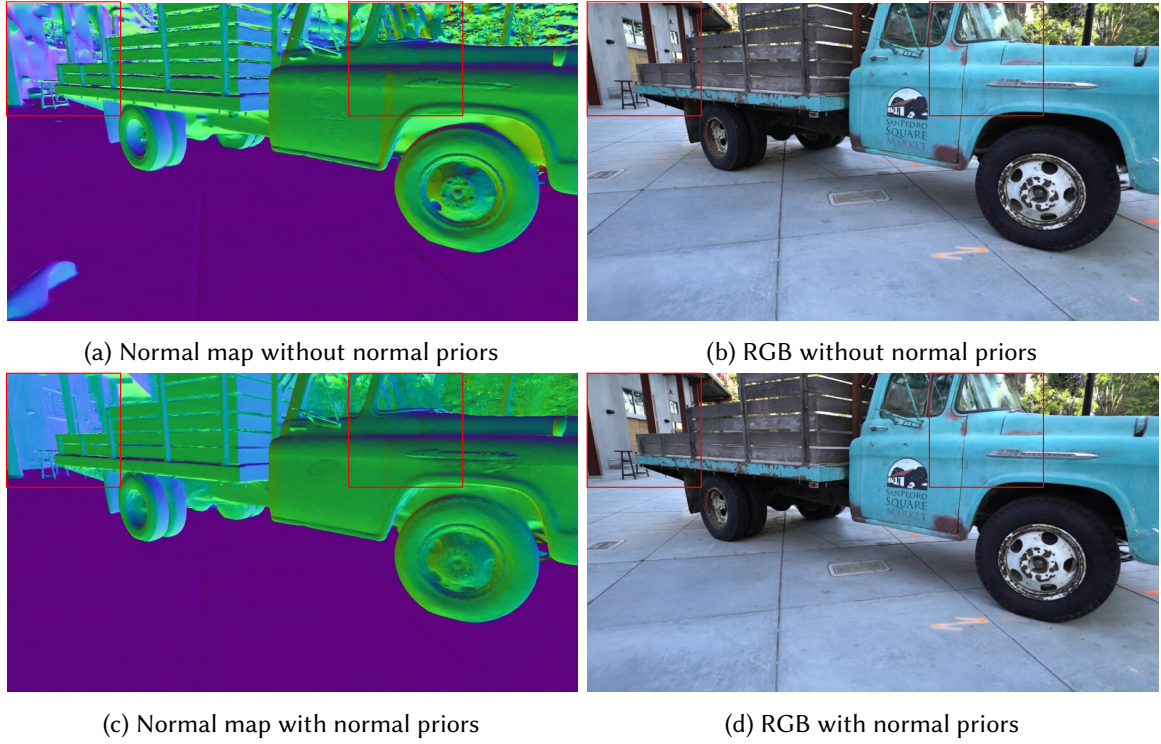
\begin{figure}
  \centering
  \setlength{\imagewidth}{.495\linewidth}
  \newcommand{\formattedgraphics}[1]{%
      \begin{tikzpicture}
        \node[anchor=south west, inner sep=0] at (0,0){\includegraphics[width=\imagewidth]{#1}};
        \draw[red] (0.6\imagewidth, 0.36\imagewidth) rectangle (0.8\imagewidth, 0.56\imagewidth);
        \draw[red] (0.00\imagewidth, 0.36\imagewidth) rectangle (0.20\imagewidth, 0.56\imagewidth);
        \end{tikzpicture}%
    }
  \subcaptionbox{\centering Normal map without normal priors}{\formattedgraphics{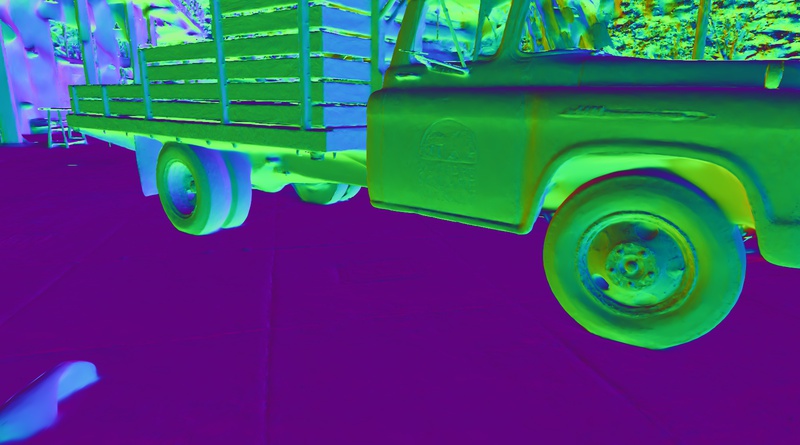}}
  \hfill
  \subcaptionbox{\centering RGB without normal priors}{\formattedgraphics{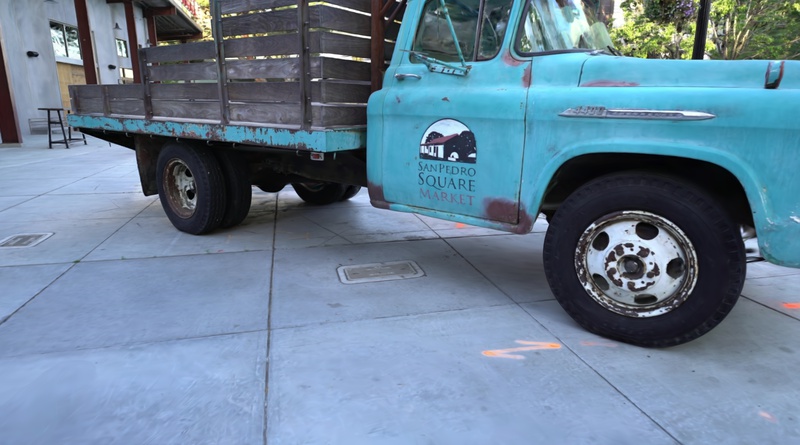}}
  \\
  \subcaptionbox{\centering Normal map with normal priors}{\formattedgraphics{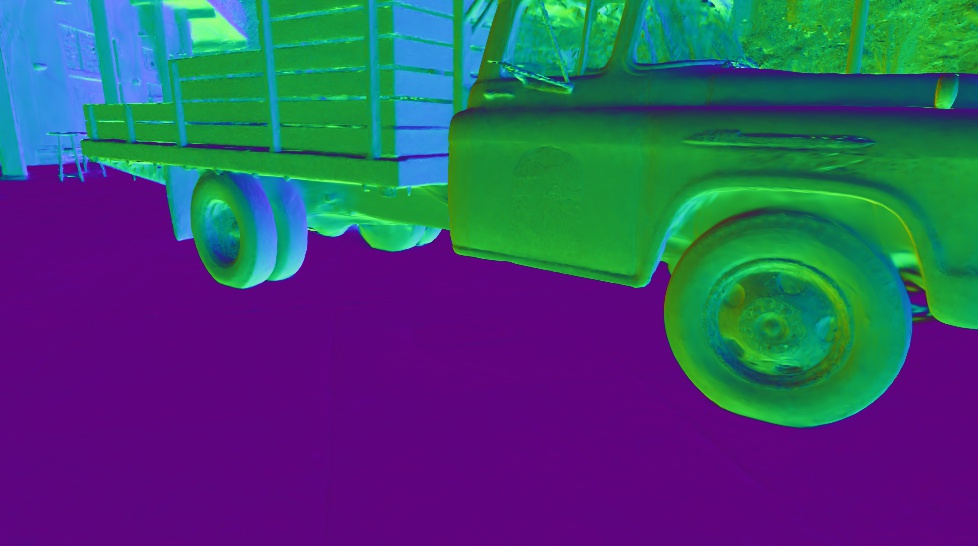}}
  \hfill
  \subcaptionbox{\centering RGB with normal priors}{\formattedgraphics{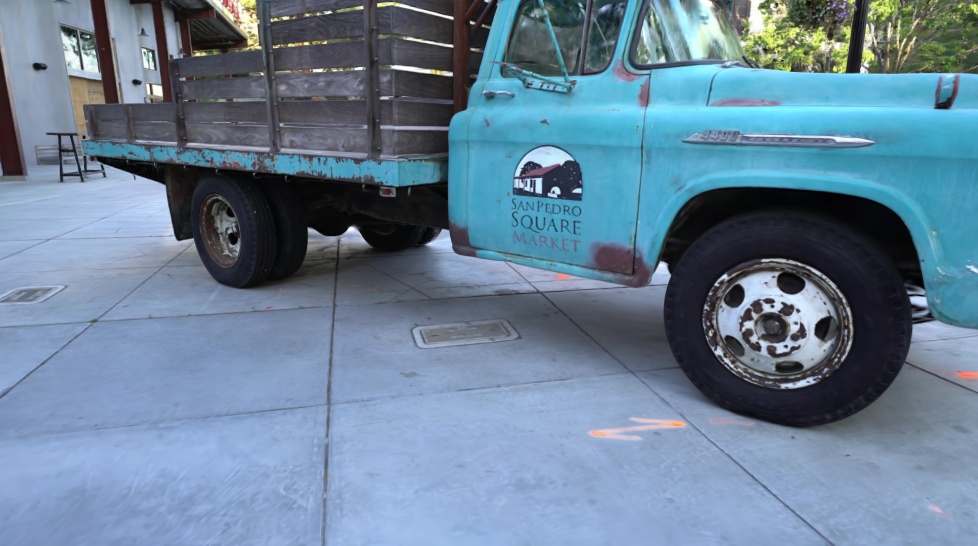}}
  \\
  \caption{
  \label{fig:normal}
  As highlighted in the red box, leveraging priors from a pretrained monocular normal estimation model leads to notable improvements in both rendering fidelity and geometric precision of PGSR for scene modeling.
  }
\end{figure}

We employ PGSR~\cite{chen2024pgsr} as a unified module for appearance and geometry reconstruction.
While PGSR achieves state-of-the-art quality, it still exhibits noticeable artifacts in texture-less regions and areas with transparent or reflective materials, such as floors, walls, and car windows (as shown in Fig.~\ref{fig:normal}).
We demonstrate that introducing normal priors from a pretrained monocular normal estimation model~\cite{bae2024rethinking} to supervise the rendered normal maps can effectively improve both the rendering fidelity and geometric accuracy of PGSR.
This further leads to higher-quality auxiliary maps, including depth maps, normal maps, and height maps, which are essential for downstream rain simulation.

\subsection{Comparison of Scene Modeling Methods}

An alternative approach to scene modeling is combining a decomposed Radiance Field (e.g., GaussianShader~\cite{jiang2024gaussianshader}) with a geometry reconstruction method (e.g., GOF~\cite{Yu2024GOF}). Specifically, GaussianShader serves as the appearance module to disentangle appearance and illumination, while GOF is employed as the geometry module to extract detailed scene structures. To maintain consistency between these two modules, multi-view depth maps generated by the GOF model are used to supervise the training of GaussianShader. The advantage of this approach lies in its Physically-Based Rendering (PBR) formulation, enabling more accurate light transport and relighting. However, this method requires training two separate models, resulting in redundancy and increased computational cost. Additionally, the high degree of flexibility in GaussianShader makes optimization challenging, often leading to suboptimal visual quality.

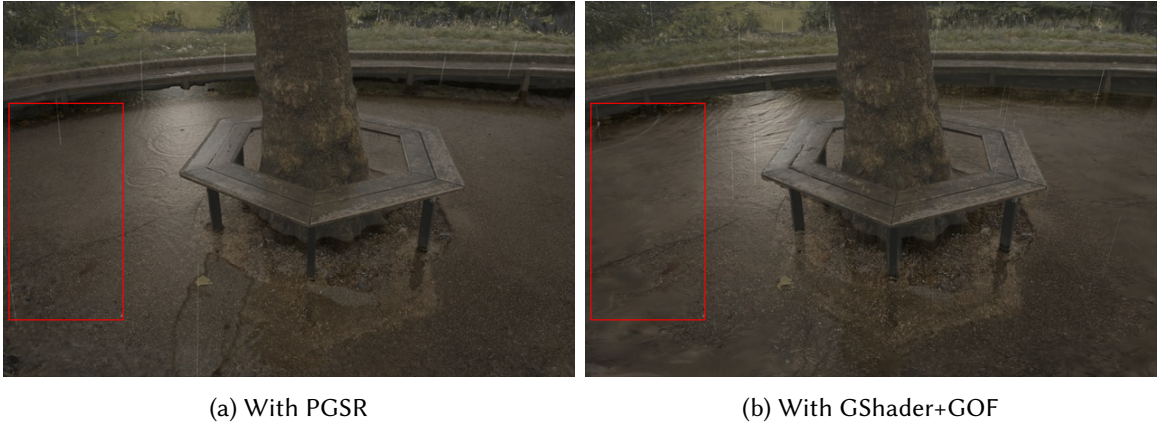
\begin{figure}
  \centering
  \setlength{\imagewidth}{.495\linewidth}
  \newcommand{\formattedgraphics}[1]{%
      \begin{tikzpicture}
        \node[anchor=south west, inner sep=0] at (0,0){\includegraphics[width=\imagewidth]{#1}};
        % 第一个红框
        \draw[red] (0.01\imagewidth, 0.10\imagewidth) rectangle (0.21\imagewidth, 0.48\imagewidth);
        % % 第二个红框
        % \draw[red] (0.4\imagewidth, 0.3\imagewidth) rectangle (0.52\imagewidth, 0.36\imagewidth);
      \end{tikzpicture}%
    }
  \subcaptionbox{\centering With PGSR}{\formattedgraphics{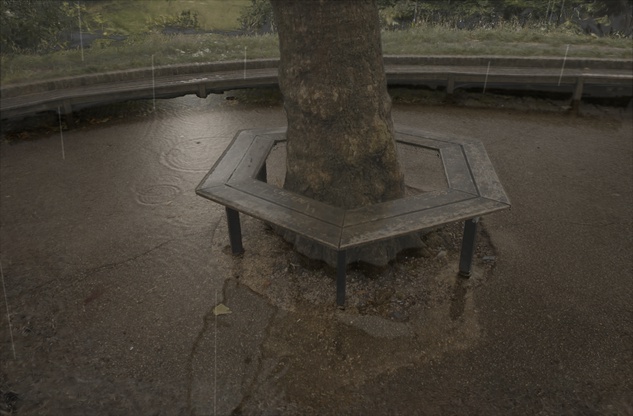}}
  \hfill
  \subcaptionbox{\centering With GShader+GOF}{\formattedgraphics{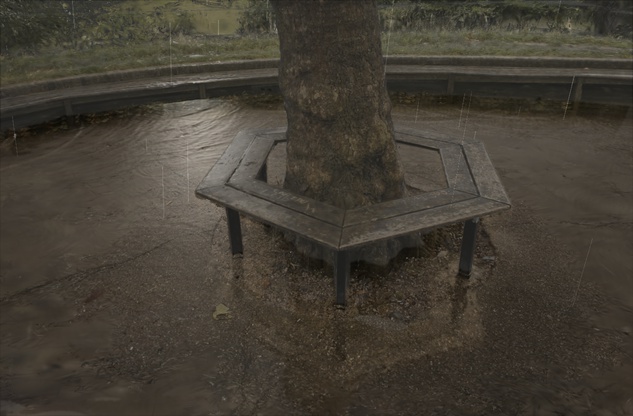}}
  \\
  \vspace*{-0.3cm}
  \caption{
  \label{fig:pgsr_gshader}
Compared to the GShader+GOF pipeline (b), using PGSR (a) achieves higher rendering quality while maintaining comparable lighting and shadow effects.
  }
    \Description{figure}
  %\vspace*{0.7cm}
\end{figure}

In contrast, PGSR adopts a unified model that achieves superior appearance and geometry quality, while remaining efficient during both training and inference. Although this approach sacrifices PBR properties in scene modeling, environment maps can still be replaced in the Water Rasterization stage. As illustrated in Fig.~\ref{fig:pgsr_gshader}, our method employing PGSR achieves comparable lighting effects to the GShader+GOF pipeline, while delivering higher rendering quality.

\section{Details of Auxiliary Map Preparation}
\label{sec:supp_auxmap}

% In Sec. 3.2 and 3.3, we model rain as spherical 3D Gaussians. For water surface Gaussians, their positions are aligned with the 3D coordinates corresponding to each pixel in the height map, with a fixed radius of 0.006. The depth of each Gaussian is defined as the distance from its center to the screen in the camera coordinate system, while its normals are computed from the normal map derived from the height map. This enables the extraction of depth and normal maps for specific viewpoints using 3DGS rasterization.
% %
% We further model the rain streaks as densely aligned lines of 3D Gaussian spheres, each with a radius of 0.006, a near-white color of [200, 200, 200], a roughness of 0.05, and a specular reflectance of 0.5. 
We model rain as spherical 3D Gaussians for auxiliary map extraction. For water surface Gaussians, their positions are aligned with the 3D coordinates corresponding to each pixel in the height map, with a fixed radius of 0.006. The normal of each Gaussian is computed from the normal map derived from the height map. This enables the extraction of normal maps for specific viewpoints using 3DGS rasterization.
We further model the rain streaks as densely aligned lines of 3D Gaussian spheres, each with a radius of 0.006 and a near-white RGB color of [200, 200, 200].

%% file: src_sup/2_rain_sim.tex
\section{Details of Rain Simulation on Height Maps}
\label{sec:supp_rainsim}

Our simulation framework begins with the two-dimensional staggered MAC grid \cite{Harlow1965} as usual,
storing components of the velocity $\bm{u}$ at the appropriate edge midpoints and the depth $h$ at the cell centers.
Based on the philosophy of time splitting, the simulation pipeline within a time step $\Delta t$ are summarized in the following text.
\begin{enumerate}
  \item \textbf{Handling Advection.} The velocity field $\bm{u}$ and the height field $h$ are treated with the semi-Lagrangian method and the \nth{1}-order upwind scheme, respectively:
  \begin{align}
    \bm{u}^*&\gets\mathrm{SemiLagrangian}(\bm{u}^n,\Delta t,\bm{u}^n)\text{,}\\
    h^*&\gets\mathrm{FirstOrderUpwind}(\bm{u}^n,\Delta t, h^n)\text{.}
  \end{align}
  \item \textbf{Enhancing Stability.} Height values that are overly close to $0$ are clamped to avoid undershooting:
  \begin{equation}
    h^*\gets0\text{,\qquad if }h^*<10^{-6}\text{.}
  \end{equation}
  \item \textbf{Constructing heights.} The total height field $\eta$ is calculated by
  \begin{equation}
    \eta\gets H+h^*\text{.}
  \end{equation}
  \item \textbf{Maintaining Rains.} Existing raindrops evolve, and new raindrops are generated.
  \item \textbf{Applying pressure.} The influence of pressure force is taken into account:
  \begin{equation}
    \bm{u}^*\gets\bm{u}^*-\Delta t\,g\,\bm{\nabla}\eta\text{,}
  \end{equation}
  in which $g$ denotes the gravitational acceleration.
  \item \textbf{Extrapolating Velocities.} It is important to extrapolate velocities from wet regions into dry regions, in order to handle boundary conditions.
\end{enumerate}

%% file: src_sup/3_ssr.tex
\section{Details of Screen-space Reflection}
\label{sec:supp_ssr}

Screen-Space Reflection (SSR) \cite{mcguire2014ray} is a real-time rendering technique that approximates reflective surfaces by reusing the existing information available in the screen space, such as the depth map and rendered color (RGB) map. The SSR avoids the computational complexity of full ray-tracing by confining reflection computation to the visible scene, whose process can be divided into several stages:
\begin{enumerate}
  \item \textbf{Ray Casting.}
  The algorithm begins by casting rays from the view position of each reflective pixel. The ray direction is determined by the surface normal at the pixel and the direction to the camera, adhering to the law of reflection:
  \begin{equation}
    \bm{R}=2(\bm{N}\cdot\bm{V})\bm{N}-\bm{V}\text{,}
  \end{equation}
  where $\bm{R}$ is the reflection direction, $\bm{N}$ is the surface normal. and $\bm{V}$ is the view vector.
  \item \textbf{Ray Marching.}
  Once the reflection direction is determined, it will be projected onto the screen space to form a 2D directional vector $\bm{d}$. Then, ray marching is performed in screen space from the reflective pixel along this direction. We use the digital differential analyzer (DDA) algorithm to iteratively sample the projected ray.
  \item \textbf{Recovering Points.}
  It is important to note that values save in the depth map are actually the $z$ component of positions in the camera coordinate system.
  Every time we sample a point from the projected ray, its corresponding point on the original ray should be recovered. We use $D$ to denote the value saved in the depth map at this pixel and use $Z$ to denote the $z$ component of the corresponding ray point in camera coordinates.
  \item \textbf{Collision Detection.}
  The ray is assumed to collide objects if and only if the following condition met:
  \begin{equation}
    D\le Z<D+\varepsilon\text{,}
  \end{equation}
  in which the parameter $\varepsilon$ is used to determine the surface thickness.
  For most of our test scenarios, we find that $\varepsilon=3\times10^{-2}$ is a good setting.
  \item \textbf{Fetching Reflective Colors.} When a collision is detected, the color saved in the pixel where the collision occurs is fetched. Consequently, the process of ray marching terminates.
  Otherwise, if no collision is detected and the projected ray is outside the screen space, the program also exists and sampled the color from the environment map using direction $\bm{R}$ as a fallback.
\end{enumerate}

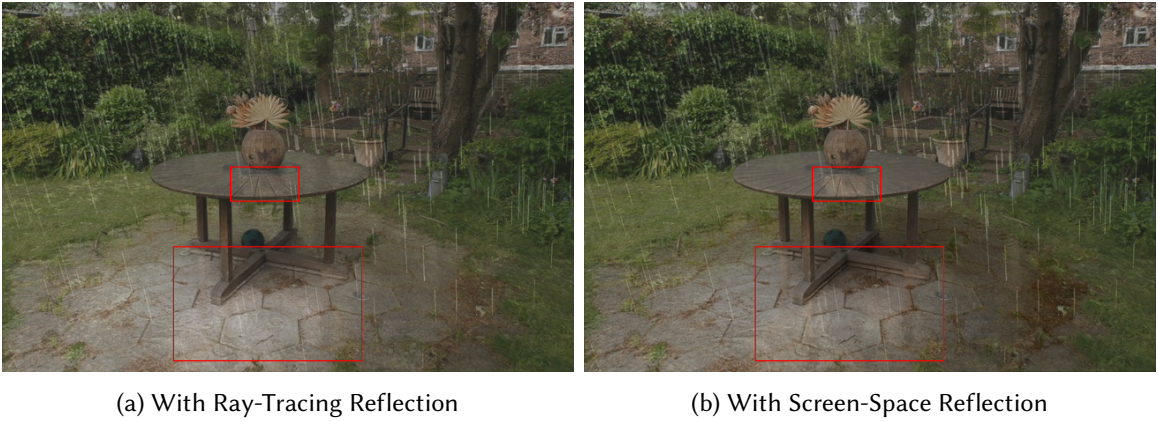
\begin{figure}
  \centering
  \setlength{\imagewidth}{.495\linewidth}
  \newcommand{\formattedgraphics}[1]{%
      \begin{tikzpicture}
        \node[anchor=south west, inner sep=0] at (0,0){\includegraphics[width=\imagewidth]{#1}};
        % 第一个红框
        \draw[red] (0.3\imagewidth, 0.02\imagewidth) rectangle (0.63\imagewidth, 0.22\imagewidth);
        % 第二个红框
        \draw[red] (0.4\imagewidth, 0.3\imagewidth) rectangle (0.52\imagewidth, 0.36\imagewidth);
      \end{tikzpicture}%
    }
      \subcaptionbox{\centering With Ray-Tracing Reflection}{\formattedgraphics{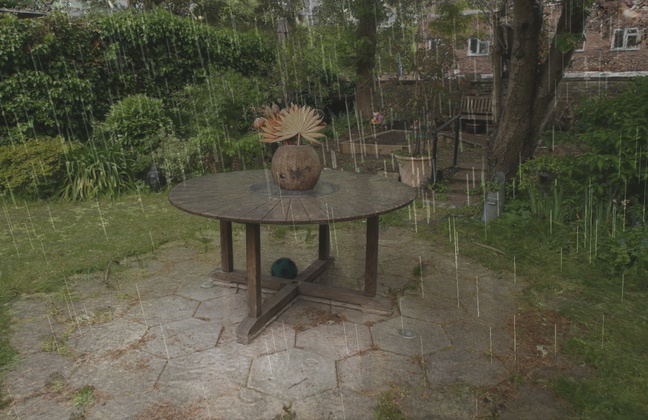}}
  \hfill
  \subcaptionbox{\centering With Screen-Space Reflection}{\formattedgraphics{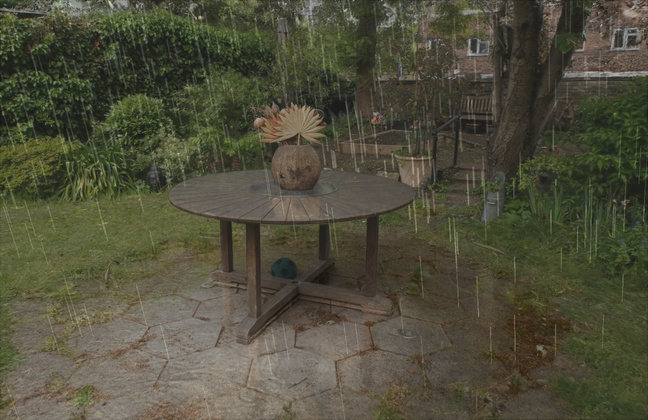}}
  \\
  \vspace*{-0.3cm}
  \caption{
  \label{fig:rt_vs_ssr}
  \textbf{Garden Scenario.}
  Compared to ray-traced reflection rendering (a), SSR-based reflection rendering (b) achieves nearly identical visual quality while offering significantly higher efficiency and reduced memory consumption. The reflection regions are highlighted with red boxes.
  }
    \Description{figure}
  %\vspace*{0.7cm}
\end{figure}

\textbf{Discussion} To evaluate the performance of ray tracing (RT) versus screen space reflection (SSR) in reflection computation, we adopt the approach described in~\cite{R3DG2023}. Specifically, we trace the reflected light paths of water surface Gaussians, identify intersections with original scene Gaussians, and apply alpha blending to compute their reflection colors. These results were then rasterized to generate a reflection map ($I_{highl}$ in the main text), which is integrated into our rendering pipeline. The final output combines this reflection map with the original image and refraction, as depicted in Fig.~\ref{fig:rt_vs_ssr} (a).

As illustrated in (b), our SSR implementation achieves a reflection quality comparable to RT. Moreover, as shown in Table 1, our method delivers a 60× speedup and a 1.65× reduction in peak memory usage. These performance gains are attributed to the computational efficiency of SSR, which approximates reflections by stepping through a limited number of screen-space pixels. In contrast, the RT approach requires time-consuming traversal of a BVH tree constructed from scene Gaussians and reordering of Gaussians along the light path. 

% \noindent
% \textbf{The value of \texttt{\textbackslash textwidth}:} \the\textwidth\par
% \textbf{The value of \texttt{\textbackslash textheight}:} \the\textheight\par
% \the\imageheight\par

%% file: src_sup/4_results.tex
\clearpage

\begin{figure*}
    \centering
    \setlength{\tabcolsep}{1pt}
    \setlength{\imagewidth}{0.19\textwidth}
    \setlength{\imageheight}{0.087\textheight}
    \renewcommand{\arraystretch}{0.6}
    \newcommand{\formattedgraphics}[3]{% #1 = 图片路径, #2 = spy on参数， #3 = spy at参数
        \begin{tikzpicture}[spy using outlines={rectangle, magnification=2, connect spies}]
            \node[anchor=south west, inner sep=0] at (0,0){\includegraphics[width=\imagewidth, height=\imageheight]{#1}};
            \spy [red, size=30pt] on #2 in node at #3;
        \end{tikzpicture}%
    }
    \begin{tabular}{m{0.38cm}<{\centering} m{\imagewidth}<{\centering} m{\imagewidth}<{\centering} m{\imagewidth}<{\centering} m{\imagewidth}<{\centering} m{\imagewidth}<{\centering}}
        & \textbf{Original} & \textbf{Rain Motion} & \textbf{Runway-V2V} & \textbf{Instruct-GS2GS} & \textbf{Ours} \\
        \rotatebox{90}{\textbf{View 1}} &
        \formattedgraphics{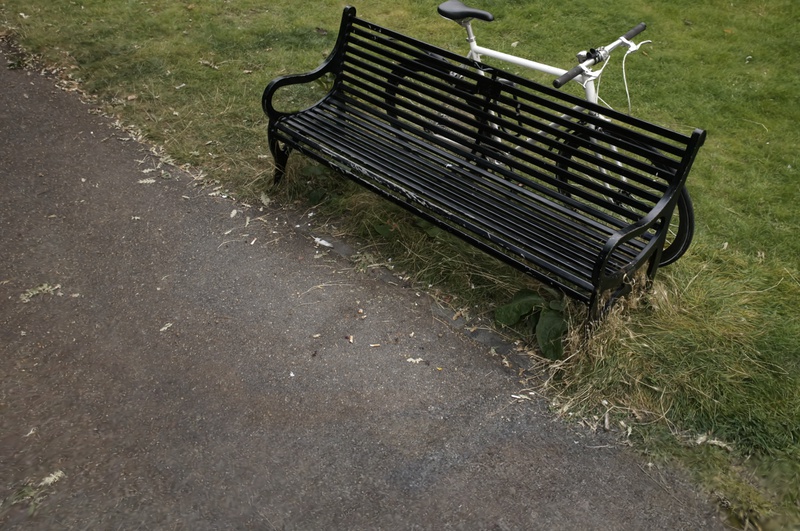}{(.41\imagewidth,.1\imagewidth)}{(.21\imagewidth,.46\imagewidth)} & 
        \formattedgraphics{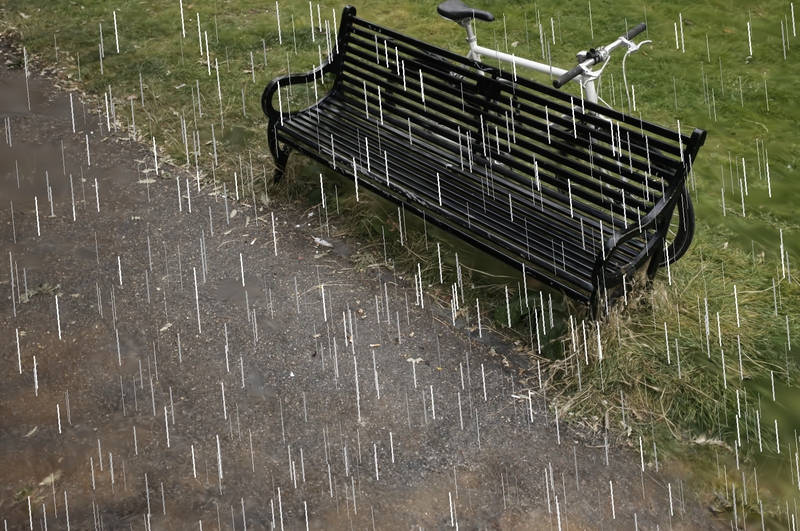}{(.41\imagewidth,.1\imagewidth)}{(.21\imagewidth,.46\imagewidth)} & 
        \formattedgraphics{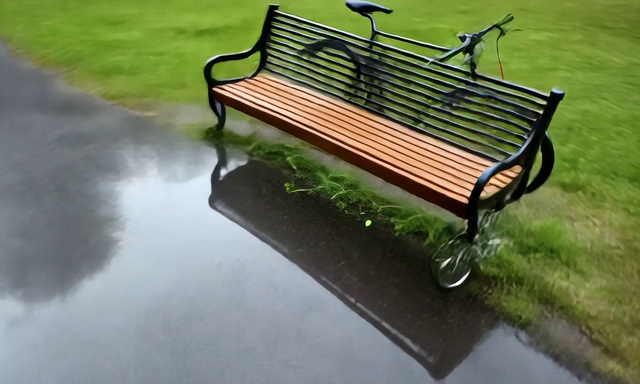}{(.41\imagewidth,.1\imagewidth)}{(.21\imagewidth,.46\imagewidth)} & 
        \formattedgraphics{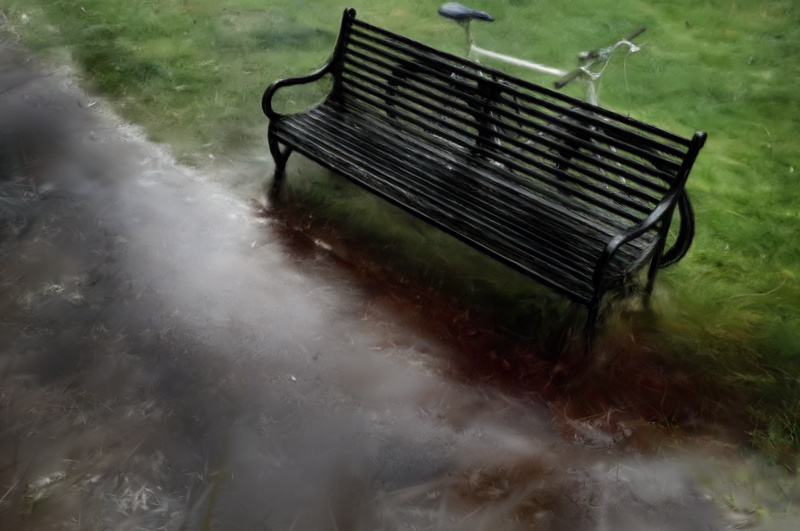}{(.41\imagewidth,.1\imagewidth)}{(.21\imagewidth,.46\imagewidth)} & 
        \formattedgraphics{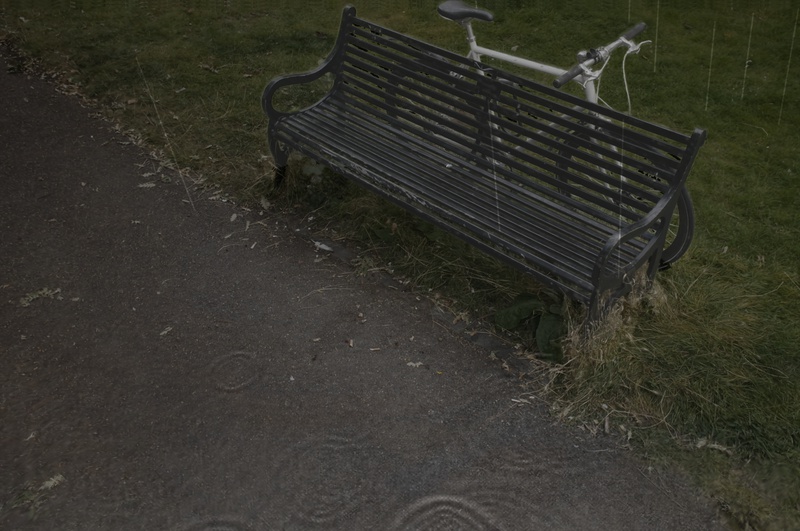}{(.41\imagewidth,.1\imagewidth)}{(.21\imagewidth,.46\imagewidth)} \\
        \rotatebox{90}{\textbf{View 2}} &
        \formattedgraphics{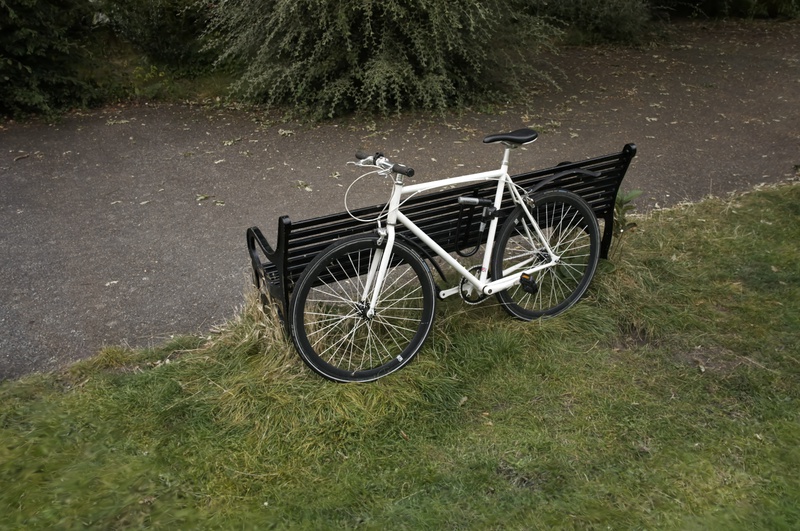}{(.22\imagewidth,.39\imagewidth)}{(.79\imagewidth,.45\imagewidth)} & 
        \formattedgraphics{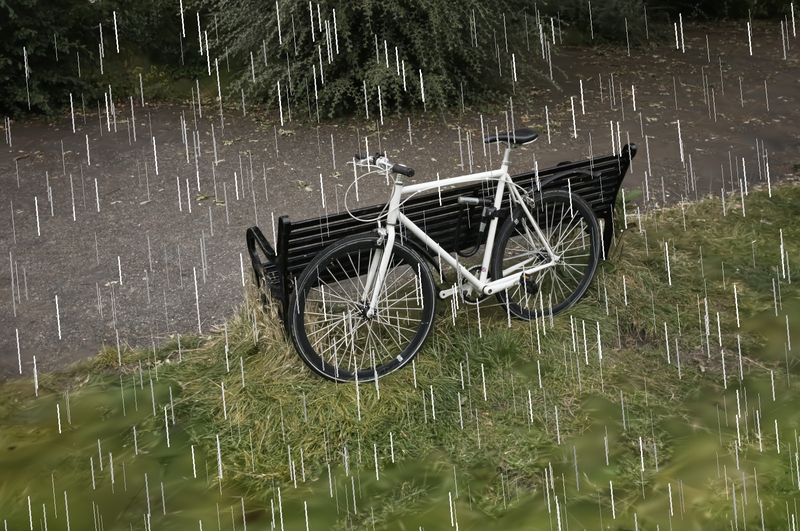}{(.22\imagewidth,.39\imagewidth)}{(.79\imagewidth,.45\imagewidth)} & 
        \formattedgraphics{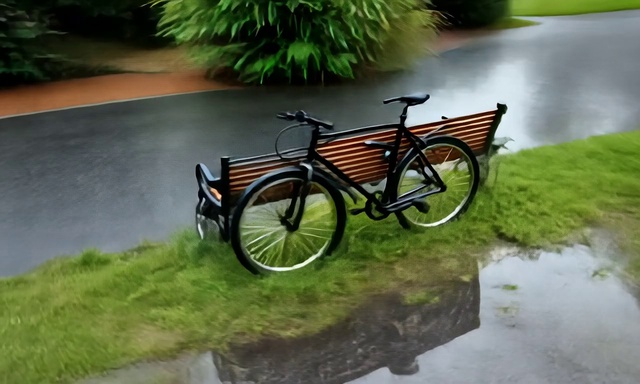}{(.22\imagewidth,.39\imagewidth)}{(.79\imagewidth,.45\imagewidth)} & 
        \formattedgraphics{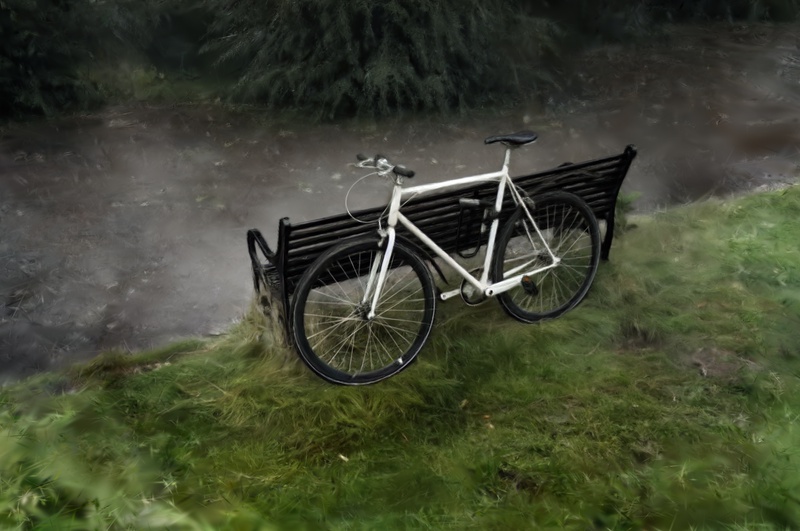}{(.22\imagewidth,.39\imagewidth)}{(.79\imagewidth,.45\imagewidth)} & 
        \formattedgraphics{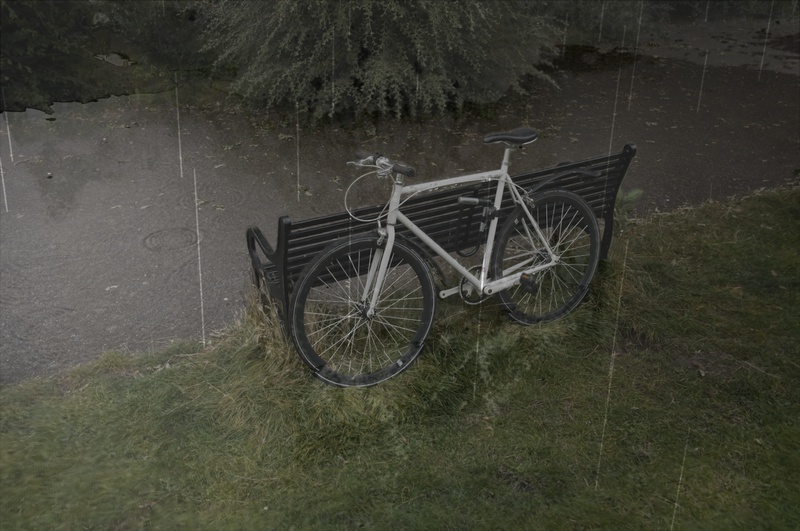}{(.22\imagewidth,.39\imagewidth)}{(.79\imagewidth,.45\imagewidth)} \\
        \rotatebox{90}{\textbf{View 3}} & 
        \formattedgraphics{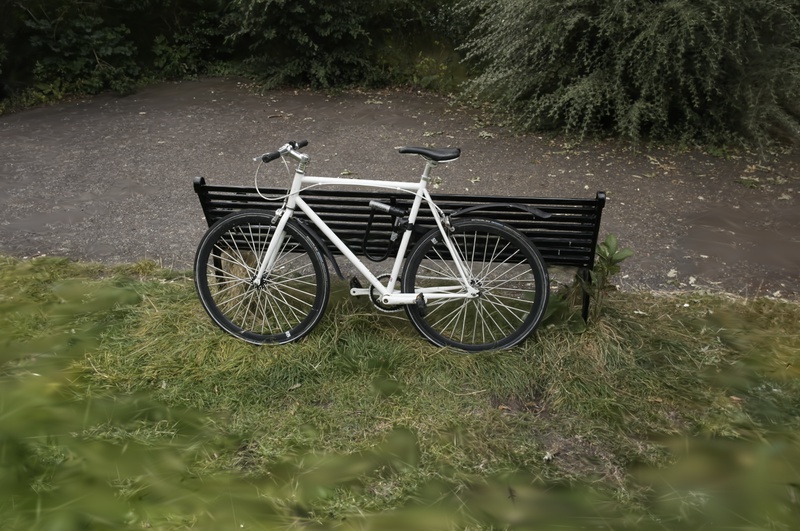}{(.54\imagewidth,.15\imagewidth)}{(.21\imagewidth,.46\imagewidth)} & 
        \formattedgraphics{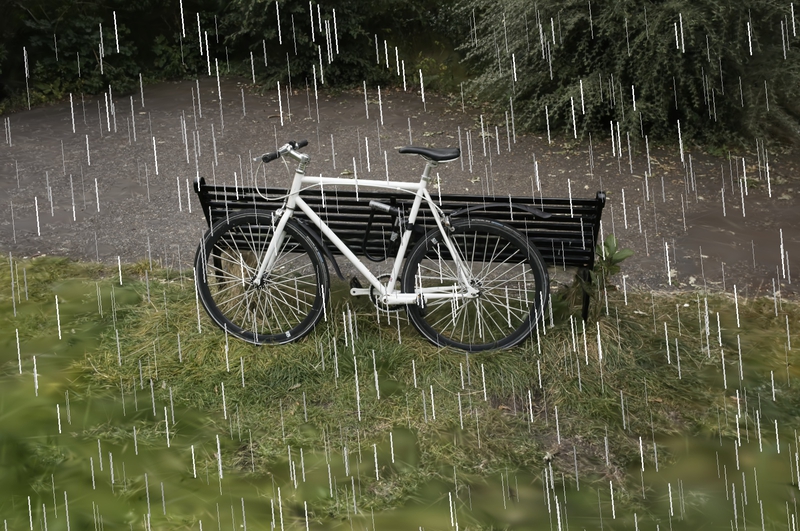}{(.54\imagewidth,.15\imagewidth)}{(.21\imagewidth,.46\imagewidth)}  & 
        \formattedgraphics{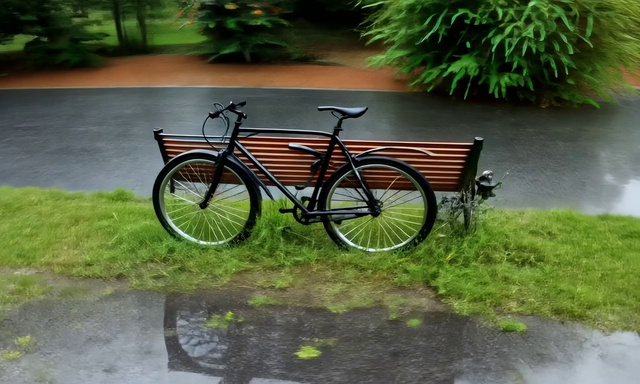}{(.54\imagewidth,.15\imagewidth)}{(.21\imagewidth,.46\imagewidth)} & 
        \formattedgraphics{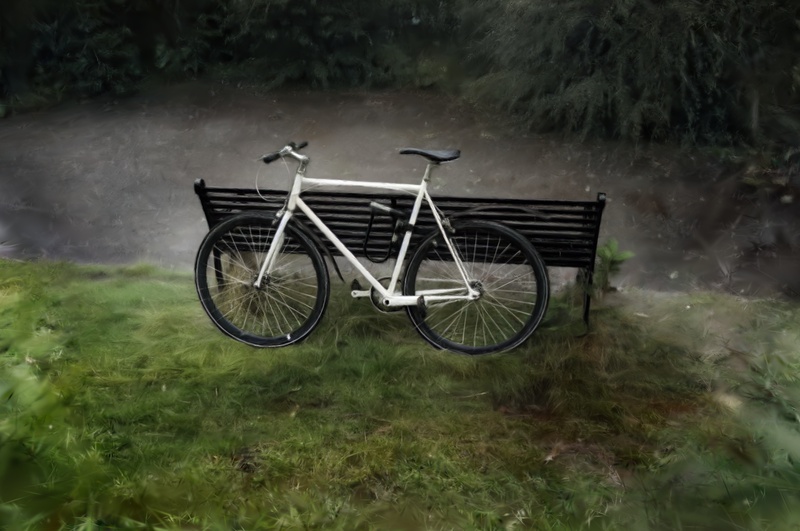}{(.54\imagewidth,.15\imagewidth)}{(.21\imagewidth,.46\imagewidth)}  & 
        \formattedgraphics{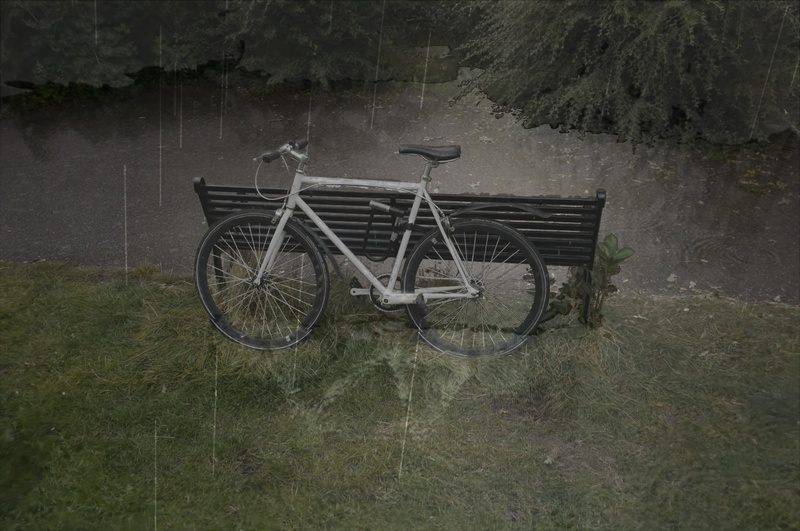}{(.54\imagewidth,.15\imagewidth)}{(.21\imagewidth,.46\imagewidth)}  \\
    \end{tabular}
    \caption{\label{fig:fixed_time_bicycle} \textbf{Bicycle Scenario.}
    \Description{figure}
    Rain synthesis results from different viewpoints at the same timestep. Runway-V2V struggles to preserve 3D consistency, while Rain Motion and Instruct-GS2GS fail to generate realistic rain streaks, puddles, and ripples. In contrast, RainyGS maintains 3D consistency and produces realistic rain streaks and water accumulation.}
    %\vspace*{1cm}
\end{figure*}
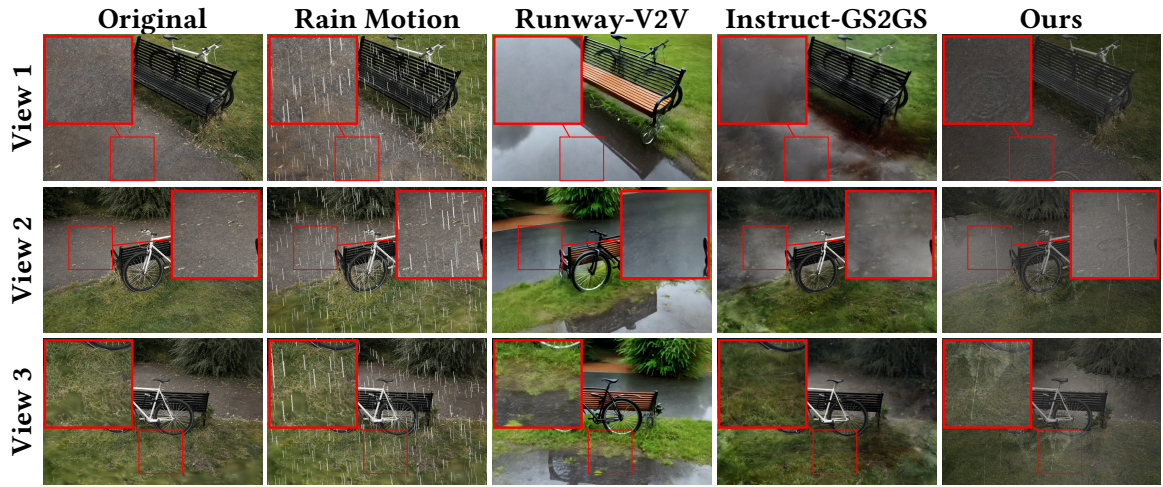

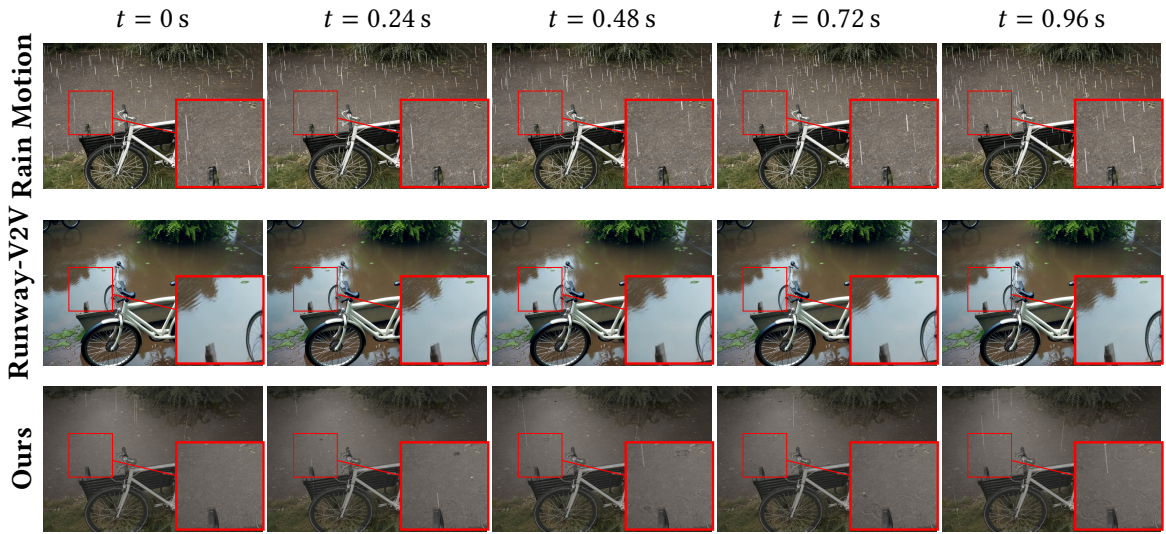
\begin{figure*}
    \centering
    \setlength{\tabcolsep}{1pt}
    \setlength{\imagewidth}{0.19\textwidth}
    \setlength{\imageheight}{0.087\textheight}
    \renewcommand{\arraystretch}{0.6}
    \newcommand{\formattedgraphics}[1]{%
      \begin{tikzpicture}[spy using outlines={rectangle, magnification=2, connect spies}]
        \node[anchor=south west, inner sep=0] at (0,0){\includegraphics[width=\imagewidth, height=\imageheight]{#1}};
        \spy [red,size=30pt] on (.22\imagewidth,.35\imagewidth) in node at (.81\imagewidth,.21\imagewidth);
        \end{tikzpicture}%
    }
    \begin{tabular}{m{0.38cm}<{\centering}m{\imagewidth}<{\centering}m{\imagewidth}<{\centering}m{\imagewidth}<{\centering}m{\imagewidth}<{\centering}m{\imagewidth}<{\centering}}
        & $t=\SI{0}{\second}$ & $t=\SI{0.24}{\second}$ & $t=\SI{0.48}{\second}$ & $t=\SI{0.72}{\second}$ & $t=\SI{0.96}{\second}$\\
        \rotatebox{90}{\textbf{Rain Motion}} &
        \formattedgraphics{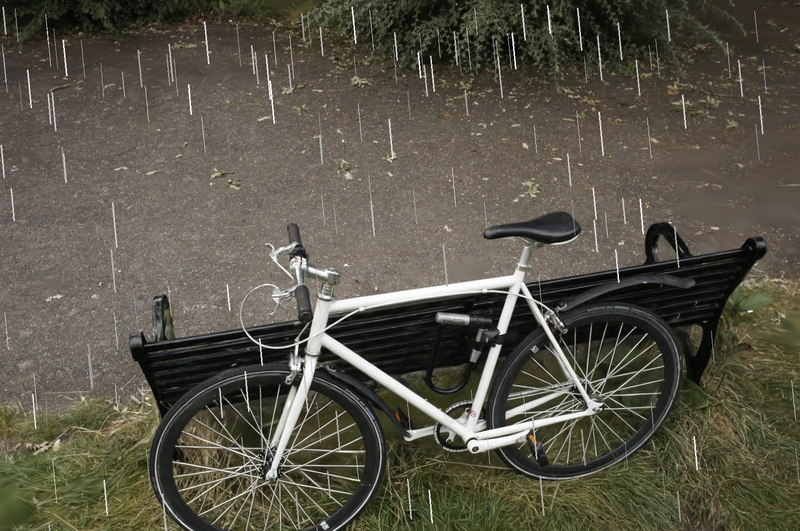} & 
        \formattedgraphics{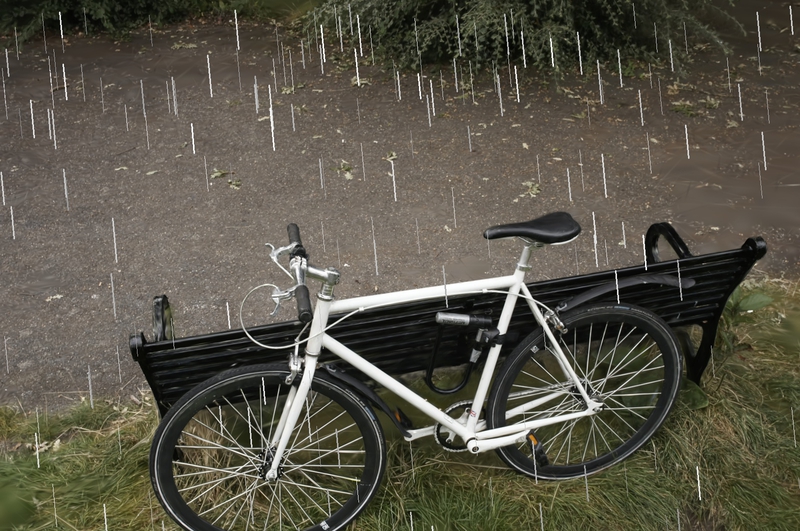} & 
        \formattedgraphics{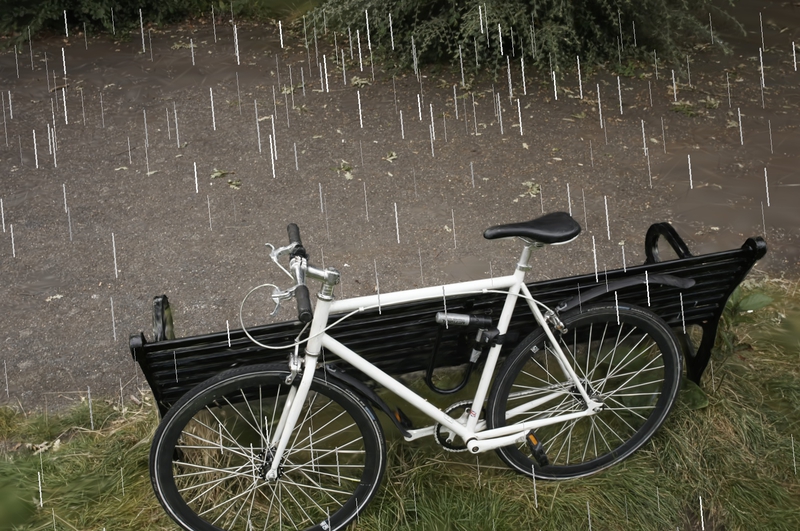} & 
        \formattedgraphics{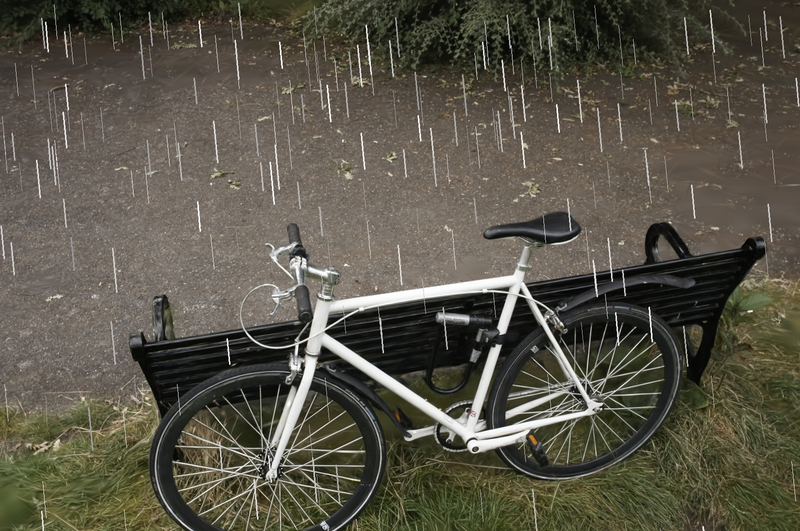} & 
        \formattedgraphics{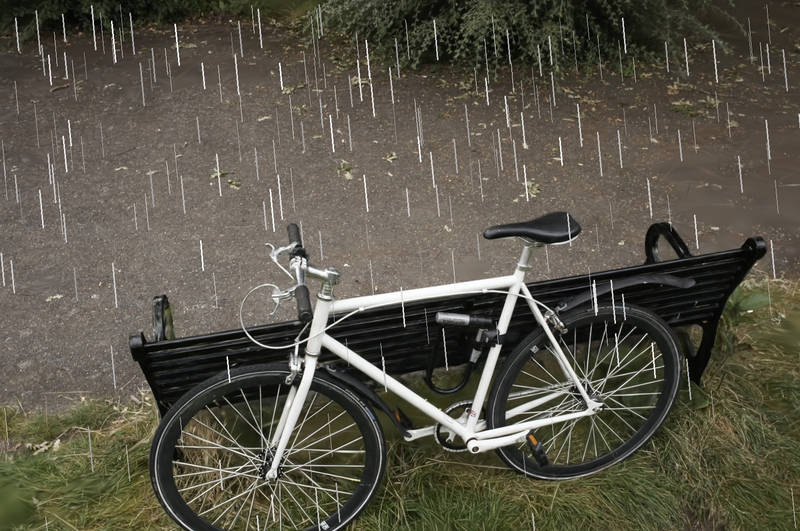}\\
        \rotatebox{90}{\textbf{Runway-V2V}} & 
        \formattedgraphics{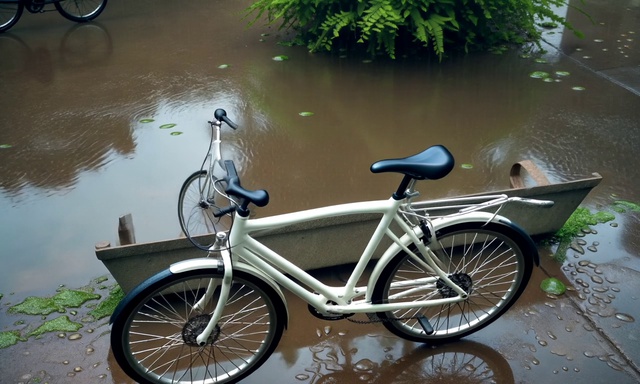} & 
        \formattedgraphics{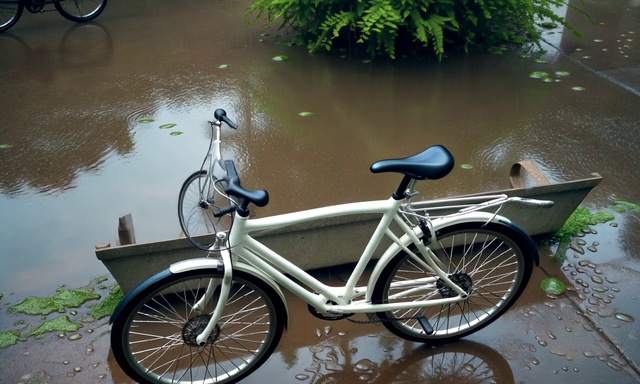} & 
        \formattedgraphics{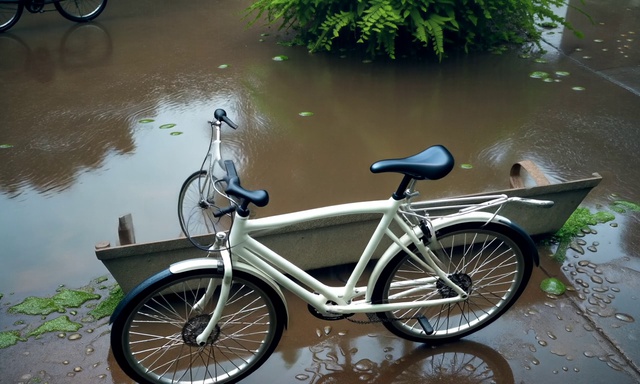} & 
        \formattedgraphics{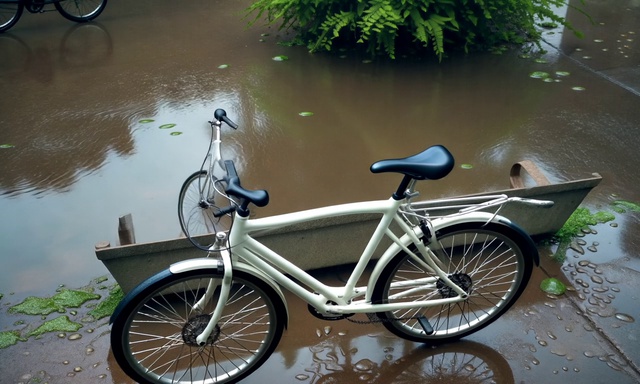} & 
        \formattedgraphics{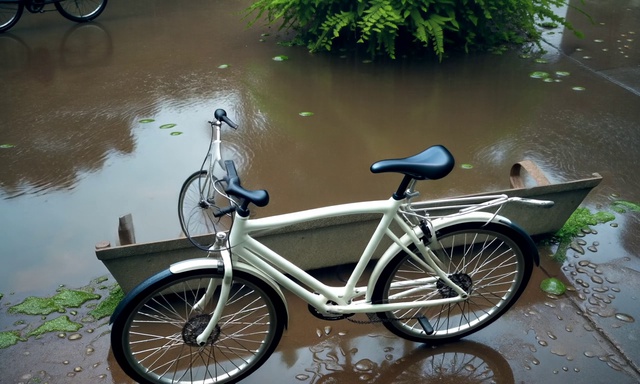}\\
        \rotatebox{90}{\textbf{Ours}} &
        \formattedgraphics{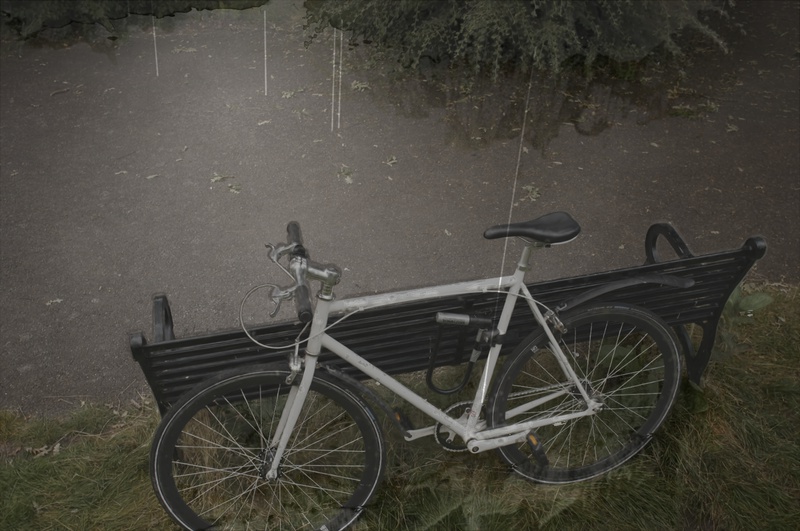} & 
        \formattedgraphics{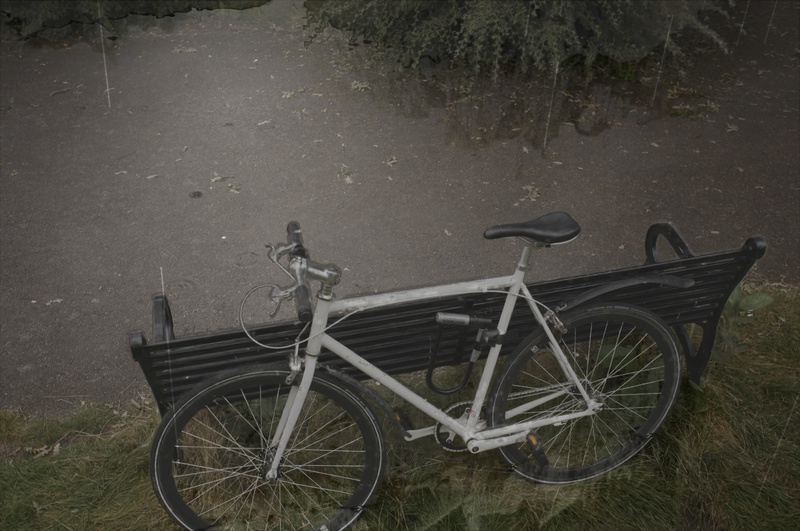} & 
        \formattedgraphics{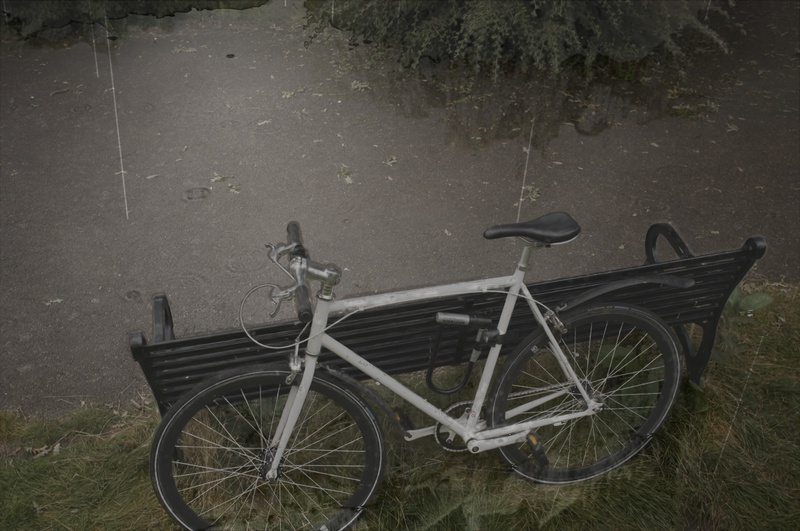} & 
        \formattedgraphics{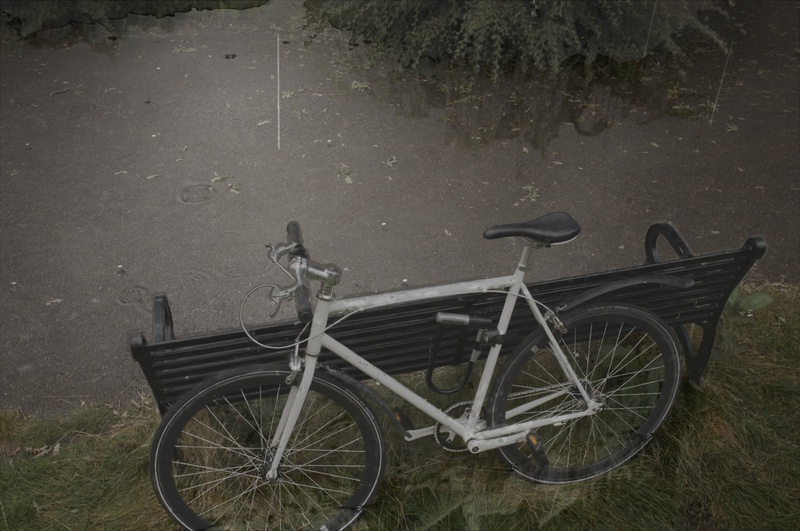} & 
        \formattedgraphics{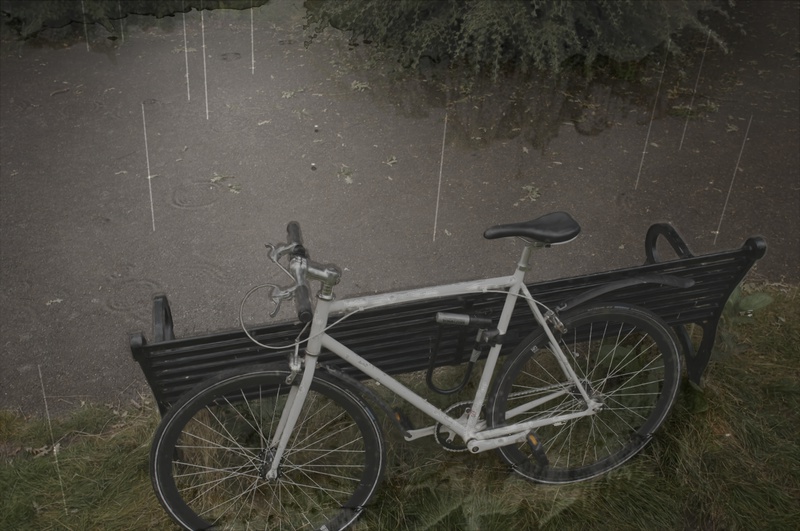}\\
    \end{tabular}
    \caption{\label{fig:fixed_view_bicycle}
    \Description{figure}
    \textbf{Bicycle Scenario.}
    Rain synthesis results from the same viewpoint at different timesteps are presented. Rain Motion generates unrealistic, random rain droplets without hydrops, while Runway-V2V generates physically inaccurate and unrealistic ripples. In contrast, RainyGS creates realistic, time-evolving hydrops and puddles.
    }
    %\vspace*{2cm}
\end{figure*}

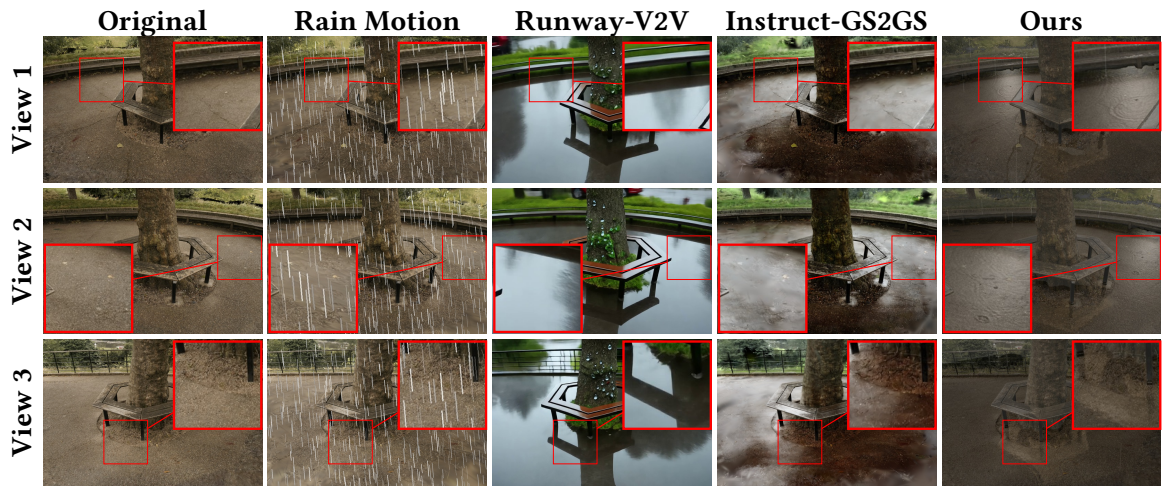
\begin{figure*}
    \centering
    \setlength{\tabcolsep}{1pt}
    \setlength{\imagewidth}{0.19\textwidth}
    \setlength{\imageheight}{0.087\textheight}
    \renewcommand{\arraystretch}{0.6}
    \newcommand{\formattedgraphics}[3]{% #1 = 图片路径, #2 = spy on参数， #3 = spy at参数
        \begin{tikzpicture}[spy using outlines={rectangle, magnification=2, connect spies}]
            \node[anchor=south west, inner sep=0] at (0,0){\includegraphics[width=\imagewidth, height=\imageheight]{#1}};
            \spy [red, size=30pt] on #2 in node at #3;
        \end{tikzpicture}%
    }
    \begin{tabular}{m{0.38cm}<{\centering} m{\imagewidth}<{\centering} m{\imagewidth}<{\centering} m{\imagewidth}<{\centering} m{\imagewidth}<{\centering} m{\imagewidth}<{\centering}}
        & \textbf{Original} & \textbf{Rain Motion} & \textbf{Runway-V2V} & \textbf{Instruct-GS2GS} & \textbf{Ours} \\
        \rotatebox{90}{\textbf{View 1}} &
        \formattedgraphics{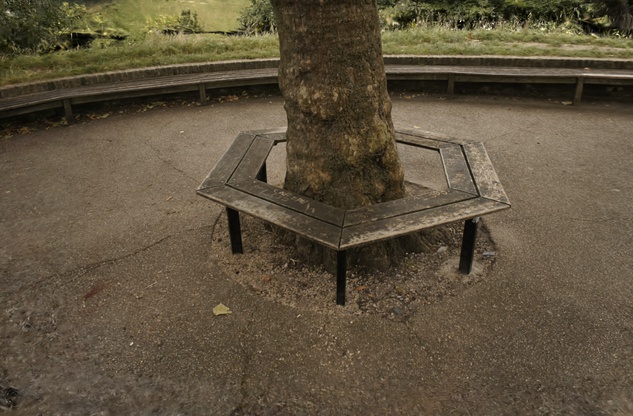}{(.27\imagewidth,.47\imagewidth)}{(.80\imagewidth,.44\imagewidth)} & 
        \formattedgraphics{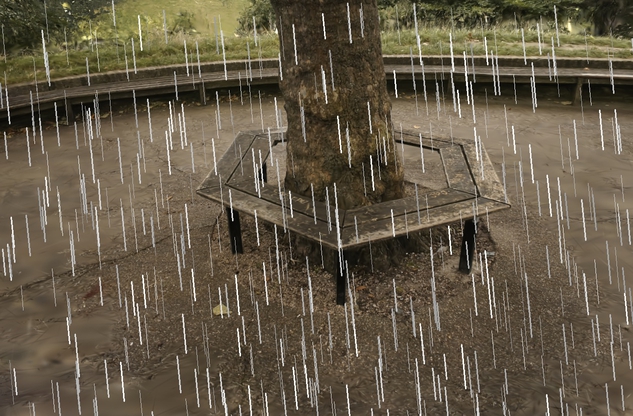}{(.27\imagewidth,.47\imagewidth)}{(.80\imagewidth,.44\imagewidth)} & 
        \formattedgraphics{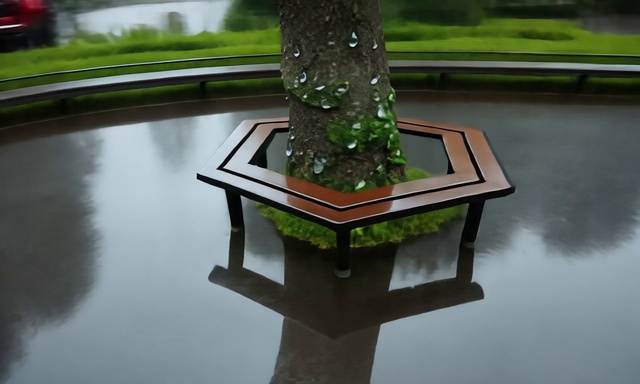}{(.27\imagewidth,.47\imagewidth)}{(.80\imagewidth,.44\imagewidth)} & 
        \formattedgraphics{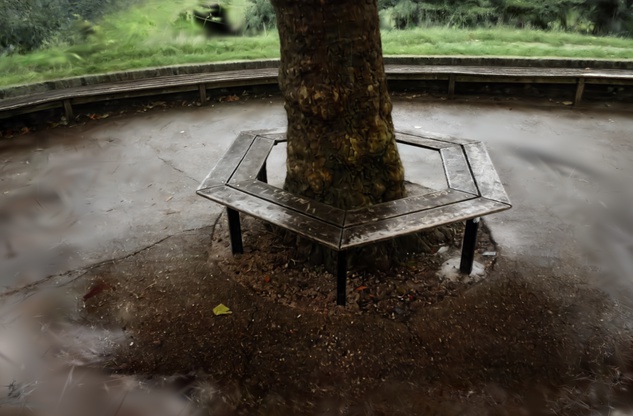}{(.27\imagewidth,.47\imagewidth)}{(.80\imagewidth,.44\imagewidth)} & 
        \formattedgraphics{img_sup/fixed_time/treehill_cr/ours/00010.png}{(.27\imagewidth,.47\imagewidth)}{(.80\imagewidth,.44\imagewidth)} \\
        \rotatebox{90}{\textbf{View 2}} &
        \formattedgraphics{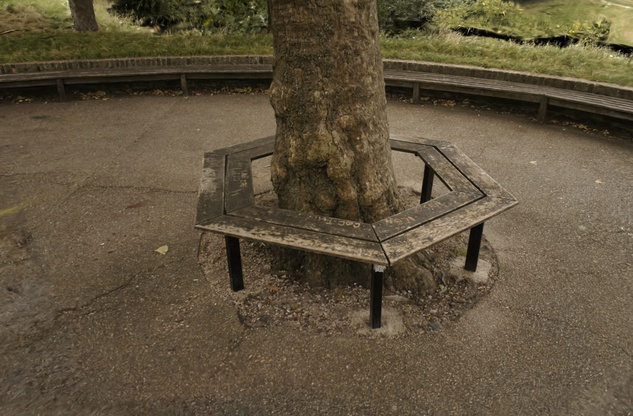}{(.90\imagewidth,.35\imagewidth)}{(.21\imagewidth,.21\imagewidth)} & 
        \formattedgraphics{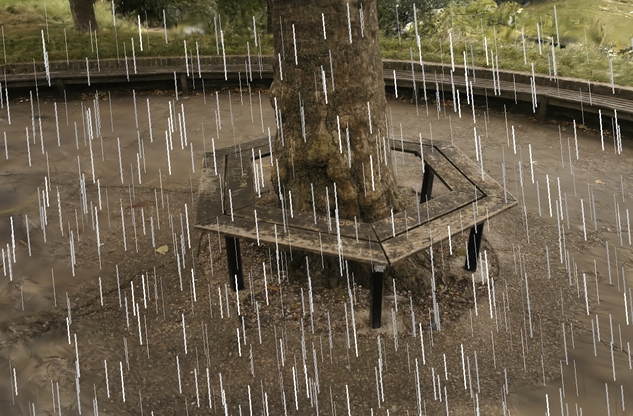}{(.90\imagewidth,.35\imagewidth)}{(.21\imagewidth,.21\imagewidth)} & 
        \formattedgraphics{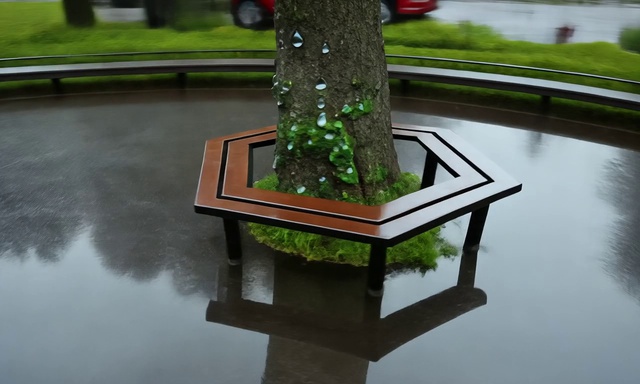}{(.90\imagewidth,.35\imagewidth)}{(.21\imagewidth,.21\imagewidth)} & 
        \formattedgraphics{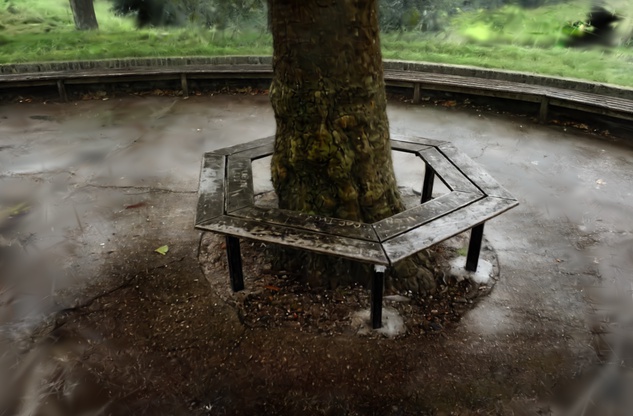}{(.90\imagewidth,.35\imagewidth)}{(.21\imagewidth,.21\imagewidth)} & 
        \formattedgraphics{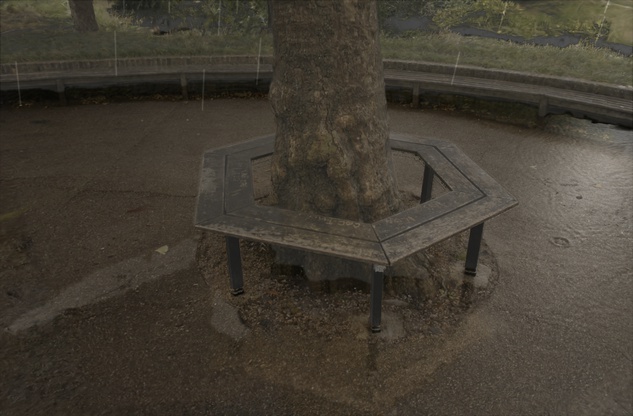}{(.90\imagewidth,.35\imagewidth)}{(.21\imagewidth,.21\imagewidth)} \\
        \rotatebox{90}{\textbf{View 3}} & 
        \formattedgraphics{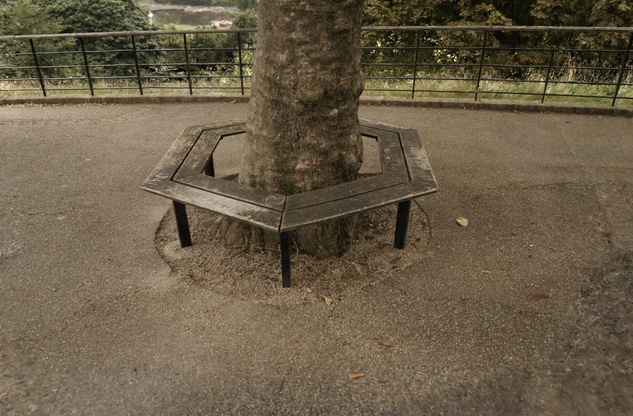}{(.38\imagewidth,.20\imagewidth)}{(.80\imagewidth,.46\imagewidth)} & 
        \formattedgraphics{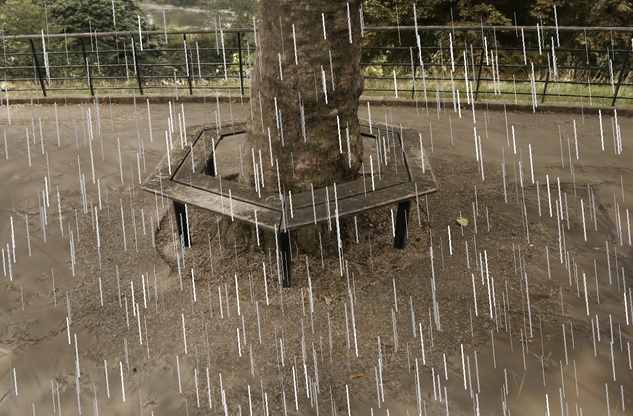}{(.38\imagewidth,.20\imagewidth)}{(.80\imagewidth,.46\imagewidth)} & 
        \formattedgraphics{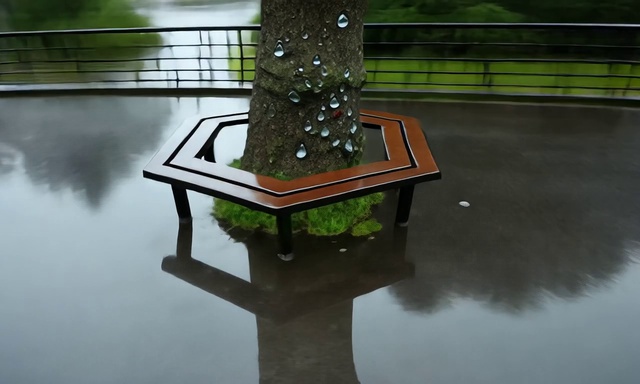}{(.38\imagewidth,.20\imagewidth)}{(.80\imagewidth,.46\imagewidth)} & 
        \formattedgraphics{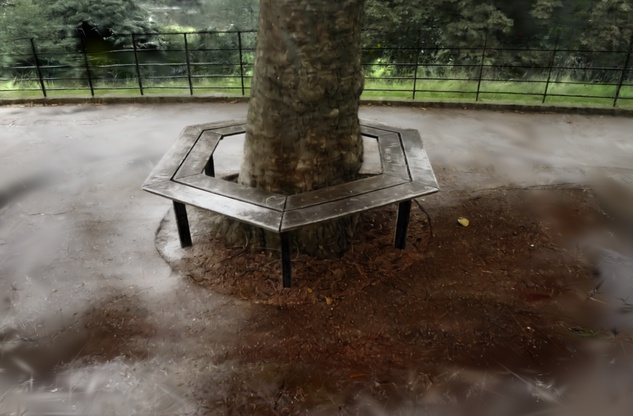}{(.38\imagewidth,.20\imagewidth)}{(.80\imagewidth,.46\imagewidth)} & 
        \formattedgraphics{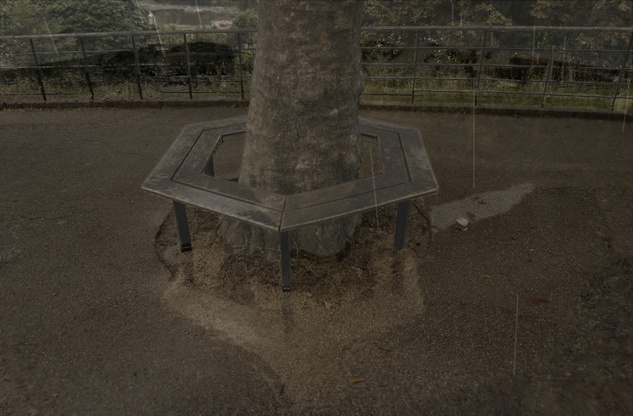}{(.38\imagewidth,.20\imagewidth)}{(.80\imagewidth,.46\imagewidth)} \\
    \end{tabular}
    \caption{\label{fig:fixed_time_treehill} \textbf{Treehill Scenario.} Comparison of rain synthesis at the same timestep from various perspectives. Runway-V2V struggles with 3D consistency, and Rain Motion and Instruct-GS2GS cannot produce realistic rain streaks, puddles, or ripples. By contrast, RainyGS achieves stable 3D consistency and faithfully renders rain streaks and water accumulation.}
    \Description{figure}
    \vspace*{1cm}
\end{figure*}

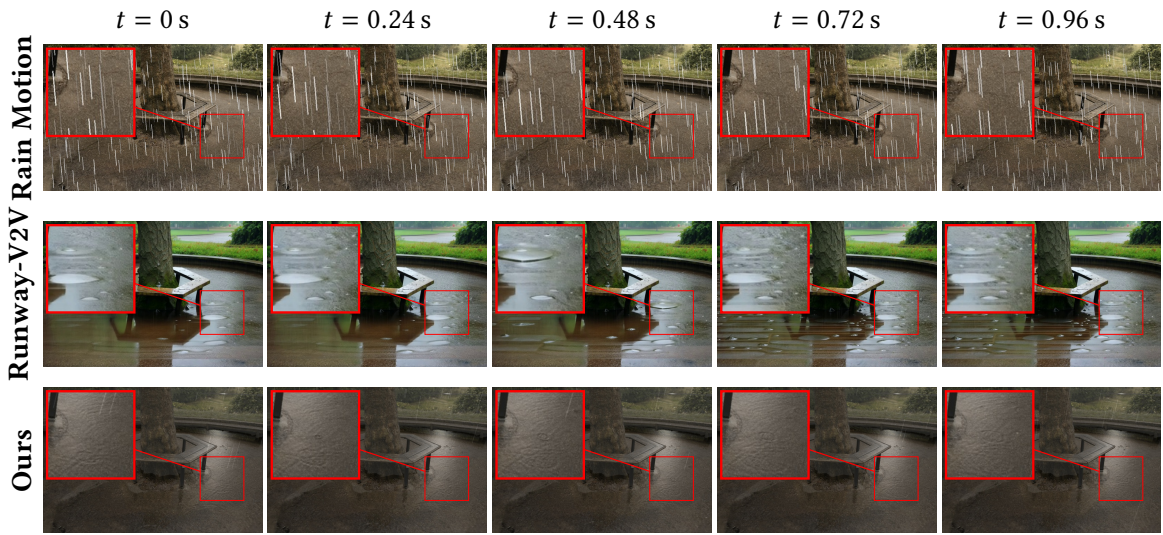
\begin{figure*}
    \centering
    \setlength{\tabcolsep}{1pt}
    \setlength{\imagewidth}{0.19\textwidth}
    \setlength{\imageheight}{0.087\textheight}
    \renewcommand{\arraystretch}{0.6}
    \newcommand{\formattedgraphics}[1]{%
      \begin{tikzpicture}[spy using outlines={rectangle, magnification=2, connect spies}]
        \node[anchor=south west, inner sep=0] at (0,0){\includegraphics[width=\imagewidth, height=\imageheight]{#1}};
        \spy [red,size=30pt] on (.82\imagewidth,.25\imagewidth) in node at (.22\imagewidth,.45\imagewidth);
        \end{tikzpicture}%
    }
    \begin{tabular}{m{0.38cm}<{\centering}m{\imagewidth}<{\centering}m{\imagewidth}<{\centering}m{\imagewidth}<{\centering}m{\imagewidth}<{\centering}m{\imagewidth}<{\centering}}
        & $t=\SI{0}{\second}$ & $t=\SI{0.24}{\second}$ & $t=\SI{0.48}{\second}$ & $t=\SI{0.72}{\second}$ & $t=\SI{0.96}{\second}$\\
        \rotatebox{90}{\textbf{Rain Motion}} &
        \formattedgraphics{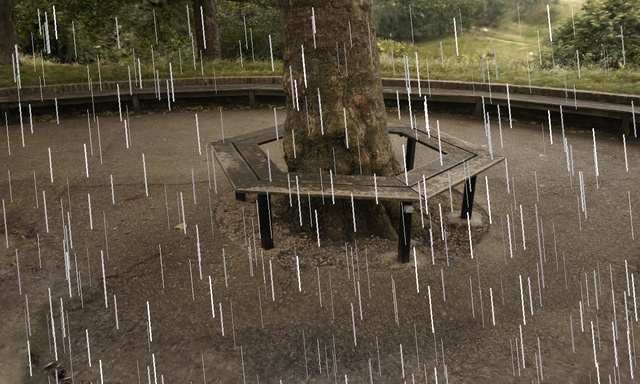} & 
        \formattedgraphics{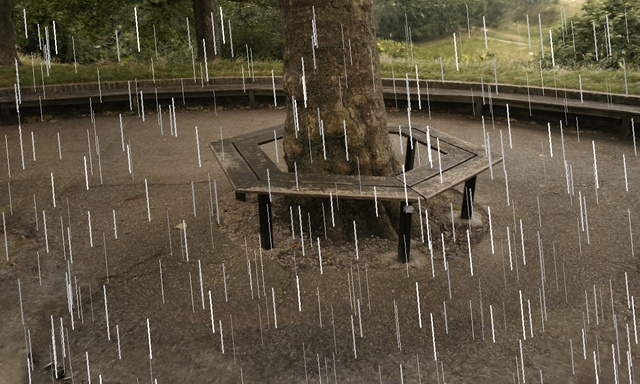} & 
        \formattedgraphics{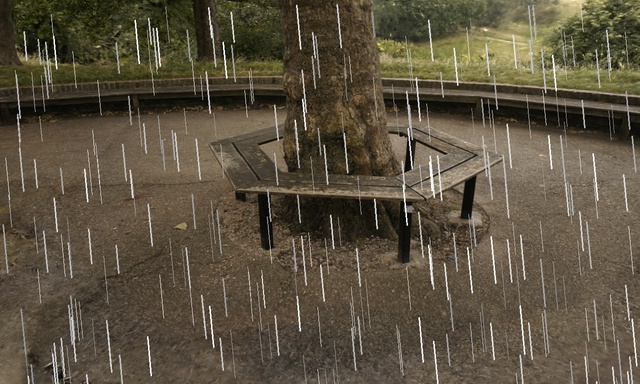} & 
        \formattedgraphics{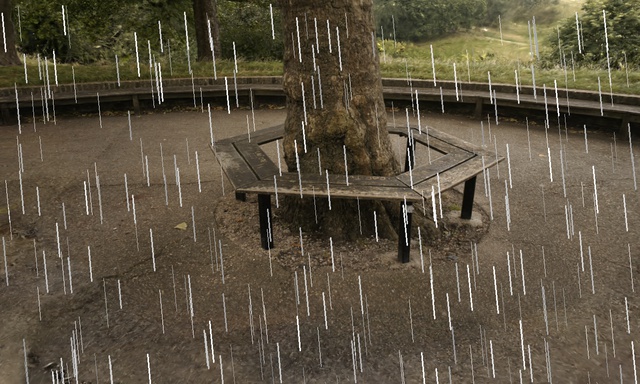} & 
        \formattedgraphics{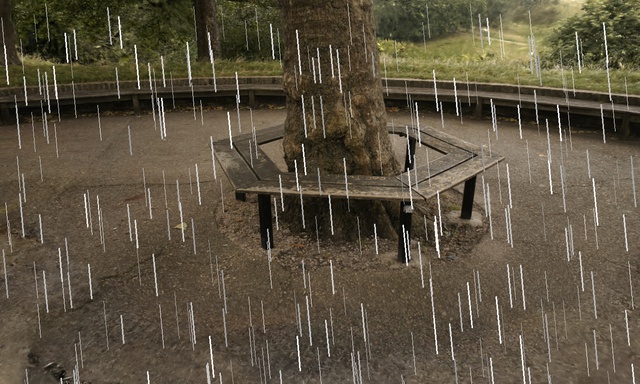}\\
        \rotatebox{90}{\textbf{Runway-V2V}} & 
        \formattedgraphics{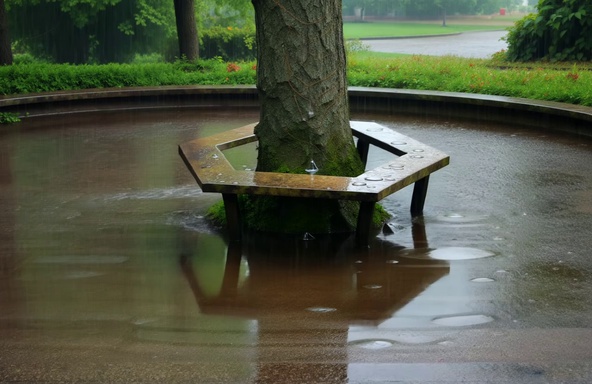} & 
        \formattedgraphics{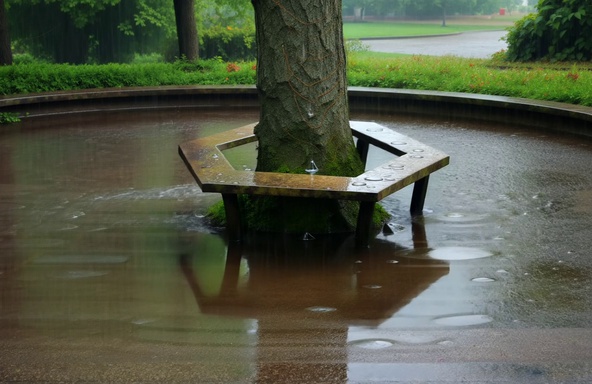} & 
        \formattedgraphics{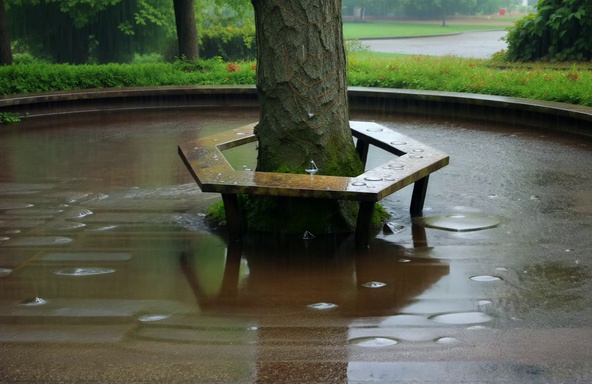} & 
        \formattedgraphics{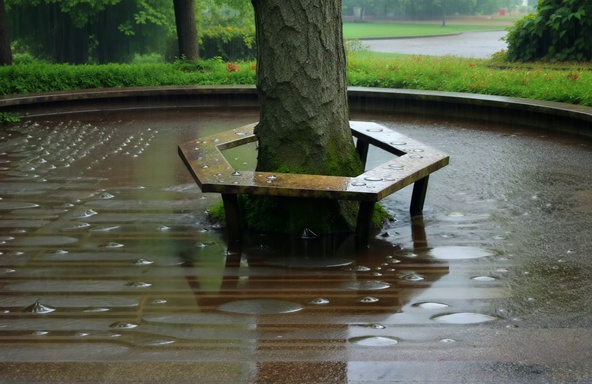} & 
        \formattedgraphics{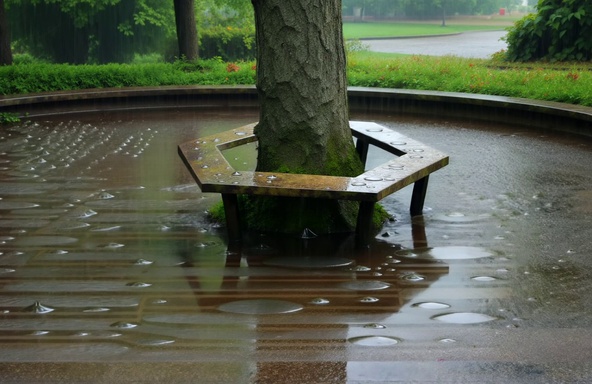}\\
        \rotatebox{90}{\textbf{Ours}} &
        \formattedgraphics{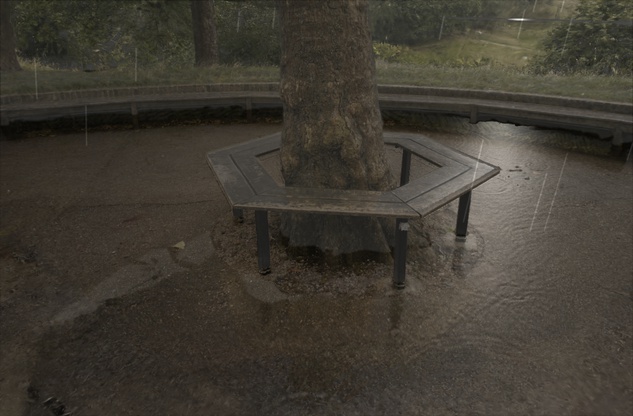} & 
        \formattedgraphics{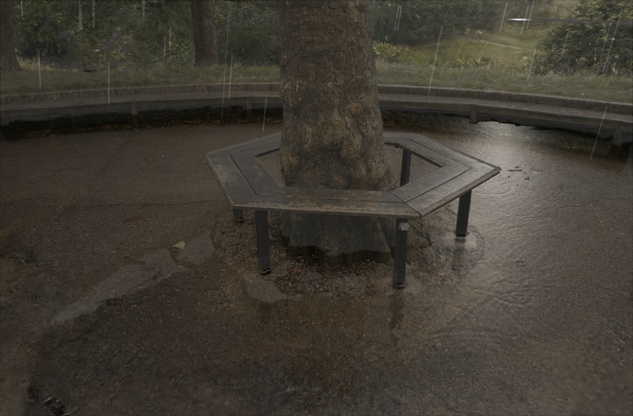} & 
        \formattedgraphics{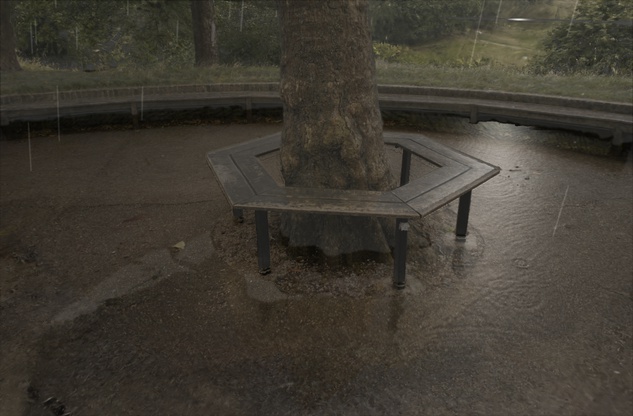} & 
        \formattedgraphics{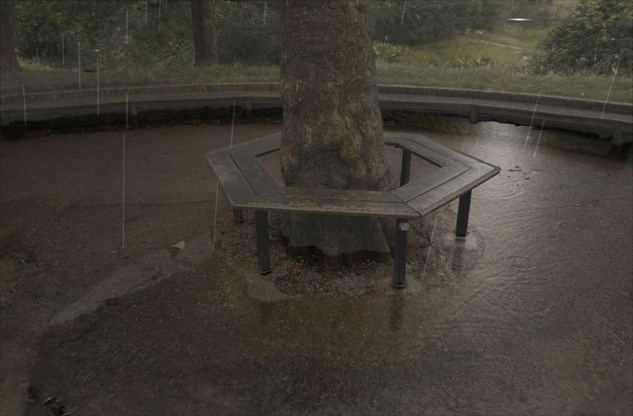} & 
        \formattedgraphics{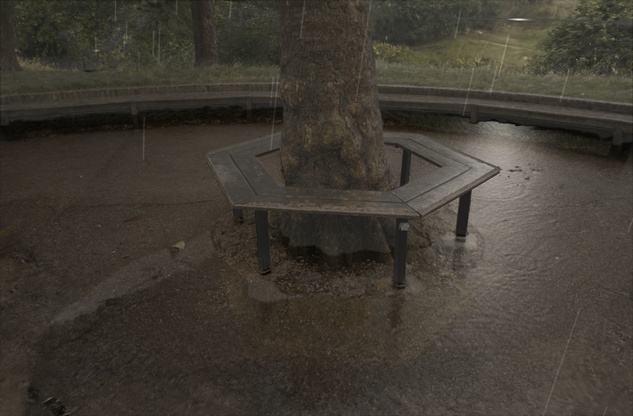}\\
    \end{tabular}
    \caption{\label{fig:fixed_view_treehill}
    \Description{figure}
    \textbf{Treehill Scenario.}
    Results of rain synthesis at different moments in time from the same viewpoint are displayed. Rain Motion fails to produce realistic rain, creating random droplets without hydrops, and Runway-V2V outputs almost static rain. RainyGS generates realistic and dynamic hydrops and puddles.
    }
    \vspace*{2cm}
\end{figure*}

\begin{figure*}
    \centering
    \setlength{\tabcolsep}{1pt}
    \setlength{\imagewidth}{0.19\textwidth}
    \setlength{\imageheight}{0.087\textheight}
    \renewcommand{\arraystretch}{0.6}
    \newcommand{\formattedgraphics}[3]{% #1 = 图片路径, #2 = spy on参数， #3 = spy at参数
        \begin{tikzpicture}[spy using outlines={rectangle, magnification=2, connect spies}]
            \node[anchor=south west, inner sep=0] at (0,0){\includegraphics[width=\imagewidth, height=\imageheight]{#1}};
            \spy [red, size=30pt] on #2 in node at #3;
        \end{tikzpicture}%
    }
    \begin{tabular}{m{0.38cm}<{\centering} m{\imagewidth}<{\centering} m{\imagewidth}<{\centering}  m{\imagewidth}<{\centering} m{\imagewidth}<{\centering} m{\imagewidth}<{\centering}}
        & \textbf{Original} & \textbf{Rain Motion} & \textbf{Runway-V2V} & \textbf{Instruct-GS2GS}& \textbf{Ours} \\
        \rotatebox{90}{\textbf{View 1}} &
        \formattedgraphics{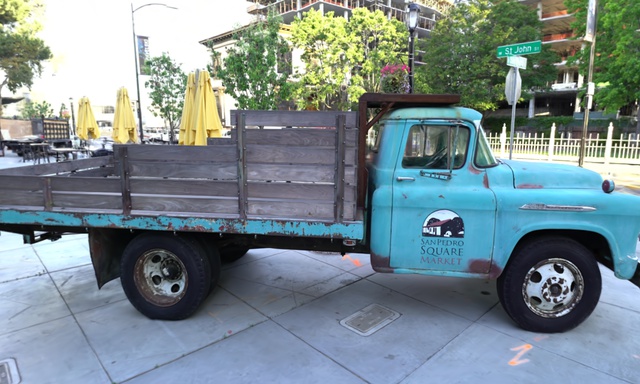}{(.85\imagewidth,.11\imagewidth)}{(.80\imagewidth,.45\imagewidth)} & 
        \formattedgraphics{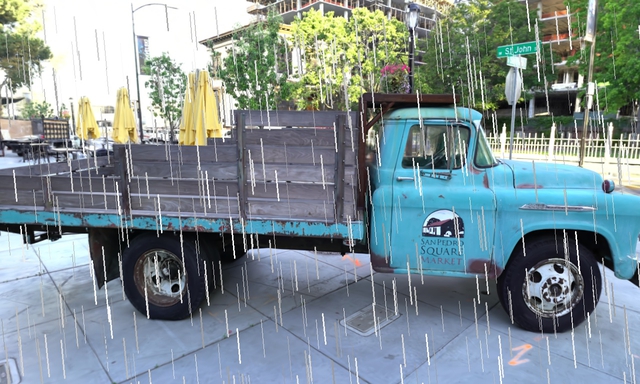}{(.85\imagewidth,.11\imagewidth)}{(.80\imagewidth,.45\imagewidth)} & 
        \formattedgraphics{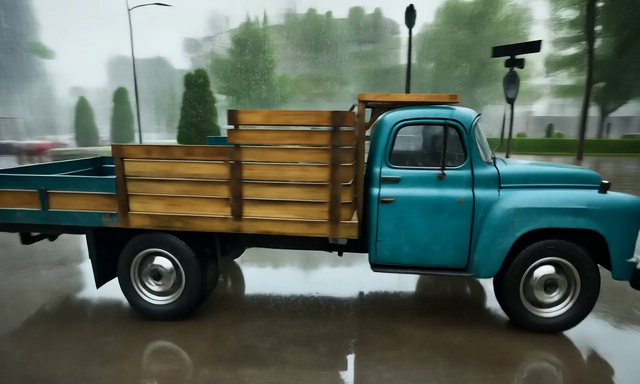}{(.85\imagewidth,.11\imagewidth)}{(.80\imagewidth,.45\imagewidth)}& 

        \formattedgraphics{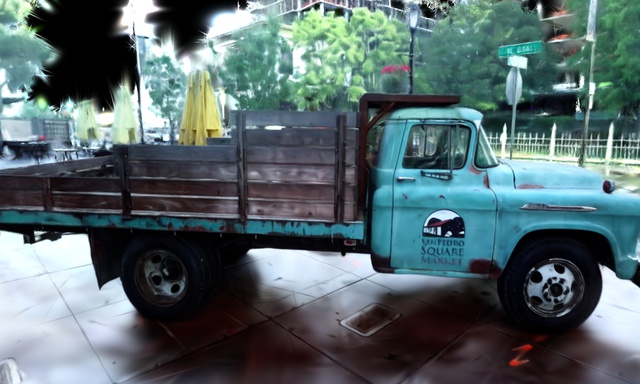}{(.85\imagewidth,.11\imagewidth)}{(.80\imagewidth,.45\imagewidth)} & 
        \formattedgraphics{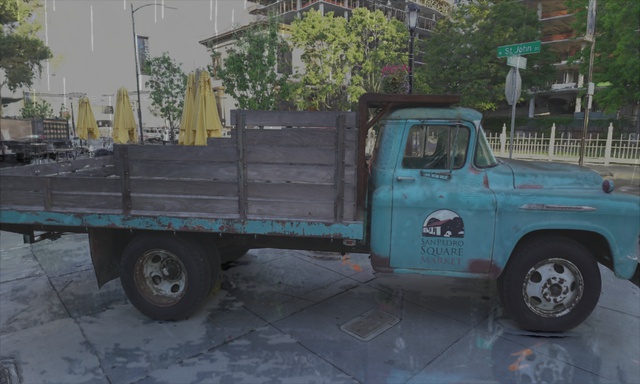}{(.85\imagewidth,.11\imagewidth)}{(.80\imagewidth,.45\imagewidth)} \\
        \rotatebox{90}{\textbf{View 2}} &
        \formattedgraphics{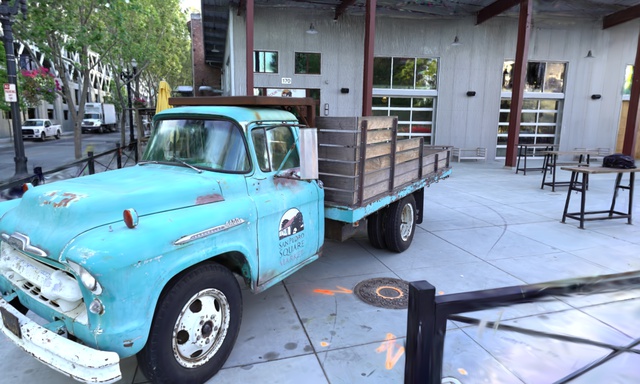}{(.55\imagewidth,.10\imagewidth)}{(.80\imagewidth,.45\imagewidth)}   & 
        \formattedgraphics{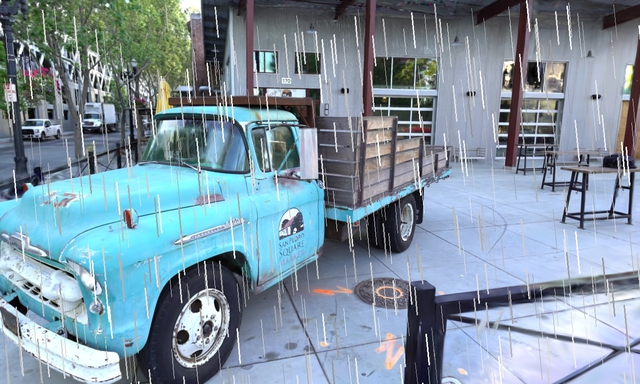}{(.55\imagewidth,.10\imagewidth)}{(.80\imagewidth,.45\imagewidth)}   & 
        \formattedgraphics{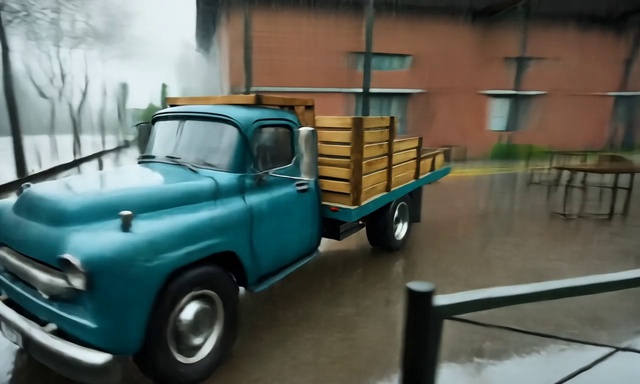}{(.55\imagewidth,.10\imagewidth)}{(.80\imagewidth,.45\imagewidth)}   & 
        \formattedgraphics{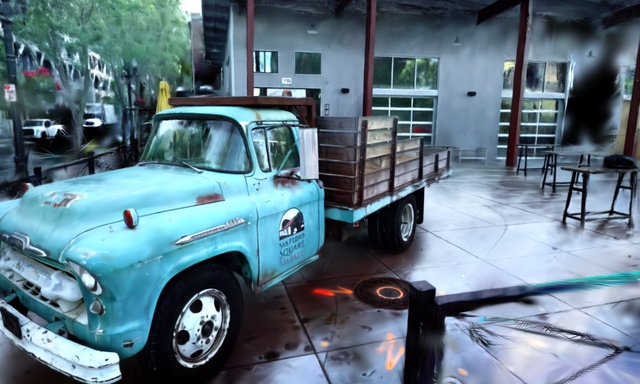}{(.55\imagewidth,.10\imagewidth)}{(.80\imagewidth,.45\imagewidth)}  & 
        \formattedgraphics{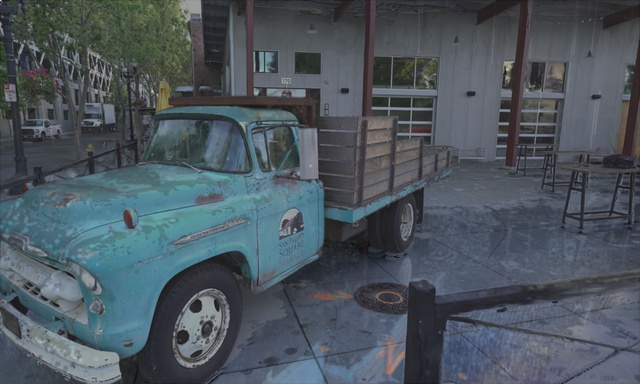}{(.55\imagewidth,.10\imagewidth)}{(.80\imagewidth,.45\imagewidth)} \\
        \rotatebox{90}{\textbf{View 3}} & 
        \formattedgraphics{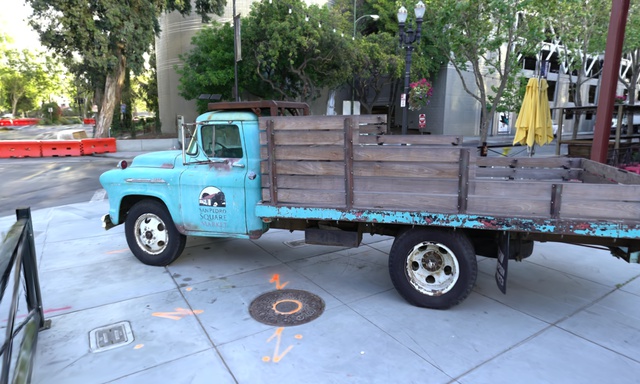}{(.68\imagewidth,.11\imagewidth)}{(.80\imagewidth,.45\imagewidth)}& 
        \formattedgraphics{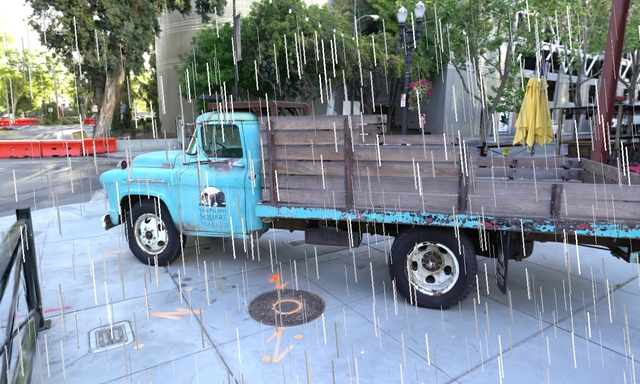}{(.68\imagewidth,.11\imagewidth)}{(.80\imagewidth,.45\imagewidth)}& 
        \formattedgraphics{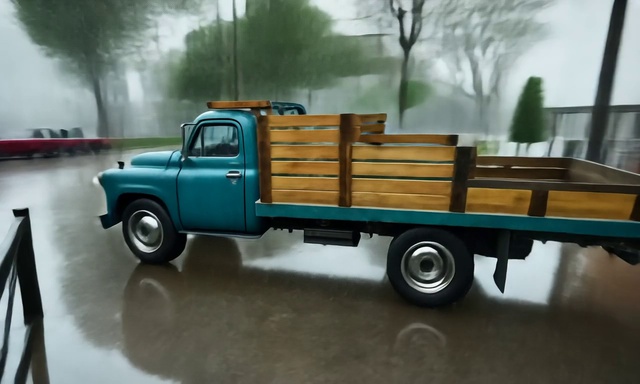}{(.68\imagewidth,.11\imagewidth)}{(.80\imagewidth,.45\imagewidth)}& 
        \formattedgraphics{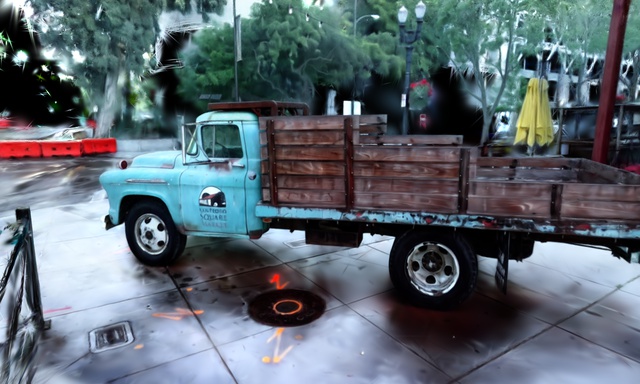}{(.68\imagewidth,.11\imagewidth)}{(.80\imagewidth,.45\imagewidth)} & 
        \formattedgraphics{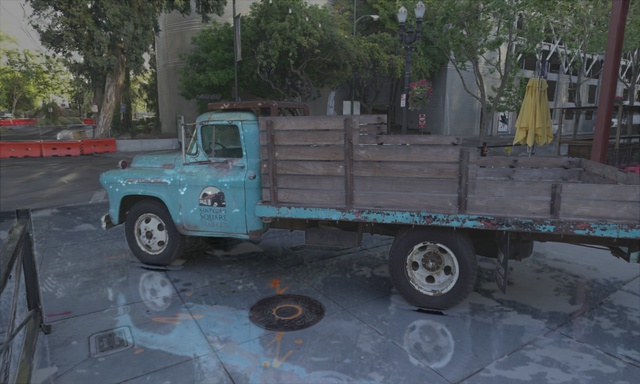}{(.68\imagewidth,.11\imagewidth)}{(.80\imagewidth,.45\imagewidth)} \\
    \end{tabular}
    \caption{\label{fig:fixed_time_truck} \textbf{Truck Scenario.} Results of rain synthesis at an identical timestep observed from multiple views. Runway-V2V fails to maintain 3D consistency, and both Rain Motion and Instruct-GS2GS lack realistic rain streaks, puddles, and ripples. In contrast, RainyGS maintains 3D consistency and generates lifelike rain streaks and puddles.}
    \Description{figure}
    \vspace*{0.5cm}
\end{figure*}
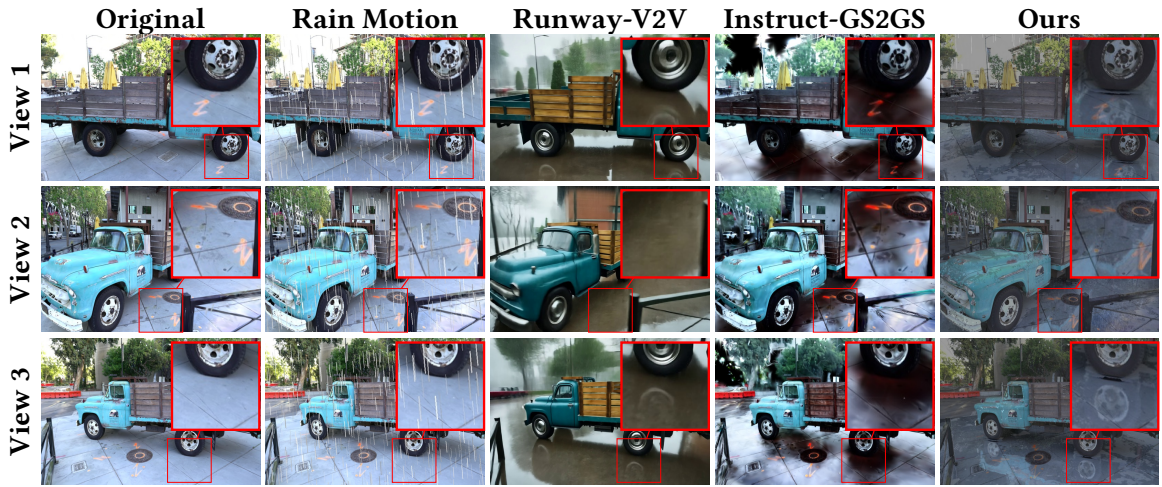

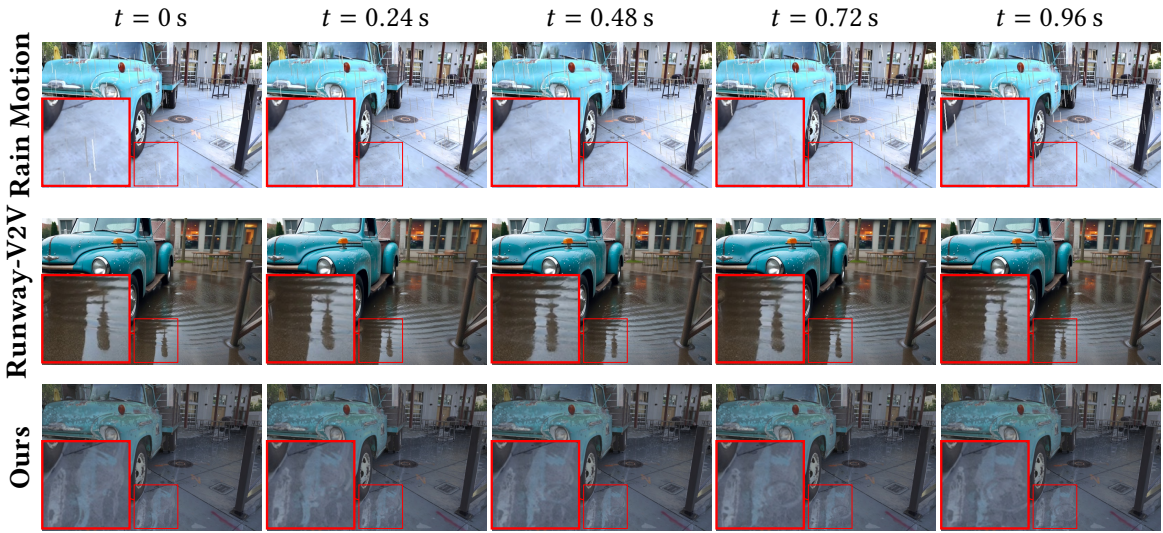
\begin{figure*}
    \centering
    \setlength{\tabcolsep}{1pt}
    \setlength{\imagewidth}{0.19\textwidth}
    \setlength{\imageheight}{0.087\textheight}
    \renewcommand{\arraystretch}{0.6}
    \newcommand{\formattedgraphics}[1]{%
      \begin{tikzpicture}[spy using outlines={rectangle, magnification=2, connect spies}]
        \node[anchor=south west, inner sep=0] at (0,0){\includegraphics[width=\imagewidth, height=\imageheight]{#1}};
        \spy [red,size=30pt] on (.52\imagewidth,.11\imagewidth) in node at (.20\imagewidth,.21\imagewidth);
        \end{tikzpicture}%
    }
    \begin{tabular}{m{0.38cm}<{\centering}m{\imagewidth}<{\centering}m{\imagewidth}<{\centering}m{\imagewidth}<{\centering}m{\imagewidth}<{\centering}m{\imagewidth}<{\centering}}
        & $t=\SI{0}{\second}$ & $t=\SI{0.24}{\second}$ & $t=\SI{0.48}{\second}$ & $t=\SI{0.72}{\second}$ & $t=\SI{0.96}{\second}$\\
        \rotatebox{90}{\textbf{Rain Motion}} &
        \formattedgraphics{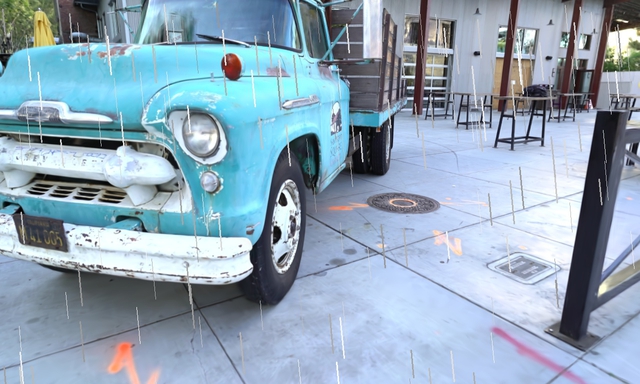} & 
        \formattedgraphics{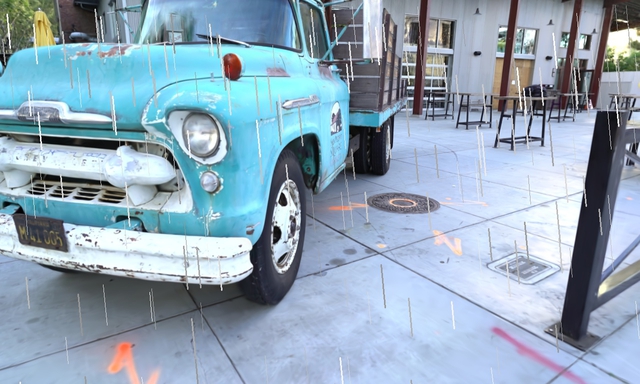} & 
        \formattedgraphics{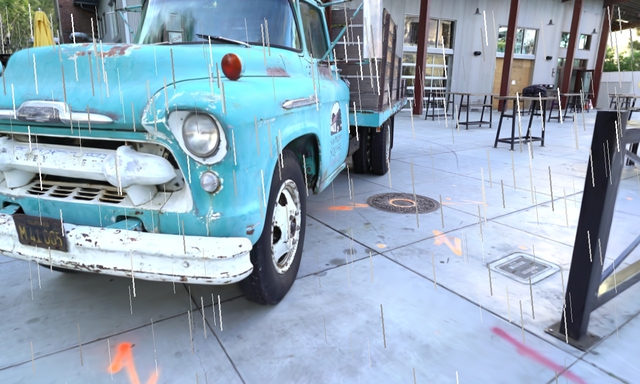} & 
        \formattedgraphics{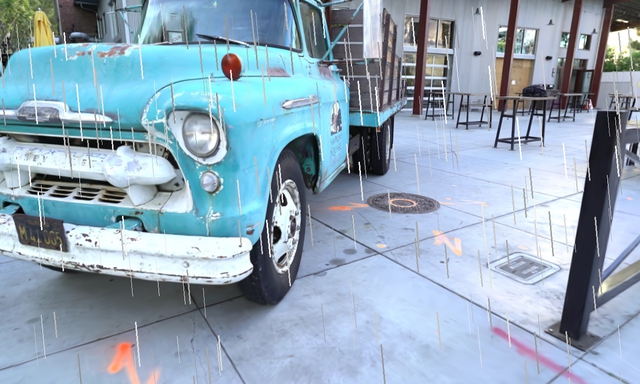} & 
        \formattedgraphics{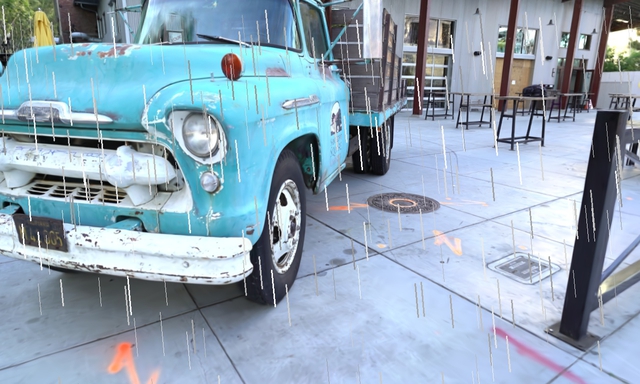}\\
        \rotatebox{90}{\textbf{Runway-V2V}} & 
        \formattedgraphics{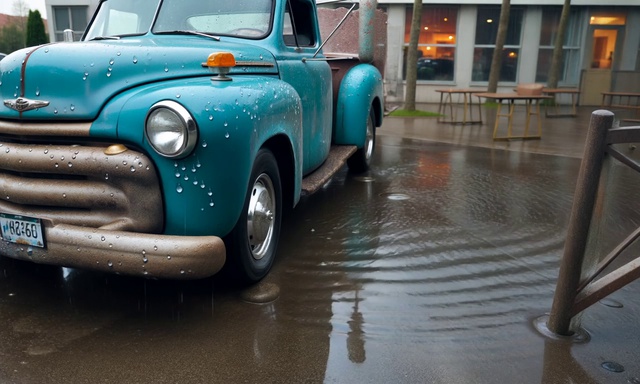} & 
        \formattedgraphics{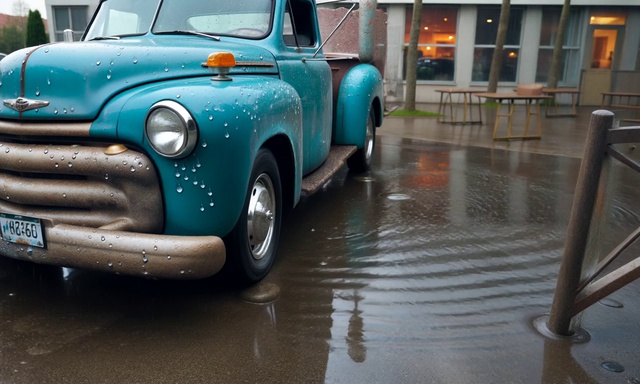} & 
        \formattedgraphics{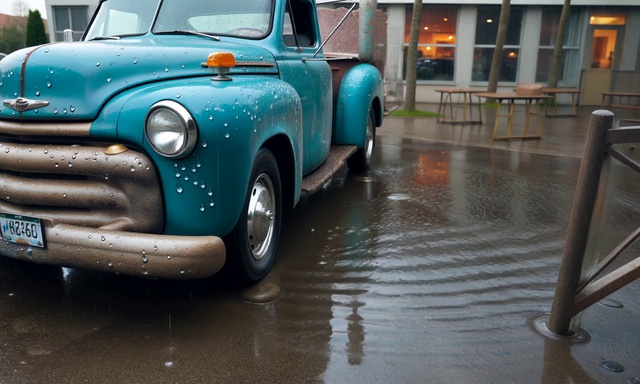} & 
        \formattedgraphics{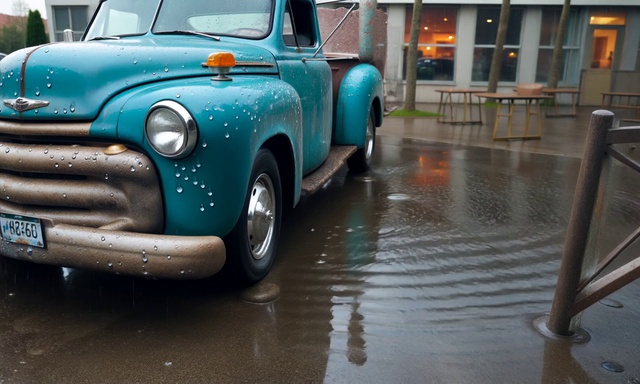} & 
        \formattedgraphics{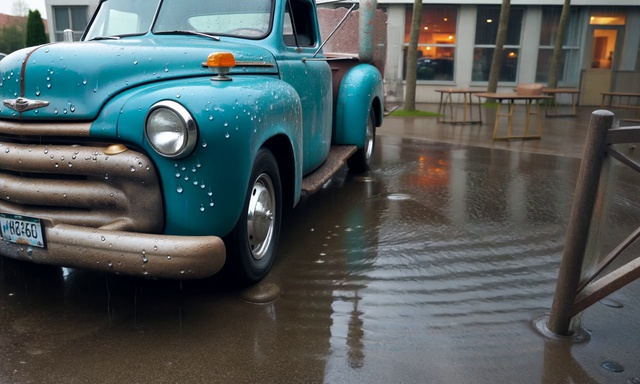}\\
        \rotatebox{90}{\textbf{Ours}} &
        \formattedgraphics{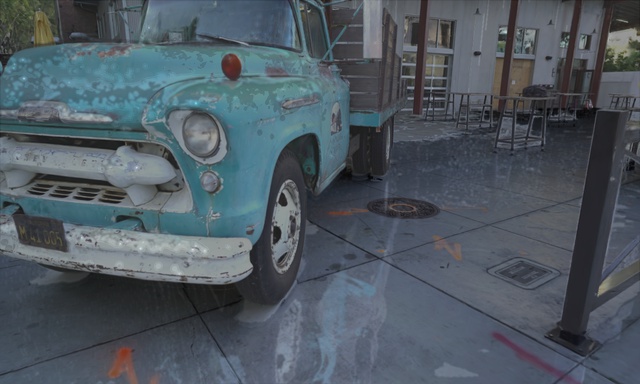} & 
        \formattedgraphics{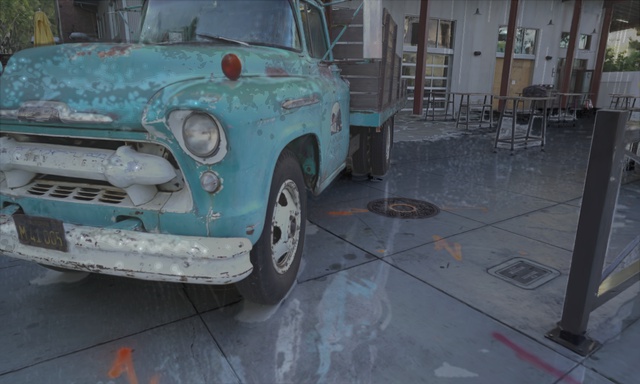} & 
        \formattedgraphics{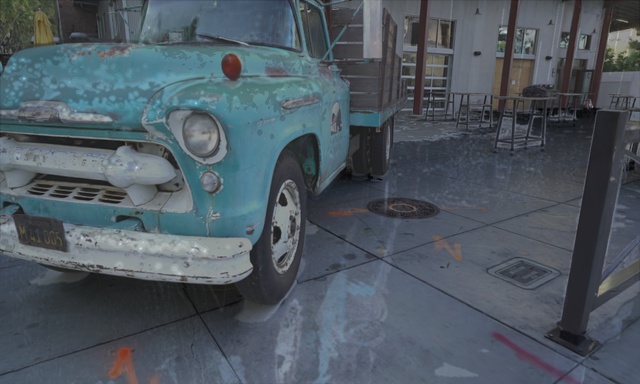} & 
        \formattedgraphics{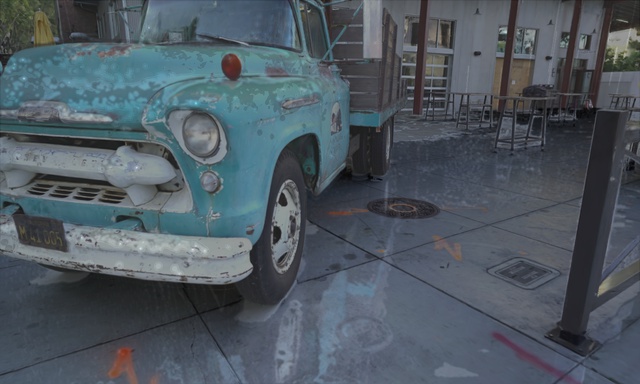} & 
        \formattedgraphics{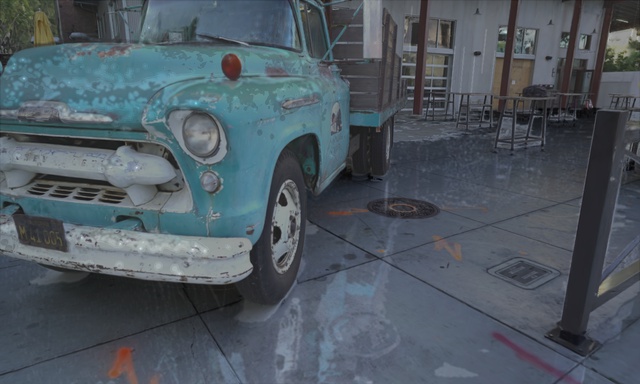}\\
    \end{tabular}
    \caption{\label{fig:fixed_view_truck}
    \Description{figure}
  \textbf{Truck Scenario.}
   Rain synthesis outputs from the same perspective over different timesteps are illustrated. Rain Motion creates artificial and random droplets with no hydrops, and Runway-V2V produces ripples that are neither physically plausible nor visually realistic. RainyGS, on the other hand, generates dynamic hydrops and puddles with temporal realism.
    }
    \vspace*{2cm}
\end{figure*}

\begin{figure*}
    \centering
    \setlength{\tabcolsep}{1pt}
    \setlength{\imagewidth}{0.19\textwidth}
    \setlength{\imageheight}{0.087\textheight}
    \renewcommand{\arraystretch}{0.6}
    \newcommand{\formattedgraphics}[3]{% #1 = 图片路径, #2 = spy on参数， #3 = spy at参数
        \begin{tikzpicture}[spy using outlines={rectangle, magnification=2, connect spies}]
            \node[anchor=south west, inner sep=0] at (0,0){\includegraphics[width=\imagewidth, height=\imageheight]{#1}};
            \spy [red, size=30pt] on #2 in node at #3;
        \end{tikzpicture}%
    }
    \begin{tabular}{m{0.38cm}<{\centering} m{\imagewidth}<{\centering} m{\imagewidth}<{\centering} m{\imagewidth}<{\centering} m{\imagewidth}<{\centering} m{\imagewidth}<{\centering}}
        & \textbf{Original} & \textbf{Rain Motion} & \textbf{Runway-V2V} & \textbf{Instruct-GS2GS} & \textbf{Ours} \\
        \rotatebox{90}{\textbf{View 1}} &
        \formattedgraphics{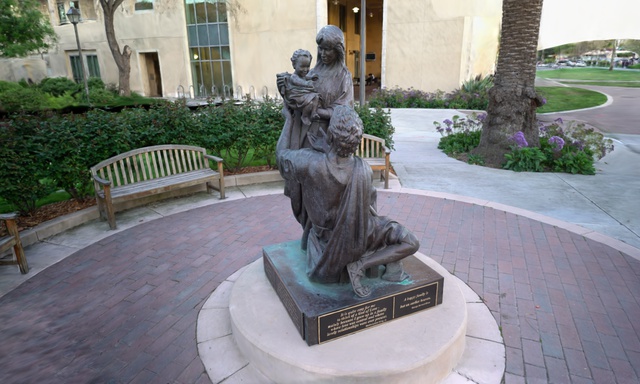}{(.82\imagewidth,.16\imagewidth)}{(.79\imagewidth,.47\imagewidth)} & 
        \formattedgraphics{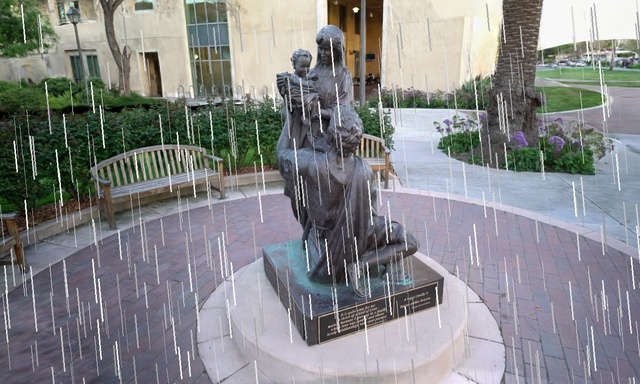}{(.82\imagewidth,.16\imagewidth)}{(.79\imagewidth,.47\imagewidth)}  & 
        \formattedgraphics{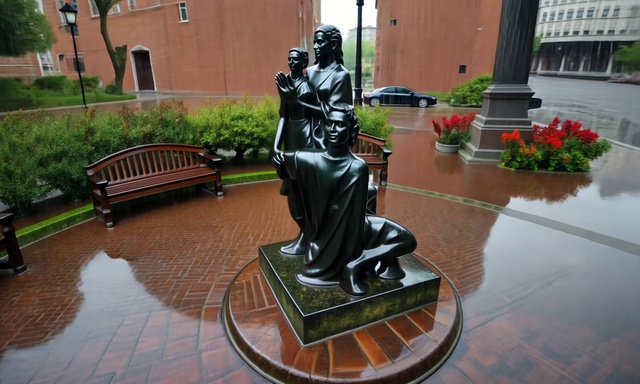}{(.82\imagewidth,.16\imagewidth)}{(.79\imagewidth,.47\imagewidth)}  & 
        \formattedgraphics{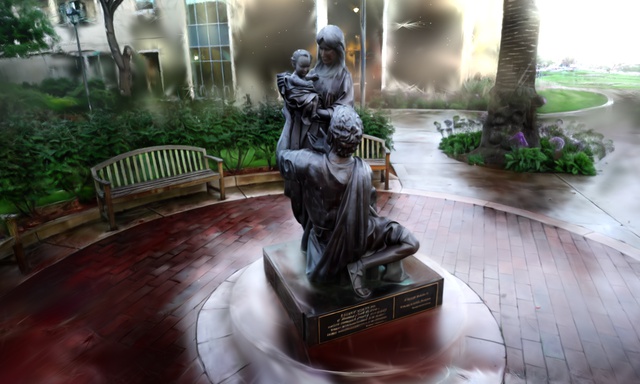}{(.82\imagewidth,.16\imagewidth)}{(.79\imagewidth,.47\imagewidth)} & 
        \formattedgraphics{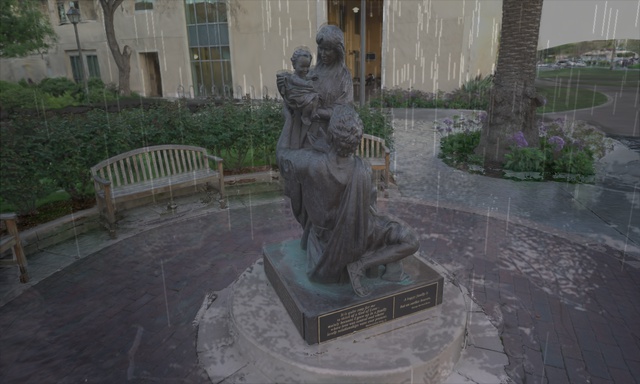}{(.82\imagewidth,.16\imagewidth)}{(.79\imagewidth,.47\imagewidth)} \\
        \rotatebox{90}{\textbf{View 2}} &
        \formattedgraphics{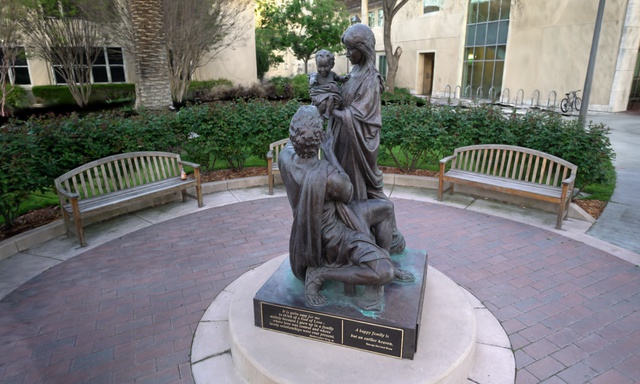}{(.29\imagewidth,.12\imagewidth)}{(.22\imagewidth,.46\imagewidth)}   & 
        \formattedgraphics{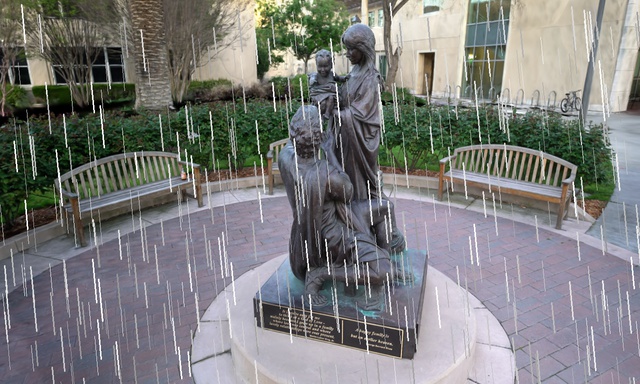}{(.29\imagewidth,.12\imagewidth)}{(.22\imagewidth,.46\imagewidth)}   & 
        \formattedgraphics{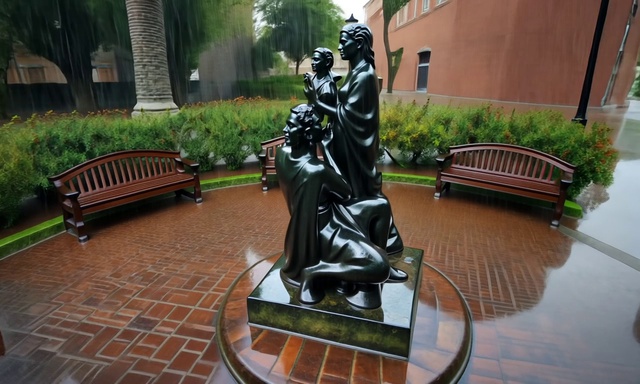}{(.29\imagewidth,.12\imagewidth)}{(.22\imagewidth,.46\imagewidth)}   & 
        \formattedgraphics{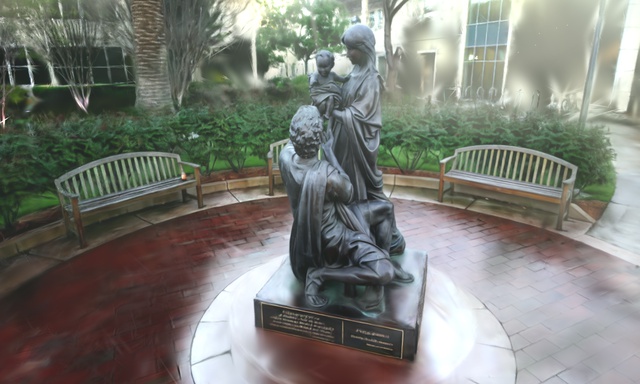}{(.29\imagewidth,.12\imagewidth)}{(.22\imagewidth,.46\imagewidth)}  & 
        \formattedgraphics{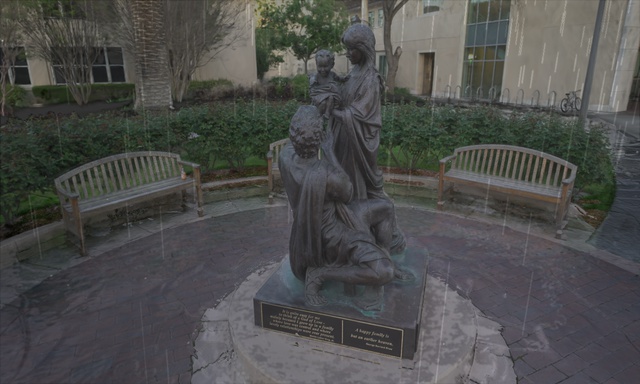}{(.29\imagewidth,.12\imagewidth)}{(.22\imagewidth,.46\imagewidth)}  \\
        \rotatebox{90}{\textbf{View 3}} & 
        \formattedgraphics{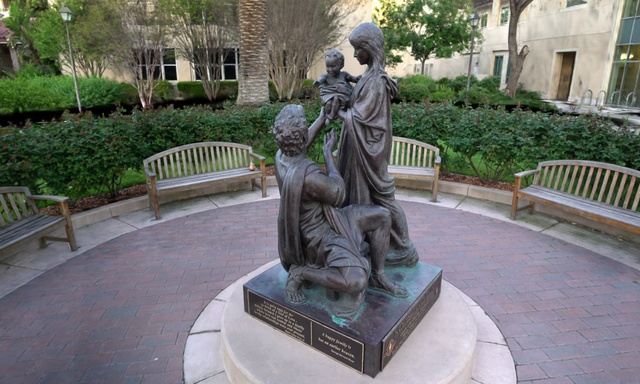}{(.29\imagewidth,.12\imagewidth)}{(.22\imagewidth,.46\imagewidth)} & 
        \formattedgraphics{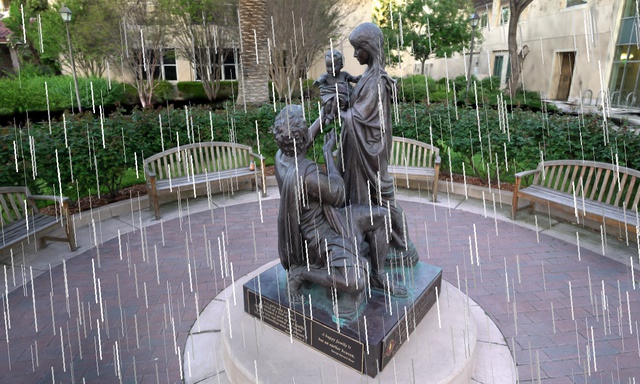}{(.29\imagewidth,.12\imagewidth)}{(.22\imagewidth,.46\imagewidth)} & 
        \formattedgraphics{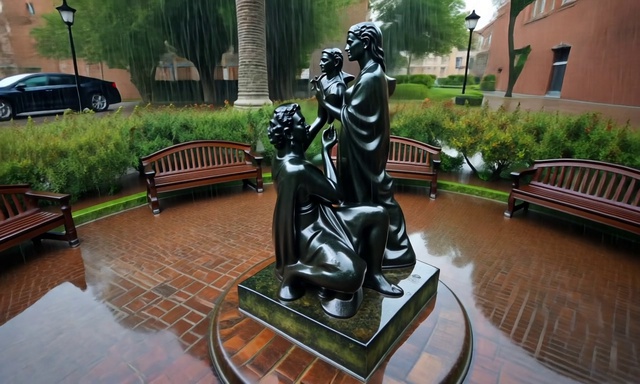}{(.29\imagewidth,.12\imagewidth)}{(.22\imagewidth,.46\imagewidth)}  & 
        \formattedgraphics{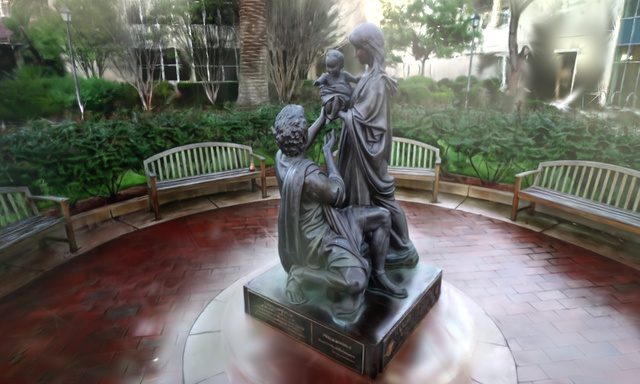}{(.29\imagewidth,.12\imagewidth)}{(.22\imagewidth,.46\imagewidth)}  & 
        \formattedgraphics{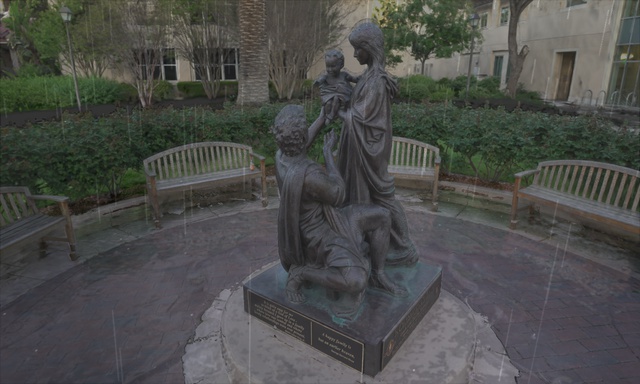}{(.29\imagewidth,.12\imagewidth)}{(.22\imagewidth,.46\imagewidth)} \\
    \end{tabular}
    \caption{\label{fig:fixed_time_family} \textbf{Family Scenario.} Rain synthesis outputs from several viewpoints at the same moment in time. Runway-V2V struggles to preserve 3D consistency, while Rain Motion and Instruct-GS2GS are unable to generate realistic rain streaks, puddles, and ripples. In contrast, RainyGS preserves 3D consistency and produces visually convincing rain streaks and water accumulation.}
    \Description{figure}
    \vspace*{1cm}
\end{figure*}
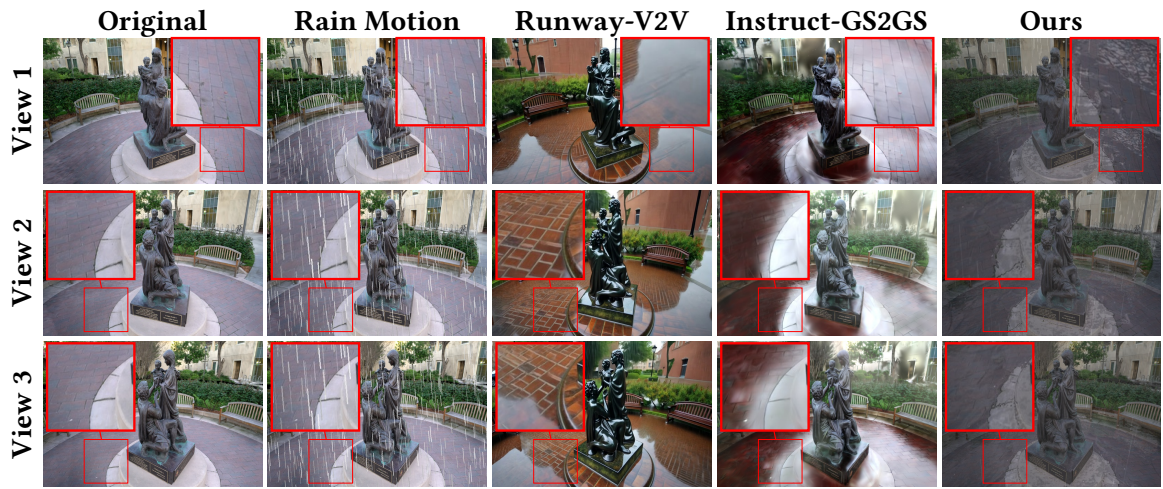

\begin{figure*}
    \centering
    \setlength{\tabcolsep}{1pt}
    \setlength{\imagewidth}{0.19\textwidth}
    \setlength{\imageheight}{0.087\textheight}
    \renewcommand{\arraystretch}{0.6}
    \newcommand{\formattedgraphics}[1]{%
      \begin{tikzpicture}[spy using outlines={rectangle, magnification=2, connect spies}]
        \node[anchor=south west, inner sep=0] at (0,0){\includegraphics[width=\imagewidth, height=\imageheight]{#1}};
        \spy [red,size=30pt] on (.29\imagewidth,.12\imagewidth) in node at (.21\imagewidth,.46\imagewidth);
        \end{tikzpicture}%
    }
    \begin{tabular}{m{0.38cm}<{\centering}m{\imagewidth}<{\centering}m{\imagewidth}<{\centering}m{\imagewidth}<{\centering}m{\imagewidth}<{\centering}m{\imagewidth}<{\centering}}
        & $t=\SI{0}{\second}$ & $t=\SI{0.24}{\second}$ & $t=\SI{0.48}{\second}$ & $t=\SI{0.72}{\second}$ & $t=\SI{0.96}{\second}$\\
        \rotatebox{90}{\textbf{Rain Motion}} &
        \formattedgraphics{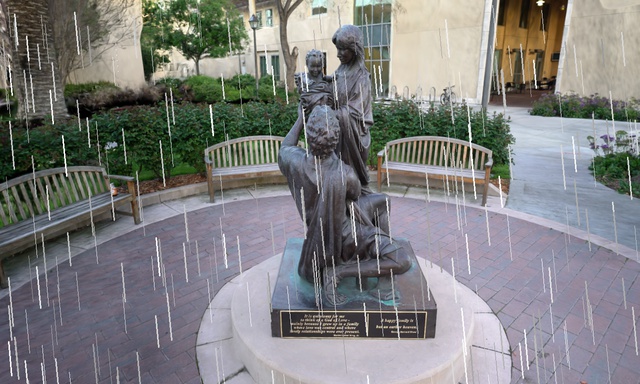} & 
        \formattedgraphics{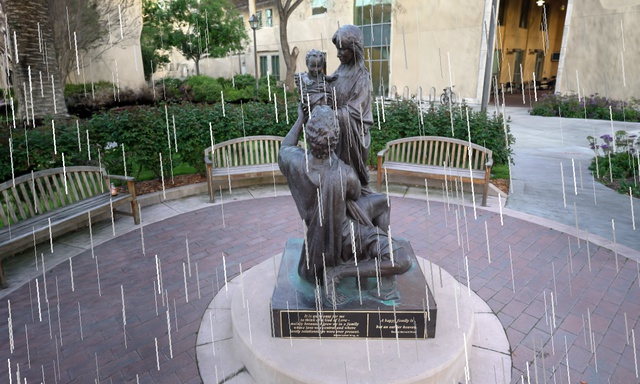} & 
        \formattedgraphics{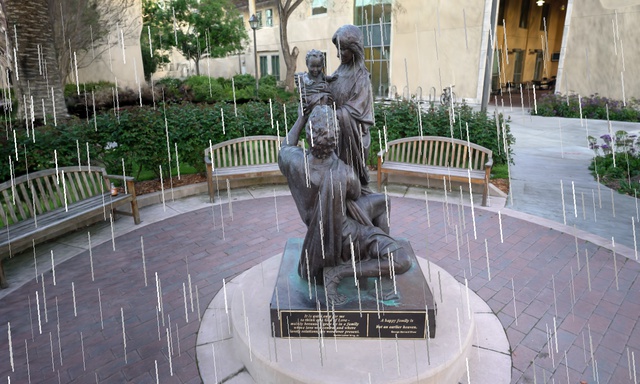} & 
        \formattedgraphics{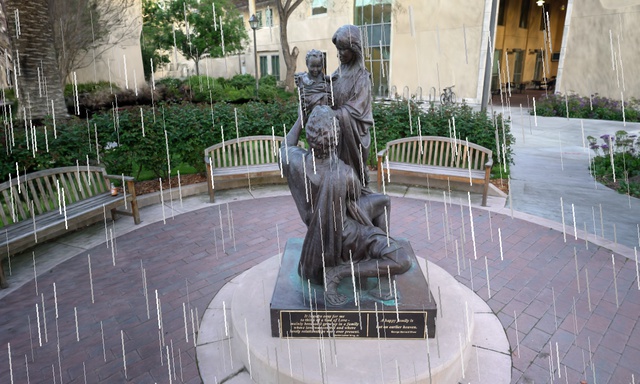} & 
        \formattedgraphics{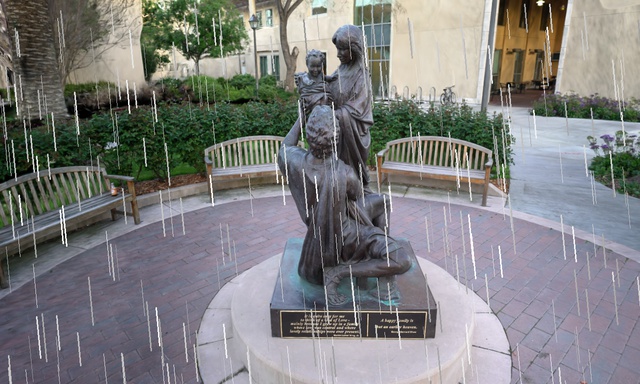}\\
        \rotatebox{90}{\textbf{Runway-V2V}} & 
        \formattedgraphics{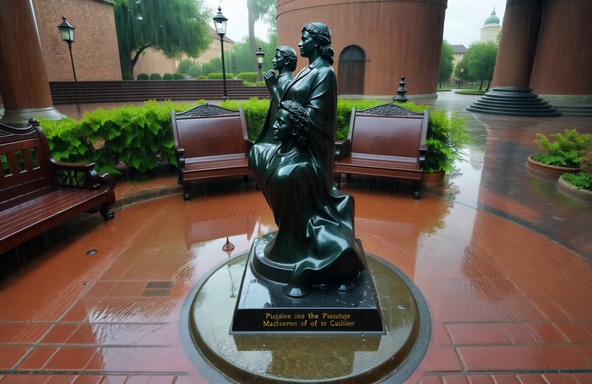} & 
        \formattedgraphics{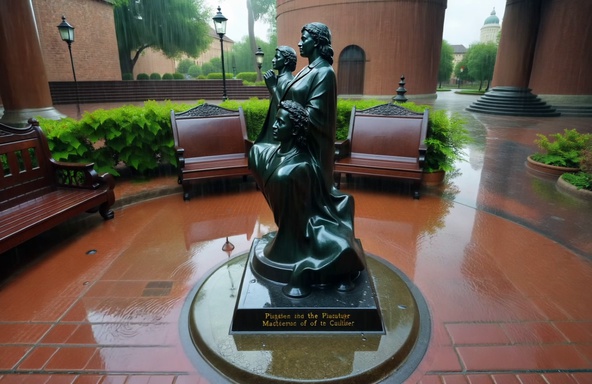} & 
        \formattedgraphics{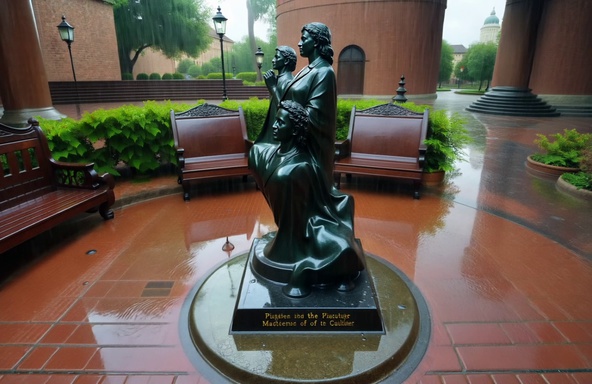} & 
        \formattedgraphics{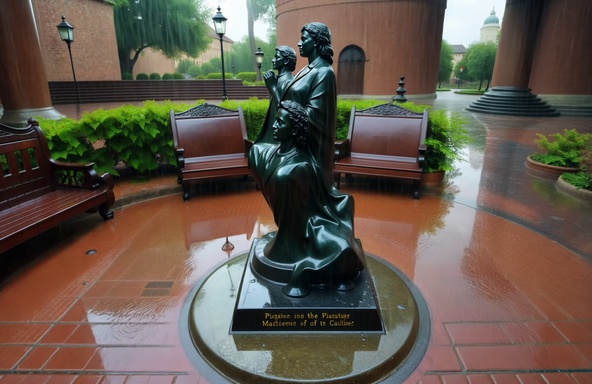} & 
        \formattedgraphics{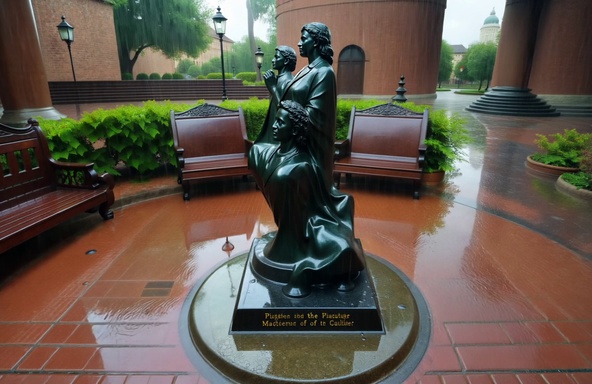}\\
        \rotatebox{90}{\textbf{Ours}} &
        \formattedgraphics{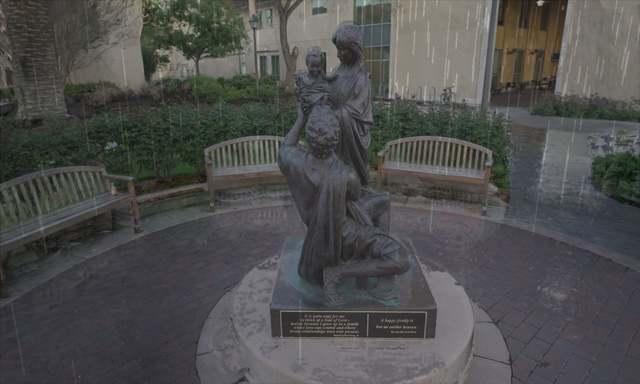} & 
        \formattedgraphics{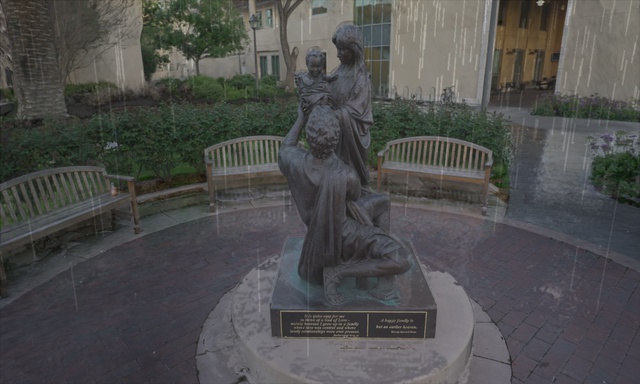} & 
        \formattedgraphics{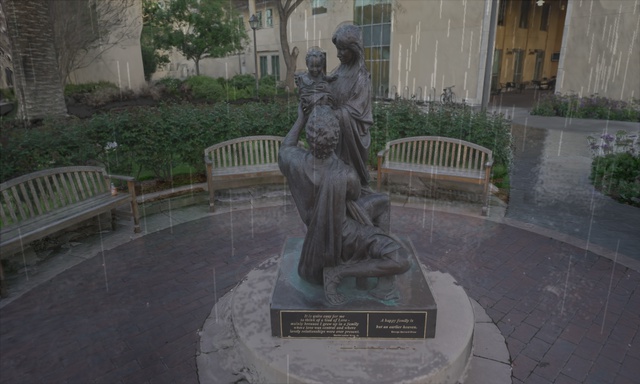} & 
        \formattedgraphics{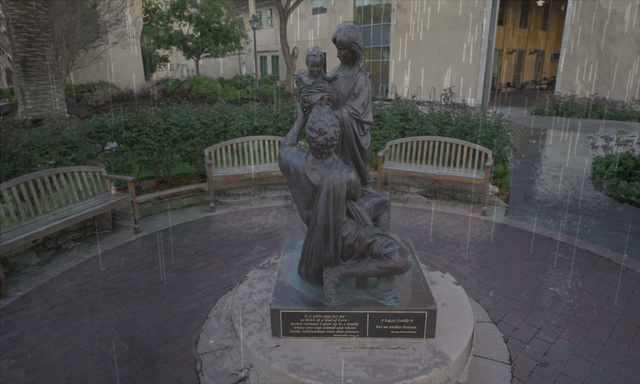} & 
        \formattedgraphics{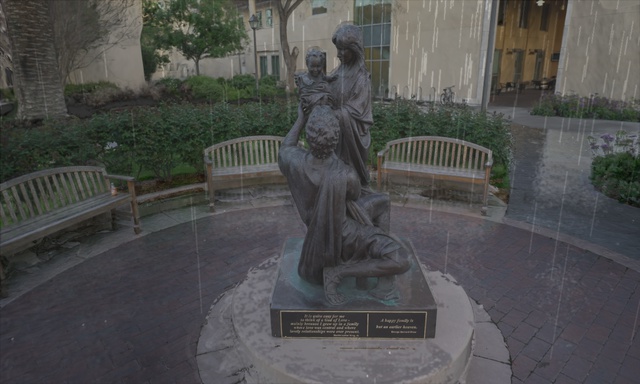}\\
    \end{tabular}
    \caption{\label{fig:fixed_view_family}
    \Description{figure}
    \textbf{Family Scenario.}
Shown are rain synthesis results at different timesteps from the same viewpoint. Rain Motion produces random, unrealistic rain droplets and lacks hydrops, while Runway-V2V yields ripples that lack physical correctness and visual authenticity. RainyGS, however, simulates realistic hydrops and puddles that evolve over time.    }
    \vspace*{1cm}
\end{figure*}

\section{More Results}
\label{sec:supp_exp}

\subsection{More Results of Comparison with Video-Based Rain Synthesis}

%\subsubsection{Qualitative Results}

We evaluate our method on the Garden, Treehill, and Bicycle scenes from the MipNeRF360 dataset, as well as the Family and Truck scenes from the Tanks and Temples dataset. For comparison, we used two baselines: Rain Motion~\cite{wang2022rethinking}, a video rain synthesis method; Runway-V2V~\cite{runway2024gen3alpha, runway2024gen3alphavideo}, a state-of-the-art commercial video-to-video generation model; and Instruct-GS2GS~\cite{igs2gs}, a recent text-driven 3DGS editing method. For Runway-V2V, we use the prompt: \textit{Raining in the scene, with water puddles on the ground, reflections of scenes and lights on wet pavement, and raindrops creating ripples on the surface, under an overcast sky}. In addition to the Garden scene presented in Fig. 5 and Fig. 6 of the main text, we further show the Bicycle, Treehill, Truck, and Family scenes in Fig.~\ref{fig:fixed_time_bicycle} - Fig.~\ref{fig:fixed_view_family}.

The results generated by Runway-V2V demonstrate a general rainy visual effect. Nevertheless, its output exhibits several shortcomings. First, the appearance and geometry of scene objects are often altered; for example, in Fig.~\ref{fig:fixed_view_bicycle}, the bench changes from dark brown to light brown, and the bicycle shifts from white to black. Second, the generated videos suffer from oversaturated colors. In addition, Runway-V2V lacks physical accuracy, resulting in artifacts such as unrealistic ripples (e.g., Fig.~\ref{fig:fixed_view_treehill}) and incorrect reflections (e.g., the wheel in Fig.~\ref{fig:fixed_time_truck}).
By contrast, Rain Motion is limited to applying simple 2D rain effects on video frames, without 3D multiview consistency or style adaptation. This leads to inferior visual fidelity and poor physical realism.
Instruct-GS2GS also falls short, as it lacks advanced rain phenomena such as water accumulation, rain streaks, and dynamic reflections, and is restricted to static edits.

In contrast, our method consistently outperforms all baselines by generating highly realistic rain effects while preserving physical plausibility, including fluid dynamics (e.g., Fig.~\ref{fig:fixed_time_bicycle}), accurate reflections (e.g., Fig.~\ref{fig:fixed_time_truck}), and refractions (e.g., Fig.~\ref{fig:fixed_time_family}). Moreover, it maintains multi-view consistency and ensures temporally coherent dynamics.

%\subsubsection{User Studies}
\begin{figure}[b]\footnotesize
  \vspace{-1em}
  \centering
  \setlength{\imagewidth}{.495\linewidth}
  \newcommand{\formattedgraphics}[1]{%
      \includegraphics[width=\imagewidth]{#1}%
    }
  \subcaptionbox{\centering  Light rain starts [low]}{\formattedgraphics{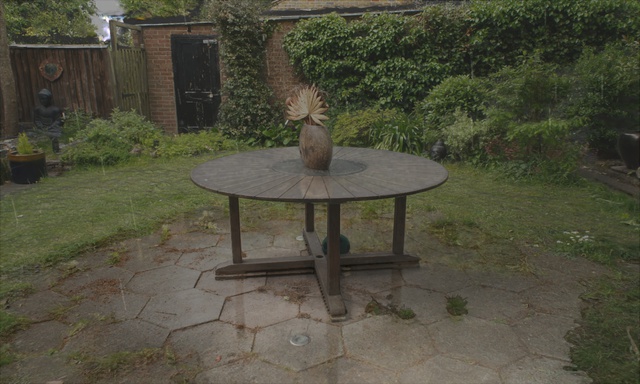}}
  \hfill
  \subcaptionbox{\centering Moderate~rain [medium] }{\formattedgraphics{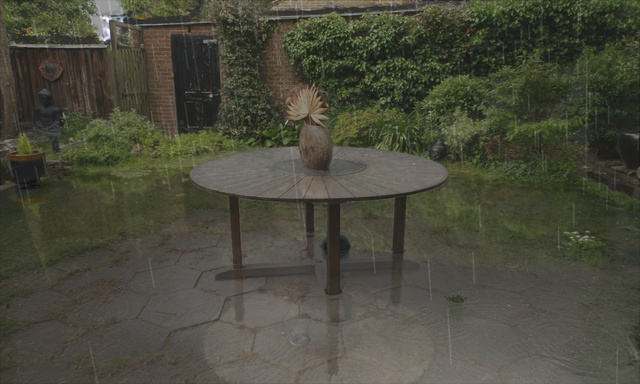}}
  \\
  \subcaptionbox{\centering Heavy rain [high]}{\formattedgraphics{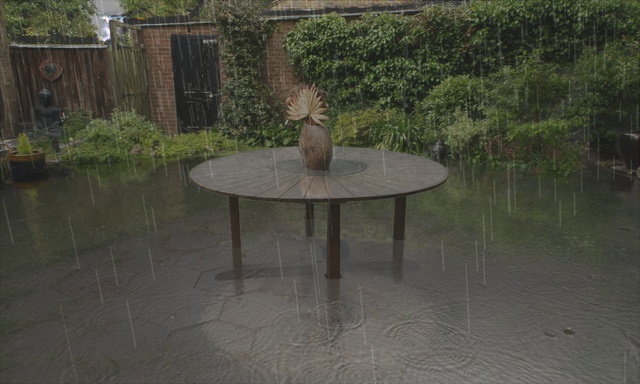}}
  \hfill
  \subcaptionbox{\centering Rain ends [high]\label{fig:r3_raining_sunny}}{\formattedgraphics{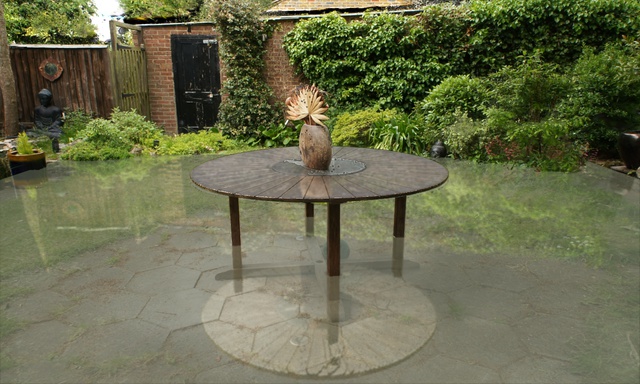}}
  \\
  \vspace{-1em}
  \caption{\small
  Rain progression from start to end, showing the control of rain intensity, water volume (indicates with []), and lighting.
  }
  \label{fig:r3_raining}
  \vspace{-1em}
\end{figure}

\subsection{User Study}

Tab.~\ref{table:user_study} summarizes the results of a user study conducted on Amazon Mechanical Turk, designed to evaluate the perceptual quality of our method compared to existing baselines. A total of 54 participants were recruited, each asked to judge 30 randomly selected examples consisting of 15 images and 15 videos. For each example, users indicated their preferred result in a pairwise comparison setting. The results show a clear and consistent preference for our method across both image and video modalities, highlighting its effectiveness and perceptual superiority over competing approaches.

% Tab.~\ref{table:user_study} shows an Amazon Mechanical Turk user study, involving 54 users, each judging 30 examples (15 images + 15 videos). Results demonstrate a consistent preference of our method. % over baselines 

% \begin{table}
% \centering
% \caption{User Study Results (\% Ours is preferred).}
% \label{table:user_study}
% \vspace{-0.5em}
% \renewcommand{\arraystretch}{1.2} % 增加行间距
% \begin{tabular}{l|c|c}
%     \hline
%     & \textbf{Image} & \textbf{Video} \\
%     \hline
%     vs Rain Motion & 70.9 & 67.5\\
%     vs Runway-V2V & 70.7 & 72.0\\
%     vs Instruct-GS2GS & 65.2 & 65.7\\
%     \hline
% \end{tabular}
% \vspace{-0.5em}
% \end{table}
\begin{table}[t]\footnotesize
\centering
\caption{User Study Results on Amazon Mechanical Turk, reporting the percentage (\%) of cases where users preferred our method over each baseline for both image and video comparisons.}
\label{table:user_study}
\setlength{\tabcolsep}{8pt}
\renewcommand{\arraystretch}{1.2}
\begin{tabular}{lcc}
\specialrule{.15em}{.1em}{.1em}
& \textbf{Image} & \textbf{Video} \\
\hline
vs Rain Motion & 70.9 & 67.5 \\ \hline
vs Runway-V2V & 70.7 & 72.0 \\ \hline
vs Instruct-GS2GS & 65.2 & 65.7 \\ 
\hline
\specialrule{.1em}{.05em}{.05em}
\end{tabular}
\vspace*{-0.3cm}
\end{table}

\subsection{Additional Results of Precise User Control}

RainyGS supports precise, physics-based user control over key parameters, including rain intensity, water level, and scene brightness, allowing flexible adjustment to meet various simulation needs.
As illustrated in Fig.~\ref{fig:r3_raining}, we present the rain progression over time on MipNeRF360-Garden, showcasing dynamic variations in rain intensity, gradual water accumulation, and changing lighting conditions, all rendered consistently from the same scene and camera viewpoint.

\subsection{More Results and Analysis of Performance}

Our experiments are performed on a platform with a Nvidia Geforce RTX 3090 Graphics Card, whose memory capacity is 24\,GB.
Here we report the computational statistics for the results given in the main text, as well as those provided in Fig.\ref{fig:fixed_time_bicycle} - Fig.~\ref{fig:fixed_view_family}, shown in Tab.~\ref{table:time_cost}.
Generally speaking, our method, namely RainyGS, takes the most of time in simulation and transferring data, which can be pre-computed.
As to the rendering stage, the speed is around 30\,FPS. The memory usage is roughly linear to the number of Gaussians, which fits the capacity of personal computers.

% \begin{table}[t]\footnotesize
% \centering
% \caption{The computational statistics for different scenes. The time here is measured per frame averagely. Here, the simulation time includes that cost for data transferring from height maps to Gaussians.}  
% \setlength{\tabcolsep}{3pt}
% \begin{tabular}{cccccc}
% \specialrule{.15em}{.1em}{.1em}
% \textbf{Scenes} &\textbf{\# Gaussians} &
% %\textbf{Precomp. Time/Frame} &
% \textbf{Simulation Time} & \textbf{Rendering Time} & \textbf{Peak Video Memory}  \\
% \hline
% Garden & \SI{5.120}{M} &
% %\SI{0}{\second} &
% \SI{0.341}{\second} & \SI{0.092}{\second} & 8.088\,GB\\
% \hline
% Truck & \SI{5.337}{M} &
% %\SI{0}{\second} &
% \SI{0.237}{\second} & \SI{0.104}{\second} & 10.874\,GB\\
% \hline
% Treehill & \SI{8.155}{M} &
% %\SI{0}{\second} &
% \SI{0.133}{\second} & \SI{0.115}{\second} & 7.372\,GB\\
% \hline
% Family & \SI{7.151}{M} &
% %\SI{0}{\second} &
% \SI{0.357}{\second} & \SI{0.103}{\second} & 9.626\,GB\\
% \hline
% Bicycle & \SI{10.393}{M} &
% %\SI{0}{\second} &
% \SI{0.206}{\second} & \SI{0.134}{\second} & 12.394\,GB\\
% \hline
% \specialrule{.1em}{.05em}{.05em}
% \end{tabular}
% \vspace*{-0.3cm}
% \label{table:time_cost}
% \end{table} 

\begin{table}[t]\footnotesize
\centering
\caption{The computational statistics for different scenes. The time here is measured per frame averagely. Here, the simulation time includes that cost for data transferring from height maps to Gaussians.}  
\setlength{\tabcolsep}{3pt}
\begin{tabular}{cccccc}
\specialrule{.15em}{.1em}{.1em}
\textbf{Scenes} &\textbf{\# Gaussians} &
%\textbf{Precomp. Time/Frame} &
\textbf{Precomputed Time} & \textbf{Rendering Time} & \textbf{Peak Video Memory}  \\
\hline
Garden & \SI{4.290}{M} &
%\SI{0}{\second} &
\SI{0.013}{\second} & \SI{0.032}{\second} & 8.561\,GB\\
\hline
Truck & \SI{4.368}{M} &
%\SI{0}{\second} &
\SI{0.022}{\second} & \SI{0.035}{\second} & 9.044\,GB\\
\hline
Treehill & \SI{8.169}{M} &
%\SI{0}{\second} &
\SI{0.049}{\second} & \SI{0.039}{\second} & 18.049\,GB\\
\hline
Family & \SI{8.524}{M} &
%\SI{0}{\second} &
\SI{0.050}{\second} & \SI{0.039}{\second} & 19.166\,GB\\
\hline
Bicycle & \SI{10.687}{M} &
%\SI{0}{\second} &
\SI{0.053}{\second} & \SI{0.041}{\second} & 23.511\,GB\\
\hline
\specialrule{.1em}{.05em}{.05em}
\end{tabular}
\vspace*{-0.3cm}
\label{table:time_cost}
\end{table} 

\begin{figure*}
    \centering
    \setlength{\tabcolsep}{1pt}
    \setlength{\imagewidth}{0.30\textwidth}
    \renewcommand{\arraystretch}{0.6}
    \newcommand{\formattedgraphics}[1]{%
      \begin{tikzpicture}[spy using outlines={rectangle, magnification=2, connect spies}]
        \node[anchor=south west, inner sep=0] at (0,0){\includegraphics[width=\imagewidth]{#1}};
        \end{tikzpicture}%
    }
    \begin{tabular}{m{0.38cm}<{\centering}m{\imagewidth}<{\centering}m{\imagewidth}<{\centering}m{\imagewidth}<{\centering}m{\imagewidth}<{\centering}m{\imagewidth}<{\centering}}
        & $t_0$ & $t_1$ & $t_2$ \\
        \rotatebox{90}{\textbf{Inputs}} &
        \formattedgraphics{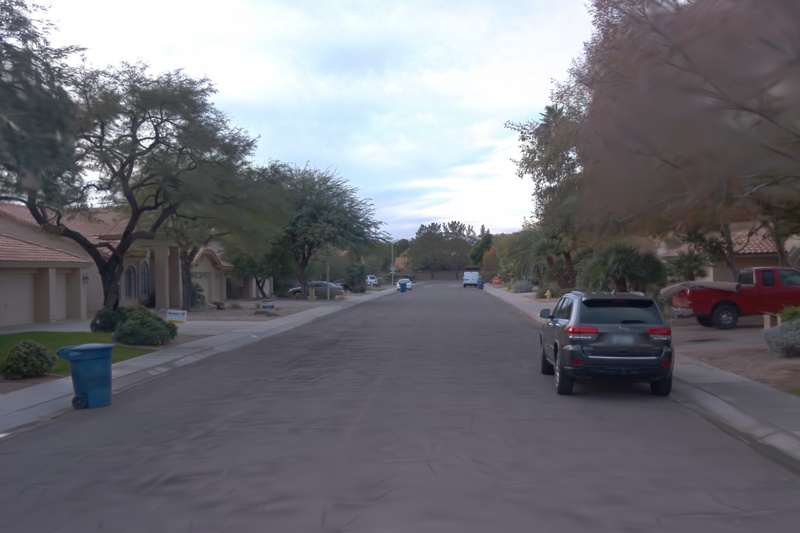} & 
        \formattedgraphics{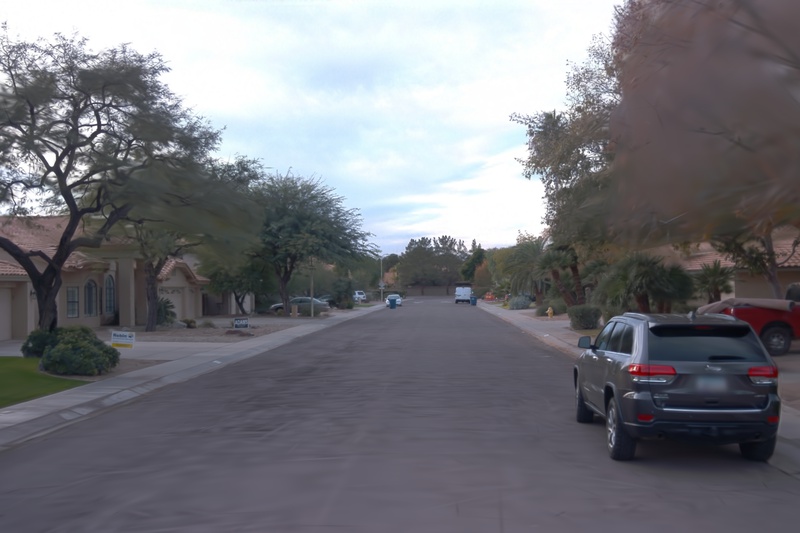} & 
        \formattedgraphics{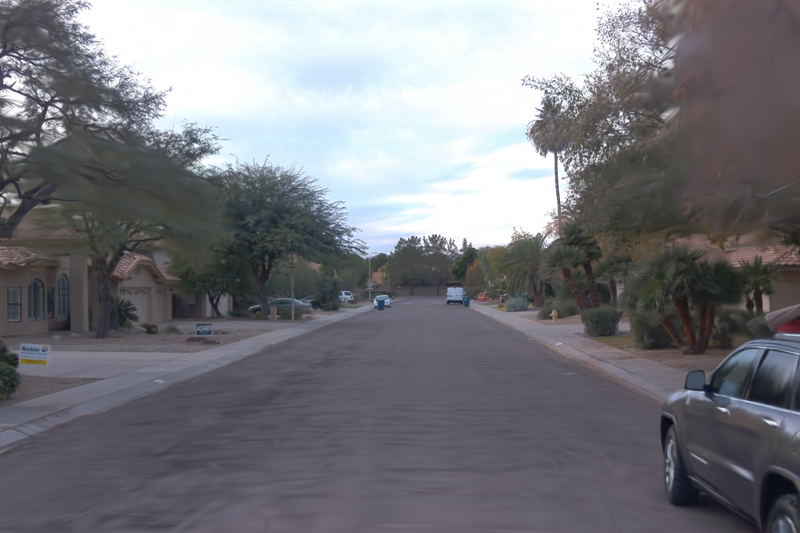}\\
        \rotatebox{90}{\textbf{Rainy}} & 
        \formattedgraphics{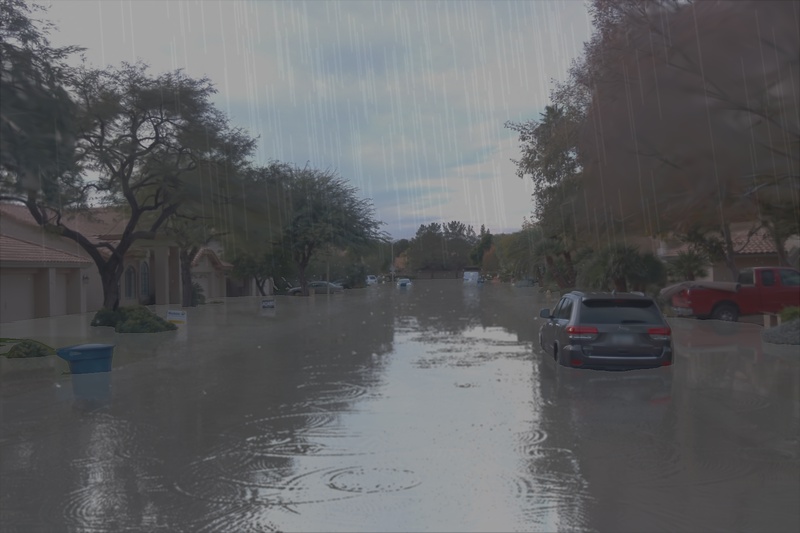} & 
        \formattedgraphics{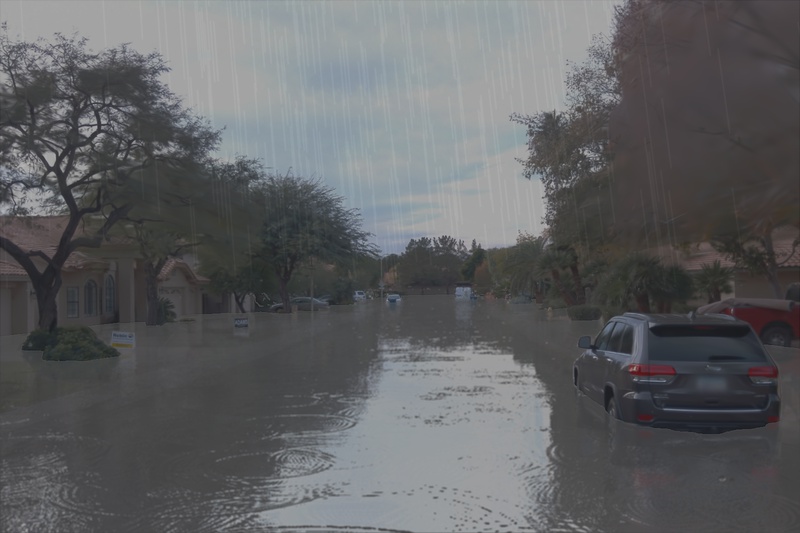} & 
        \formattedgraphics{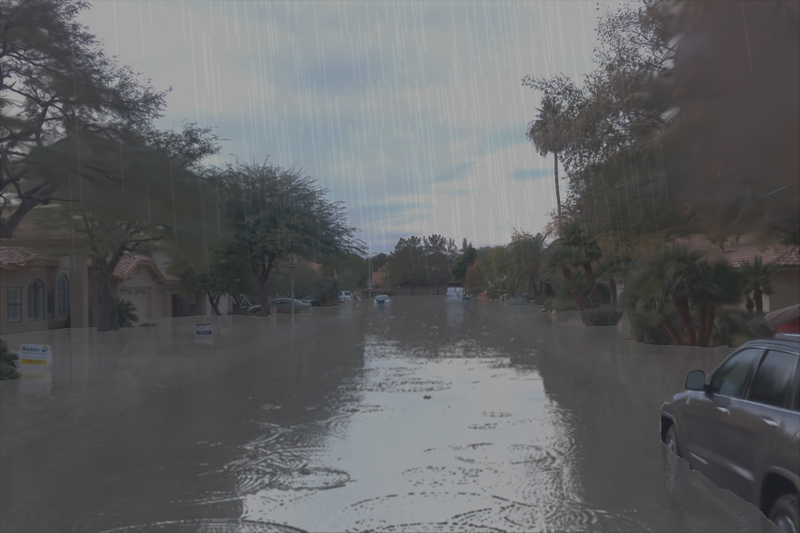}\\
    \end{tabular}
    \caption{\label{fig:waymo}
    Downstream Application: Rainy Weather Synthesis for Waymo Scenes.
    Our method facilitates the efficient transformation of continuous autonomous driving scenes into rainy weather conditions, enabling robustness testing of autonomous driving perception algorithms. Additionally, it offers a novel framework for synthesizing adverse weather data in autonomous driving.
    }
    \Description{figure}
    \vspace*{-0.3cm}
\end{figure*}
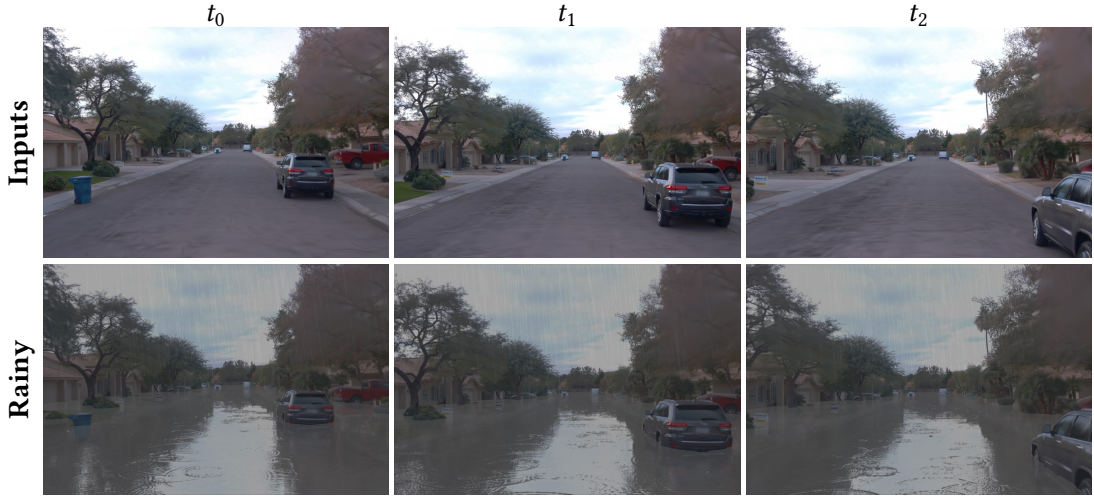

\subsection{Additional Results of Downstream Tasks}
As shown in Fig. ~\ref{fig:waymo}, RainyGS can be utilized for downstream tasks in autonomous driving. For instance, it efficiently transforms continuous driving scenarios, such as Waymo scenes, into rainy weather conditions. This capability facilitates robustness testing of detection algorithms and provides a new method for simulating adverse weather and generating synthetic data for autonomous driving research.
% We observe certain artifacts on the road, which can be attributed to the suboptimal surface reconstruction of the road. Addressing these issues will be left as part of future work.

%% file: sec_supp/X_suppl.tex
\section{Details of Auxiliary Map Preparation}
\label{sec:supp_auxmap}

\section{Details of Rain Simulation on Height Maps}
\label{sec:supp_rainsim}

\section{Details of Screen-space Reflection}
\label{sec:supp_ssr}

\section{More Results}
\label{sec:supp_exp}

\subsection{More Results of Comparison with Video-Based Rain Synthesis}

\subsubsection{Qualitative Results}

\subsubsection{User Studies}

\subsubsection{Quantitative Results of Video Quality Assessment Model}

\subsection{More Results and Analysis of Performance}

\subsection{Additional Results of Downstream Tasks}

\section{Failure Cases}
\label{sec:supp_failure}